\documentclass[11pt,a4paper]{article}
\pdfoutput=1   % required for arXiv submission if pdflatex is used
\usepackage{amsmath,amsfonts,amssymb,amsbsy}
\usepackage{mathrsfs,latexsym}
\usepackage{graphicx}
\usepackage{color}
\usepackage{booktabs}
\usepackage{multirow}
\usepackage{multicol}
\usepackage{subfig}
\usepackage{placeins}
\usepackage{alpha}
\usepackage{macros_alpha}
\renewcommand{\strange}{\mathrm{strange}}
\usebiblio{latticen}
\newcommand{\cm}{{\mathrm {cm}}}
\begin{document}

%\preprintno{%
%TCDMATH 16-XX\\
%}

\title{%
	 $I=1$ and $I=2$ $\pi-\pi$ scattering 
	phase shifts from $\nf = 2+1$ lattice QCD  
}

%\collaboration{\includegraphics[width=2.8cm]{alpha_color_60}}

\author[trin]{John~Bulava}
\author[kek]{Brendan~Fahy}
\author[trin]{Ben~H\"{o}rz}
\author[pac]{Keisuke~J.~Juge}
\author[cmu]{Colin~Morningstar}
\author[wupp]{Chik~Him~Wong}

\address[trin]{School~of~Mathematics, Trinity~College, Dublin~2, Ireland}
\address[kek]{High Energy Accelerator Research Organization (KEK), Ibaraki 305-0801, Japan}
\address[pac]{Department~of~Physics, University~of~the~Pacific, Stockton, CA~95211, USA}
\address[cmu]{Department~of~Physics, Carnegie~Mellon~University, 
Pittsburgh, PA~15213, USA}
\address[wupp]{Department~of~Physics, University~of~Wuppertal, Gaussstrasse~20,
D-42119, Germany}

\begin{abstract}
	The $I=1$ $p$-wave and $I=2$ $s$-wave elastic $\pi$-$\pi$ scattering 
	amplitudes are 
calculated from a first-principles lattice QCD simulation using a single 
ensemble of gauge field configurations with $\nf = 2+1$ 
dynamical flavors of anisotropic clover-improved Wilson fermions. This ensemble has a large 
spatial volume  
$V=(3.7\mathrm{fm})^3$, pion mass 
$m_{\pi} = 230\mathrm{MeV}$, and spatial lattice spacing 
$a_s = 0.11\mathrm{fm}$. 
Calculation of the necessary temporal correlation matrices is 
efficiently performed using the stochastic LapH method, while the large volume 
enables an improved energy resolution compared to previous work.  
For this single ensemble we obtain $m_{\rho}/m_{\pi} = 3.350(24)$, $g_{\rho\pi\pi} = 5.99(26)$, and a clear signal for the $I=2$ $s$-wave. The success of 
	the stochastic LapH method in this proof-of-principle large-volume 
	calculation paves the way for quantitative study of the  
	lattice spacing effects and quark mass dependence of scattering
	amplitudes using state-of-the-art ensembles. 
\end{abstract}


\maketitle

\section{Introduction} 

Hadron-hadron scattering amplitudes are of central importance in the 
phenomenology of QCD and confining scenarios of Beyond-the-Standard Model (BSM)
physics. While Euclidean lattice gauge simulations are a 
proven first-principles approach for these theories,  
the calculation of hadron-hadron scattering on the lattice has long 
been a challenge. First and foremost, the Maiani-Testa No-Go Theorem demonstrates that 
on-shell amplitudes cannot (in general) be directly obtained from 
Euclidean space matrix elements~\cite{Maiani:1990ca}. This 
difficulty was overcome by L\"{u}scher's  
relation  between elastic scattering 
phase shifts and the deviation of finite-volume two-hadron energy spectra 
from their non-interacting values~\cite{Luscher:1990ux}. 

While this relation has been known since the early 90's, only recently are  
lattice QCD calculations of scattering amplitudes starting to have sufficient 
statistical precision and energy resolution to clearly identify resonance 
features. This delay is mostly due to the difficulty in precisely calculating 
temporal correlation functions
\begin{align}\label{e:cor}
	C_{ij}(t-t_0) = \langle \mathcal{O}_{i}(t) \bar{\mathcal{O}}_j(t_0) \rangle =
	\sum_{n} A_{ni}A^{*}_{nj} \mathrm{e}^{-E_n (t-t_0)},
\end{align}
where $\hat{\mathcal{O}}_i$ and $\hat{\mathcal{O}}_j$ are suitable 
interpolating operators with the quantum numbers of interest and the sum is over all finite-volume energy eigenstates. After calculating such correlation functions on a gauge field
ensemble, the finite-volume energies $\{E_n\}$ are extracted from their 
temporal fall-off.  

To obtain finite-volume two-hadron energies, 
correlation functions between two-hadron interpolating operators are 
required. These two-hadron correlation functions in turn typically
require the evaluation of valence-quark-line-disconnected Wick contractions\footnote{`Disconnected' Wick contractions  
are those in which quark fields at the same time are contracted.} and   
contain interpolating operators which annihilate states with definite momentum. 
After integration over the Grassmann-valued quark fields, this requires
the quark propagator from all space-time points to all 
space-time points. As the quark propagator is the inverse of the 
large-dimension and ill-conditioned Dirac matrix, these `all-to-all' propagators (and 
thus multi-hadron correlation functions) are 
naively intractable. Inversion of the Dirac matrix $M$ is performed by solving 
the linear system $M\phi = \eta$ for multiple right-hand sides and is
typically the dominant cost in calculating fermionic correlation 
functions. The solution of this system for each spacetime point is not
feasible, preventing the naive approach to all-to-all quark propagators.

However, substantial progress has been made by treating quark propagation only 
in the subspace spanned by the lowest-lying eigenmodes of the 
three-dimensional gauge-covariant Laplace operator~\cite{Peardon:2009gh}.
Apart from facilitating the evaluation of these correlation functions, this 
`distillation' procedure has the added benefit of reducing the contamination 
of unwanted excited states. It can thus be viewed as a form of `quark smearing', a 
common procedure used in lattice QCD to reduce the contribution  
of higher terms in Eq.~\ref{e:cor} by suppressing their overlaps. The spatial profile of this smearing 
wavefunction is approximately gaussian with a width controlled by the number 
of low-lying Laplacian eigenmodes retained in the projection ($N_v$). 

The cutoff 
eigenvalue therefore defines the smearing wavefunction and must be 
fixed in physical units. 
Unfortunately, if the cutoff eigenvalue is held fixed the number of eigenmodes 
in this subspace increases proportionally to the spatial volume. The
distillation approach requires a number of Dirac matrix inversions $N_D \propto N_v$ which results in an  
unfavorable volume scaling, hindering the application 
of this method to large physical volumes. Nonetheless, it has been applied 
successfully in smaller volumes~\cite{Wilson:2014cna,Lang:2014yfa,
Prelovsek:2013ela,Lang:2011mn,Dudek:2012xn,Dudek:2012gj,Lang:2012db}. 

Based on this idea, the stochastic LapH method was proposed in 
Ref.~\cite{Morningstar:2011ka} and achieves an improved scaling with the 
physical volume by introducing stochastic estimators in the low-dimensional 
subspace spanned by the Laplacian eigenmodes. The variance of these 
stochastic estimators is reduced by `dilution'~\cite{Foley:2005ac}, which 
partitions the space using $N_{\mathrm{dil}}$ complete, orthogonal projectors. 
Ref.~\cite{Morningstar:2011ka} demonstrates that the efficiency of these 
modified stochastic estimators remains constant for fixed (sufficiently 
large) $N_{\mathrm{dil}}$ as the spatial volume is increased. Since 
$N_D \propto N_{\mathrm{dil}}$ in this approach, the volume scaling is significantly 
improved. 
This scaling is tested further in this work by applying the 
stochastic LapH method for the first time to lattices with  
$L=  3.7\mathrm{fm}$, while it has been successful in smaller 
volumes~\cite{Lang:2014yfa,Helmes:2015gla}. Although the stochastic LapH 
method was designed to enable exploratory calculations of finite-volume 
spectra, we demonstrate here that it can resolve
these energies with a sufficient precision to determine elastic scattering 
phase shifts in a large spatial volume.

As a first large-volume application we treat $\pi-\pi$ scattering in the $I=1$ 
and $I=2$ channels. The lowest-lying hadronic resonance, the $\rho$-meson, 
occurs in the $\ell=1$ partial wave 
of the $I=1$ channel, resulting in significant shifts of finite-volume 
energies from their non-interacting values. In contrast 
the $I=2$, $\ell=0$ partial  
wave is considerably more weakly interacting and well-described by the 
effective range expansion. Therefore, this channel presents a 
more stringent test of the stochastic LapH method as deviations from 
non-interacting energies are generally much smaller. For example, in large 
volume the 
difference between the ground-state energy in the $I=2$ $A_{1g}^{+}$ channel (relevant for the $\ell=0$ partial wave) 
and $2m_{\pi}$ is given by
\begin{align}
	\Delta E = E_{2\pi} - 2\mpi = -\frac{4\pi a_0}{m_{\pi}L^3} + 
	\mathrm{O}(L^{-4}) 
\end{align}
where $a_0$ is the $I=2$ $s$-wave scattering length. Although additional 
statistics are accrued by summing over a large spatial 
volume, the signal in this channel also decreases with the spatial volume, 
complicating the determination of $a_0$. 

Although these two systems are benchmark tests of the efficacy of our methods,
they are  
also interesting in their own right. The quark mass 
dependence of the $\rho$-resonance pole position is an 
important input to Unitarized Chiral Perturbation Theory (see e.g. Refs.~\cite{Hanhart:2014ssa,Bolton:2015psa}) while 
the $\ell=0$ scattering length in the $I=2$ channel provides another important 
test of Chiral Perturbation Theory. 

This work is part of an ongoing effort to investigate the low-lying resonance 
spectrum of QCD. 
Preliminary work using only single-hadron interpolating operators has been 
reported in 
Refs.~\cite{Basak:2005aq,Bulava:2009jb,Bulava:2010yg} 
while development of the all-to-all propagator algorithms discussed above is detailed in 
Refs.~\cite{Peardon:2009gh,Morningstar:2011ka}. First results with multi-hadron operators are given in 
Ref.~\cite{Morningstar:2013bda} and a preliminary account of the results shown here is given in 
Refs.~\cite{Fahy:2014jxa,Bulava:2015qjz}.

During the preparation of this manuscript, a calculation of the $I=1$ $p$-wave 
scattering phase shift appeared~\cite{Wilson:2015dqa} using the same ensemble 
of gauge configurations. Rather than stochastic LapH, 
Ref.~\cite{Wilson:2015dqa} employs the full distillation method of Ref.~\cite{Peardon:2009gh}. 
Comparison of results and computational cost with Ref.~\cite{Wilson:2015dqa} is made in  Sec.~\ref{s:concl}. 

\newpage
The main results of this work are Figs.~\ref{f:i1scat} and~\ref{f:i2scat}, which show the $I=1$ $p$-wave and $I=2$ $s$-wave scattering 
phase shifts (respectively) as well as Eqs.~\ref{e:rhofit} and~\ref{e:i2fit}, which describe fits to those scattering phase shifts. 
Our methodology is described in Sec.~\ref{s:meth}, which provides details of the 
gauge field ensemble, the stochastic LapH method discussed above, 
our procedure for extracting finite-volume energies from temporal correlation functions, and the L\"{u}scher method for obtaining 
scattering phase shifts from those energies. Finally, results are described in 
Sec.~\ref{s:res} while conclusions and a comparison with previous work are in 
Sec.~\ref{s:concl}. Additional details concerning the determination of 
finite-volume energies are relegated to an appendix.

\section{Methodology}\label{s:meth}

 Here we detail technical aspects of the methods used in this work. 
 For this exploratory 
large-volume calculation, an anisotropic lattice regularization is employed 
to achieve a large spatial volume and a good temporal resolution at moderate 
computational cost. On this anisotropic lattice the ratio of the spatial and temporal lattice spacings (the renormalized anisotropy) appears in the pion dispersion relation 
and must be determined precisely. 
The required temporal correlation matrices are measured on these gauge configurations using the stochastic 
LapH method, while ground and low-lying excited-state energies are extracted from 
them using solutions of a generalized eigenvalue problem (GEVP). Finally, 
these energies are used in L\"{u}scher formulae to obtain elastic 
scattering phase shifts. 

\subsection{Ensemble details}\label{s:ens}

In order to suppress unwanted (exponential) finite-volume effects in lattice 
QCD simulations with light pions, large spatial volumes are required. These 
large volumes also increase the density of states in two-hadron channels, 
improving the energy resolution of scattering phase shifts.  
A large \emph{temporal} extent is additionally required to suppress thermal effects in 
correlation functions with periodic temporal boundary conditions. Finally, a 
good temporal resolution is needed to accurately extract 
finite-volume energies from the fall-off of temporal correlation functions. 

In order to satisfy these requirements with a moderate computational cost, 
we employ an anisotropic lattice regularization in which the spatial and 
temporal lattice spacings differ. 
Our ensemble of gauge configurations is covered in detail in 
Ref.~\cite{Lin:2008pr} and reviewed here briefly. Basic details are listed in 
Tab.~\ref{t:ens}, where the temporal lattice spacing ($a_t$) is 
determined 
\begin{table}
\centering
\begin{tabular}{|c|c|c|c|c|c|c|c|}
		\hline
		$(L/a_{s})^3\times(T/a_{t})$ & $a_tm_{K}$ & $a_tm_{\pi}$ & $m_{\pi}$ $(\mathrm{MeV})$ & $a_t (\mathrm{fm})$ & $m_{\pi}L$ & $N_{\mathrm{cfg}}$ \\
		\hline
		$32^3 \times 256$ & $0.08354(15)$ & $0.03938(19)$ & $233.0(1.2)$ &  $0.033357(59)$ & $4.3$ & $412$ \\ 
		\hline
	\end{tabular}
	\caption{\label{t:ens}Ensemble details for our $\nf = 2+1$ dynamical gauge 
		configurations.  More details on the ensemble generation 
	can be found in Ref.~\cite{Lin:2008pr}, while the scale is determined using 
$m_{K}$ as discussed in the text.}
\end{table}
by setting the mass of the kaon to $m_{K,\mathrm{phys}} = 494.2\mathrm{MeV}$. 
This physical value was obtained in Ref.~\cite{Colangelo:2010et} by taking 
the 
isospin-symmetric limit and removing QED effects. We prefer scale setting  
with $m_K$ to the method of  
Ref.~\cite{Lin:2008pr}, which uses the mass of the Omega baryon ($m_{\Omega}$), 
due to difficulties in determining $m_{\Omega}$. Still, this scale  
should be viewed as indicative as the kaon mass was not extrapolated to 
the physical light quark masses but taken on this single ensemble only. 
However, our results for dimensionful quantities are naturally
expressed in terms of $m_{\pi}$ so that the lattice spacing enters only in 
comparison with the literature in Fig.~\ref{f:rho_sum}. The determination of 
$a_tm_{\pi}$, $a_tm_{K}$, and the renormalized anisotropy $\xi$ will be 
discussed shortly.

Although these 412 configurations are separated by $20$ 
 Hybrid Monte Carlo (HMC) molecular 
dynamics trajectories of length $\tau = 1.0$, there is a small amount of 
residual autocorrelation evident in the measured correlation 
functions.\footnote{In lattice QCD the largest autocorrelations are typically 
observed for `smoothed' observables such as the topological charge and smoothed action~\cite{Schaefer:2010hu}, which are not examined here.} 
In 
order to mitigate effects of this autocorrelation on estimates of 
statistical uncertainties, we average measurements on pairs of subsequent 
configurations.
Statistical errors are estimated using the bootstrap 
technique~\cite{efron1986} on this rebinned ensemble with 
$N_{B} = 800$ bootstrap samples.

In this anisotropic setup the lattice regulator is fully 
specified by the temporal lattice spacing $a_t$ and renormalized anisotropy 
$\xi = a_{s}/a_{t}$. The anisotropy is determined from the (continuum) pion 
dispersion relation
\begin{align}\label{e:disp}
	\left[a_{t}E_{\pi}(\boldsymbol{d}^2)\right]^2 = (a_tm_{\pi})^2 + \left(\frac{2\pi a_s}{\xi L}\right)^2 \boldsymbol{d}^2,
\end{align}
where $\boldsymbol{d} \in \mathbb{Z}^3$ is the 
quantized finite-volume three-momentum of the pion.  

Determination of $\xi$ requires the single-pion energies $a_{t}E_{\pi}(\boldsymbol{d}^2)$ in 
Eq.~\ref{e:disp}. Periodic temporal boundary conditions are used for this ensemble, potentially 
complicating the extraction of finite-volume energies from the 
fall-off of temporal correlation functions.  
In particular, a single zero-momentum 
pion correlation function has the `cosh' form in the limit 
of ground-state saturation
\begin{align}\label{e:teff1}
	\lim_{ {t \gg 1/E_1,} \atop {T-t\gg 1/E_1} } C_{\pi}(t) = A\mathrm{e}^{-m_{\pi}t}\left(1 + \mathrm{e}^{-m_{\pi}(T-2t)}\right),  
\end{align}
where $E_1$ is the relevant first excited-state energy.
Two-pion correlation functions with zero total momentum have the more 
complicated form (ignoring small energy shifts due to pion interactions) 
\begin{align}\label{e:teff2}
	\lim_{ {t \gg 1/E_1,} \atop {T-t\gg 1/E_1} } C_{2\pi}(t) = A\mathrm{e}^{-2m_{\pi}t}\left(1 + 
	\mathrm{e}^{-2m_{\pi}(T-2t)} + B\mathrm{e}^{-m_{\pi}(T-2t)}\right),  
\end{align}
while two-pion correlation functions with non-zero total momenta 
have a similar but more complicated additional exponential term.

Since our finite-volume energies are extracted from fits of temporal 
correlation functions to an exponential form, these additional terms 
add potentially significant complication as has been discussed in (e.g.) 
Ref.~\cite{Dudek:2012gj}. Fortunately, the large temporal extent of our 
lattice ($m_{\pi}T \approx 10$) suppresses such terms below the statistical accuracy
of the energy levels. This can be demonstrated by performing two-parameter 
correlated-$\chi^2$ 
fits of the single zero-momentum pion correlation function to both a single 
exponential and the cosh of Eq.~\ref{e:teff1}. The
second exponential in Eq.~\ref{e:teff1} is larger than or equal to the 
additional problematic exponential terms which appear in two-pion correlation 
functions, apart from small hadronic interaction effects. 

\begin{figure}
	\centering
	\includegraphics[width=0.49\textwidth]{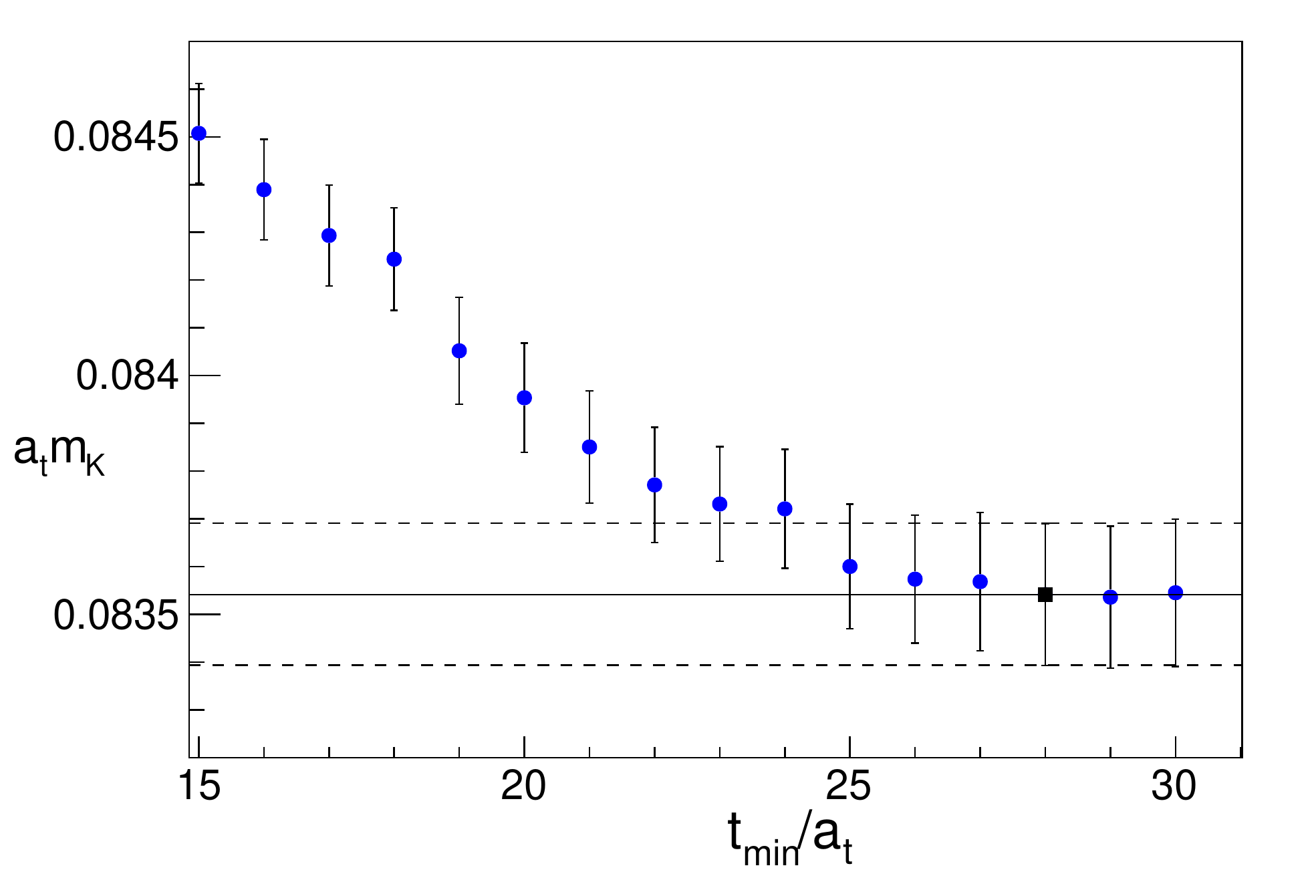}
	\includegraphics[width=0.49\textwidth]{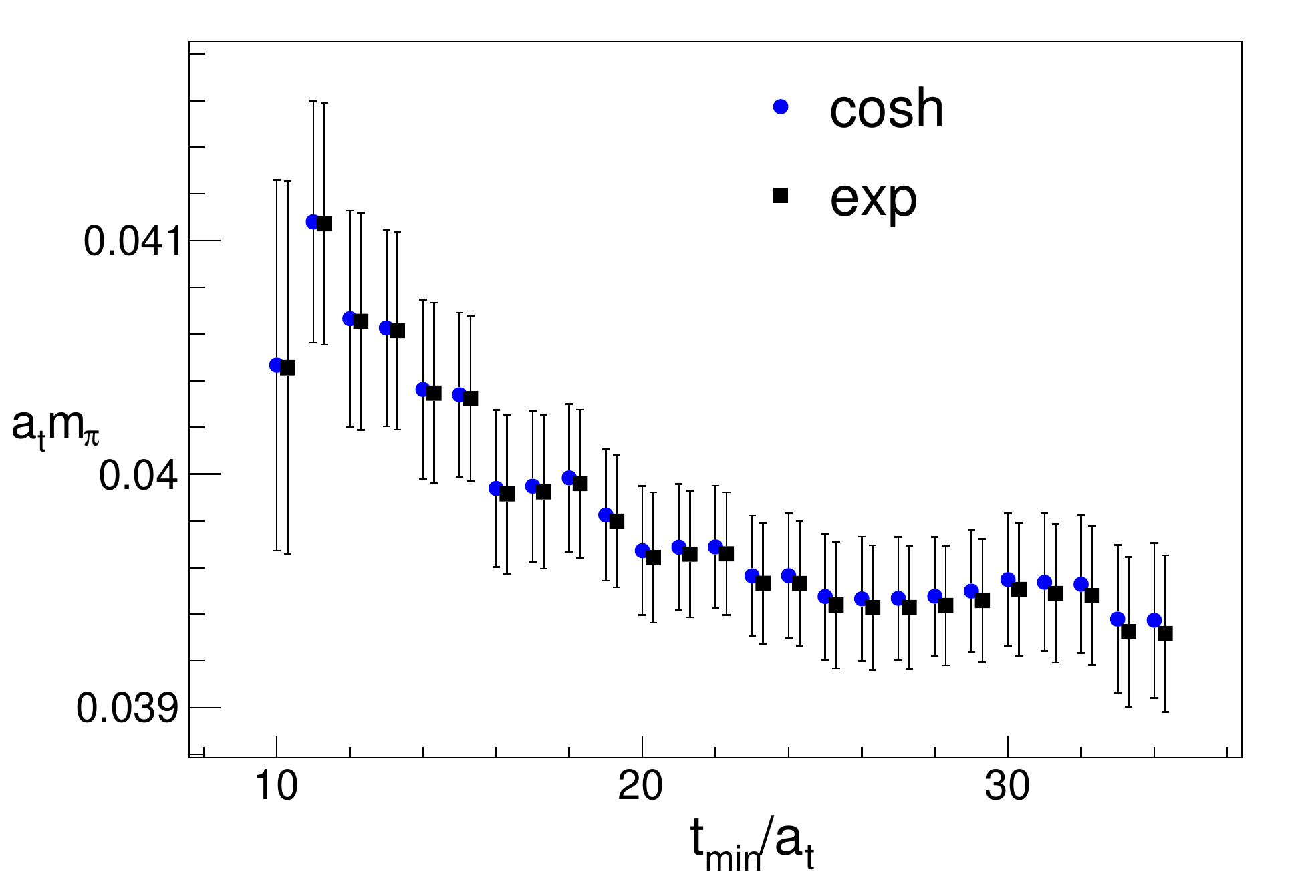}
	\caption{\label{f:cexp} \textbf{Left}: $\tmin$-plot (defined in the text) 
		for two-parameter single-exponential correlated-$\chi^2$ fits used in 
		the determination of $a_tm_{K}$ from which $a_t$ is set. The solid and dashed lines 
		show the mean value and $1\sigma$ errors (respectively) for the chosen fit 
		range, which is also indicated by a black square. 
		\textbf{Right}: Comparison of $t_{\mathrm{min}}$-plots for 
		single-exponential fits (denoted `exp') and 
		 cosh fits according to Eq.~\ref{e:teff1}
	for the single-pion correlation function with zero total momentum. Fit ranges
	are shown with $t_{\mathrm{max}}=38a_t$ and varying $t_{\mathrm{min}}$. The
	consistency of these two fit forms for our most precisely determined 
	correlation function demonstrates that `thermal' effects due to 
the finite temporal extent may be neglected. }
\end{figure}

As the single pion at rest is our most precisely determined 
correlation function, it is most sensitive to these thermal effects. 
The absence of these effects, such as the second exponential in 
Eq.~\ref{e:teff1}, indicates that additional exponentials in 
two-pion correlation functions may be neglected. The comparison of 
single-exponential and cosh fits for various fitting ranges is shown in 
Fig.~\ref{f:cexp}. Clearly, no effect from the finite temporal extent is 
evident for these temporal separations. All subsequent correlated-$\chi^2$ fits to temporal correlation 
functions will thus ignore finite-$T$ effects. The extraction of 
these energies will be discussed in more detail in Sec.~\ref{s:ener}. 

The fits to the single-pion zero-momentum correlation function shown in 
Fig.~\ref{f:cexp}, as well as all other fits to correlated data in this 
work, minimize a correlated-$\chi^2$ to properly treat the covariance between
observables measured on the same ensemble of gauge configurations. The covariance matrix is obtained using the bootstrap estimator 
\begin{align}\label{e:cov}
	\mathrm{Cov}(t,t') &= \frac{1}{N_B - 1} \sum_{n=1}^{N_B} (\langle C(t) \rangle_n - \langle\langle C(t) \rangle\rangle)  (\langle C(t') \rangle_n - \langle\langle C(t') \rangle\rangle),
\end{align}
where $N_B = 800$ is the number of bootstrap samples, 
$\langle C(t) \rangle_n$ is the bootstrap replicum of $C(t)$ on the $n$th 
sample, and $\langle\langle C(t) \rangle\rangle$ is the average over all 
bootstrap replica. 
The covariance matrix is taken as identical across each bootstrap sample's 
determination of the fit parameters. 

Apart from effects due to the finite temporal extent, the 
range of timeslices $[\tmin,\tmax ]$ over which the fit is performed is 
another source of systematic error. In particular, fitted values exhibit a
marked sensitivity to $\tmin$ due to the influence of 
higher-lying exponentials in Eq.~\ref{e:cor}. In this work we employ
`$\tmin$-plots' to ensure that 
this systematic error is smaller than the statistical error on the fit 
parameters. The guidelines for selecting a $\tmin$ satisfying this criterion 
are given in Eq.~\ref{e:tmin}. These plots show the fitted values for many $\tmin$ 
with a fixed $\tmax$, and are exemplified in Fig.~\ref{f:cexp} which shows 
$\tmin$-plots for $a_tm_{K}$ and $a_tm_{\pi}$.  
With stochastically-estimated correlation functions, $\tmin$-plots are 
preferable to effective masses $m_{\mathrm{eff}}(t) = \ln [C(t)/C(t+1)]$ for 
determining the range of times over which a single exponential 
dominates. Stochastically-estimated effective masses 
typically have larger error than the corresponding fitted 
energies and are thus not useful to assess systematic errors from the choice 
of $\tmin$. 

\begin{figure}
	\includegraphics[width=0.49\textwidth]{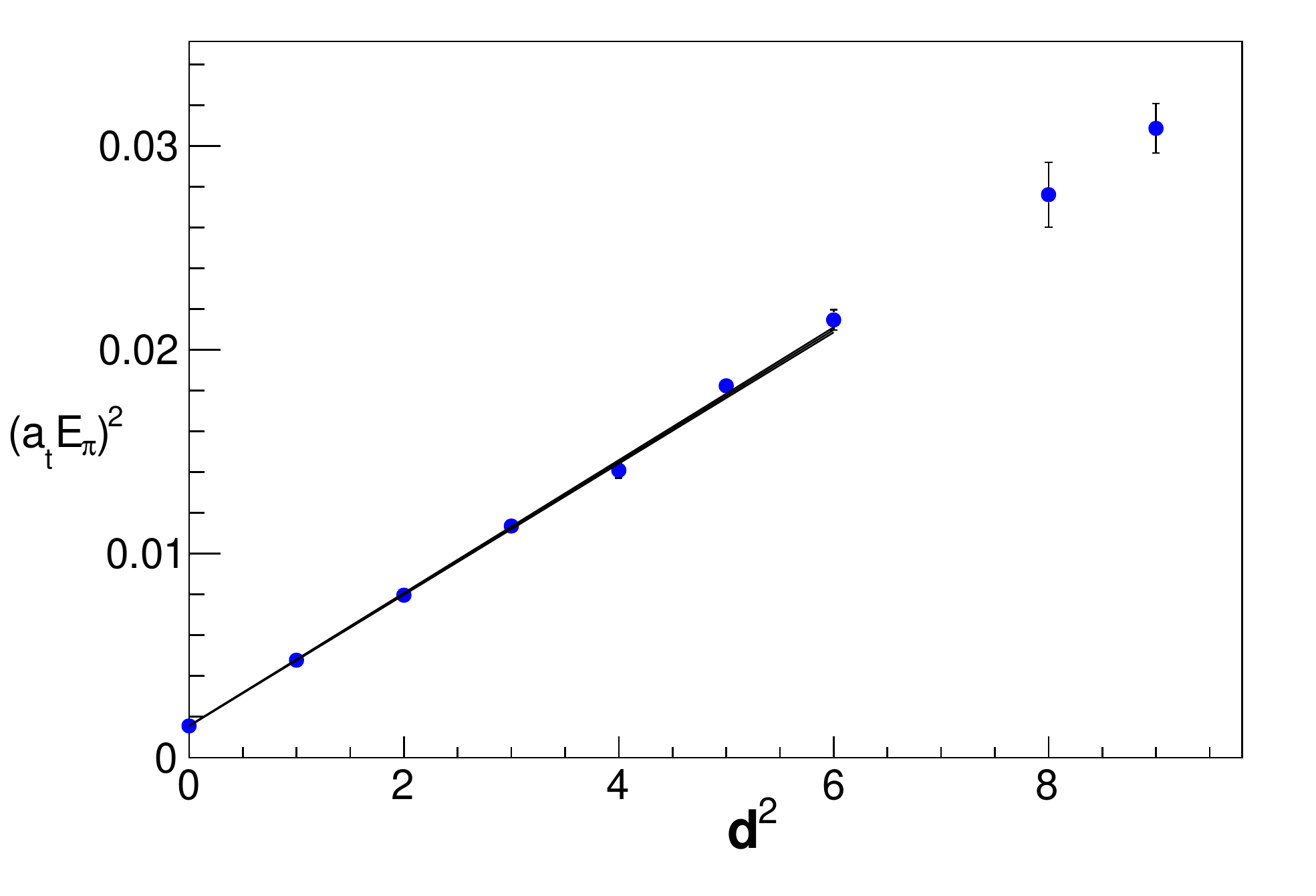}
	\includegraphics[width=0.49\textwidth]{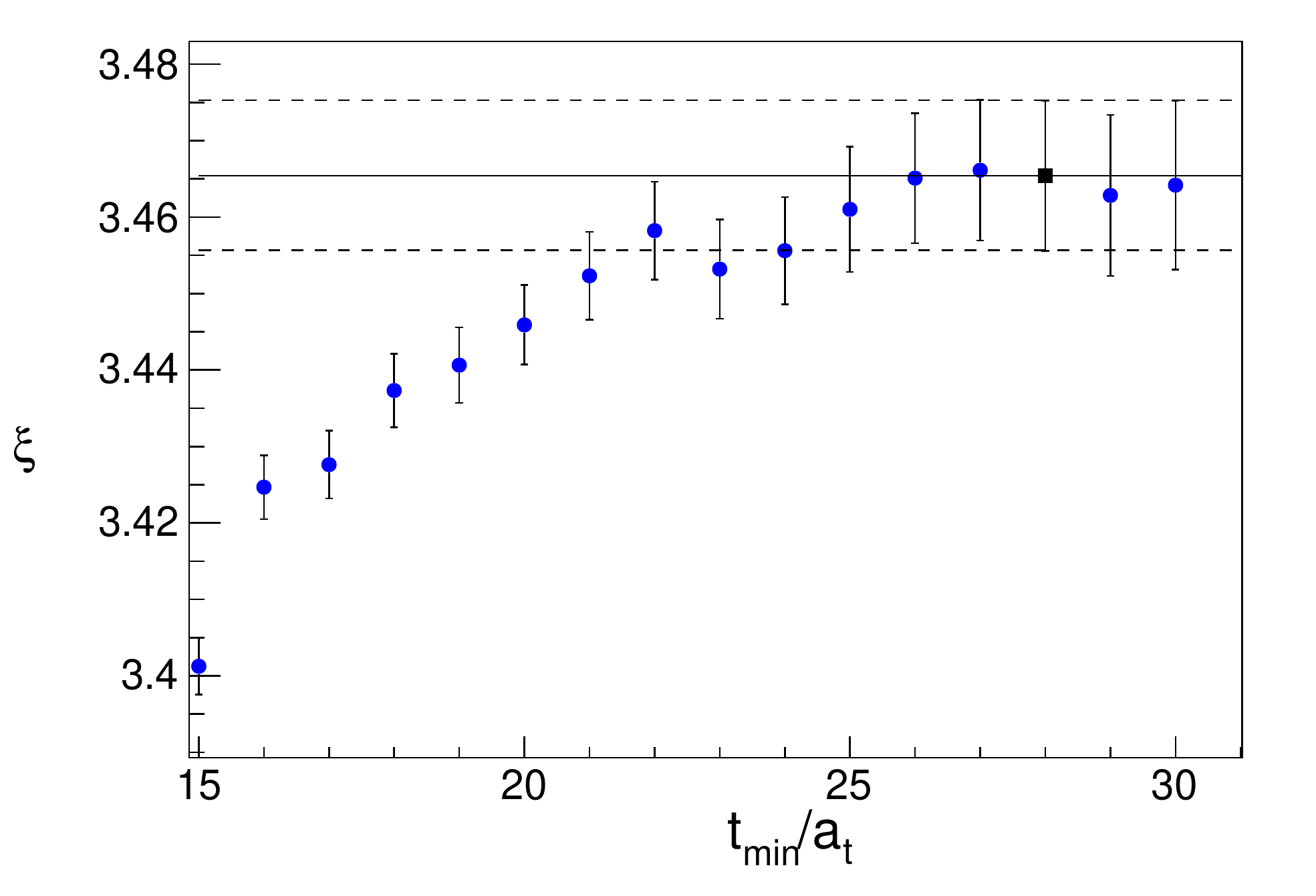}
	\caption{\label{f:fart}Two strategies to determine $\xi$. 
		\textbf{Left}: (Strategy 1) Single pion energies at various 
		momenta together with a linear fit to Eq.~\ref{e:disp}. 
		\textbf{Right}: (Strategy 2) $t_{\mathrm{min}}$-plot for 
		a simultaneous fit to all pion correlation functions to 
		Eq.~\ref{e:xi2} together with the chosen fit range indicated by the solid 
	and dotted lines.} 
\end{figure}

We now discuss two determinations of $\xi$. In the first, single-pion 
energies at various total momenta\footnote{Pion correlation functions are 
	averaged over 
equivalent momenta before fitting.} are first obtained from this fitting 
procedure and then used in Eq.~$\ref{e:disp}$.
The $t_{\mathrm{min}}$-plots for single-exponential 
fits to these correlation functions together with the fitted energies used in 
our analysis are given in Fig.~\ref{f:pi_tmins} of App.~\ref{a:pi}.
 Generally, these energies are chosen somewhat conservatively so that 
 systematic errors due to excited-state contamination are small in comparison to the statistical errors.
 They are summarized in Fig.~\ref{f:fart} together with a fit to Eq.~\ref{e:disp} for $\boldsymbol{d}^2 \le 6$.
 Correlation exists among the fitted energies;  
their covariance is estimated using the bootstrap method of Eq.~\ref{e:cov}
and fixed on each bootstrap sample. 
 Evidently
 the continuum dispersion relation describes the single-pion energies 
 up to large total momenta, suggesting that lattice spacing effects are 
 under control here. The pion 
 mass and $\xi$ determined from this linear fit (denoted `Strategy 1') are 
 given in Tab.~\ref{t:xi}.  

An alternative determination (denoted `Strategy 2') fits all single-pion
correlation functions simultaneously to the ansatz 
\begin{align}\label{e:xi2}
	C_{\boldsymbol{d}^2}(t) = A_{\boldsymbol{d}^2} \times \mathrm{e}^{-\frac{t}{a_{t}}\sqrt{(a_tm_{\pi})^2 + \left(\frac{2\pi a_s}{\xi L}\right)^2 \boldsymbol{d}^2}}
\end{align}
where the $\{A_{\boldsymbol{d}^2}\}$, $m_{\pi}$ and $\xi$ are free 
parameters. The covariance between all correlation functions at all time
separations is explicitly taken into account in these correlated-$\chi^2$ fits.
The results for $\xi$ from this fit are shown in Fig.~\ref{f:fart} for 
various $t_{\mathrm{min}}$ (identical for all correlation functions) together with the chosen fit range. This fit is 
also given in Tab.~\ref{t:xi}, where it is denoted `Strategy 2'. 

Although the continuum dispersion relation fits the data well, we additionally 
perform fits like Strategy 2 but using the lattice-modified dispersion relation 
\begin{align}\label{e:lat}
	(a_tE_{\pi})^2 = (a_tm_\pi)^2 + 4\sum_{i=1}^{3} \sin^2 \left(\frac{\pi a_s}{\xi L} d_i\right). 
\end{align}
The results of this fit are also consistent and shown in Tab.~\ref{t:xi} as 
`Strategy 3'. 
For this 
work we take $\xi$ from Strategy 1 as it is the most conservative estimate, 
although the final results have little dependence on this choice. 
\begin{table}
\centering
\begin{tabular}{|c|c|c|c|}
	\hline
	Strategy & $a_{t}m_{\pi}$ & $\xi$ & $\chi^2/d.o.f$ \\
	\hline 
	1  & 0.03938(19) & 3.451(11) & 1.4 \\
	2  & 0.03978(19) & 3.4654(98) & 1.19 \\
	3  & 0.03978(19) & 3.4649(98) & 1.20 \\
	\hline
\end{tabular}
\caption{\label{t:xi}Results for $\xi$ from linear fits of moving pion 
	energies to Eq.~\ref{e:disp} (Strategy 1), a simultaneous fit 
	of all pion correlation functions to Eq.~\ref{e:xi2} (Strategy 2), and a 
simultaneous fit to Eq.~\ref{e:lat} (Strategy 3).}   	
\end{table}

\subsection{Correlation function calculation}\label{s:corr} 

Because of the finite spatial extent and lattice spacing, the 
symmetry group of lattice QCD is $O^{D}_h$, the double 
cubic point group. Irreducible representations of this group (or the relevant 
little group for a particular momentum) together with  
total isospin and $G$-parity fully specify the quantum numbers of our energy 
eigenstates. Therefore, operators which transform irreducibly under these symmetries are employed.
The procedure for constructing such operators is well 
known. Here we are concerned only with the ground state and $2-3$  
low-lying excited states in the relevant irreducible representations 
(irreps).
However, this work is part of a broader program to explore many higher-lying
resonances in QCD. Interpolating operators with large overlap onto these 
higher-lying resonances are more complicated and require non-trivial
spatial structures as in Refs.~\cite{Basak:2005aq,Morningstar:2013bda}. Such operators are not used here, but rather only 
(smeared) single-site interpolators for each hadron. 

Correlation matrices are typically required to obtain excited-state energies. 
In order to build correlation matrices in each of the irreps, we 
examine the expected non-interacting single-$\rho$ and two-pion levels.
Generally, an interpolator for each of these levels below the 
inelastic threshold $E_{\cm}/m_{\pi}=4$ is included while additional 
two-pion operators are used as a check of systematic effects.

As discussed above, these multi-hadron correlation matrices 
require all-to-all quark propagators. We use the method of 
Ref.~\cite{Morningstar:2011ka} which  
introduces noise in the subspace spanned by low-lying eigenmodes of the 
gauge-covariant Laplace operator. This noise can be diluted~\cite{Foley:2005ac} in time (T), spin (S), and 
Laplacian eigenvector (L) indices, each of 
which can be fully diluted (`F') or have some number of dilution projectors 
`interlaced'  (`I$n$') uniformly throughout the space. Note that the distillation method 
of Ref.~\cite{Peardon:2009gh} is recovered in the maximal dilution limit (TF, 
SF, LF).  

\begin{table}
\centering
\begin{tabular}{|c|c|c|c|c|c|}
	\hline
	$N_{v}$ & line type & $N_{r}$      & scheme & $N_{t_0}$  & $N_{D}$  \\
\hline 
    264    & fixed     & 5           & (TF, SF, LI8)    & 8 &   1280     \\
						 &  relative     & 2           & (TI16, SF, LI8) & - & 1024 \\  
\hline
\end{tabular}
\caption{\label{t:dil} The number of eigenvectors ($N_v$), noise sources ($N_{r}$),
	source times ($N_{t_0}$), and Dirac 
	matrix inversions per configuration ($N_D$), together with the 
dilution schemes for fixed and relative quark lines.}
\end{table}

On this anisotropic ensemble it is beneficial to choose different dilution schemes for quark 
propagators between different times (so-called `fixed' 
quark lines) and for quark propagators starting and ending at the same time 
(`relative' quark lines). 
These different dilution schemes are specified in 
Tab.~\ref{t:dil} together with the number of required Dirac matrix inversions ($N_D$) 
and low-lying Laplacian eigenvectors defining the LapH subspace ($N_v$).
In order to ensure an unbiased estimate of correlation functions, each quark line requires an independent stochastic source. The total number of such sources 
used per configuration ($N_r$) is shown in Tab.~\ref{t:dil} together with the 
number of source times ($N_{t_0}$) used to reduce statistical errors. 
It should be noted that only $N_{r} = 4$ fixed lines and 
$N_r=1$ relative lines (for a minimum $N_D=640$)
are required to ensure unbiased estimates of the required correlation functions.
However, additional source times and 
noise sources are employed here to increase statistics. While the additional 
noise sources are required for 
other systems, different noise combinations provide additional stochastic 
estimates and are thus averaged over. The required Wick contractions are 
enumerated in Ref.~\cite{Morningstar:2011ka}.

\subsection{Finite-volume energies}\label{s:ener}

After constructing the correlation functions as described in Sec.~\ref{s:corr}, the 
method for extracting finite-volume energies from them is now 
discussed. For this work we aim to utilize not only the ground state in each 
irreducible representation, but several excited states as well. In order to 
reliably extract these excited-state energies, solutions of a generalized 
eigenvalue problem are employed. 

In each channel, a correlation matrix is formed 
 consisting of a single-site $\rho$ interpolating operator 
(if present)
together with the relevant two-pion operators. These two-pion operators are 
chosen to match the expected non-interacting states and all such operators 
below inelastic threshold are included. 

For each of these correlation matrices ($C(t)$) we solve the generalized eigenvalue 
problem 
\begin{align}\label{e:gevp}
	C(t_d)v(t_0,t_d) = \lambda(t_0,t_d)C(t_0)v(t_0,t_d)  
\end{align}
for a particular set of $(t_0,t_d)$. The eigenvectors $\{v_{n}(t_0,t_d)\}$ are 
used to define correlation functions between 
`optimal' interpolators~\cite{Michael:1982gb} 
\begin{align}\label{e:rcor}
	\hat{C}_{ij}(t) = \left( v_{i}(t_0,t_d), C(t)v_j(t_0,t_d)\right)
\end{align}
where the outer parentheses denote an inner product over GEVP indices. Although 
these optimal interpolators are constructed to have maximal overlap with a 
single Hamiltonian eigenstate, the off-diagonal elements of 
$\hat{C}_{ij}(t)$ are not exactly zero resulting in a
source of systematic error that 
must be assessed. It should be noted that this is a different approach 
to Refs.~\cite{Luscher:1990ck,Blossier:2009kd} 
which require the solution of the GEVP at different $(t_0,t_d)$, possibly 
introducing ambiguities between closely spaced levels at different times,
but guaranteeing that the eigenvalues approach the desired exponential fall-off.  

To extract energies in a particular channel we solve the GEVP of 
Eq.~\ref{e:gevp} and form the rotated correlation matrix of 
Eq.~\ref{e:rcor}. The GEVP diagonalization is not performed on each bootstrap sample, due to similar ambiguities identifying closely spaced levels on 
different bootstrap samples. 
We first perform 
two-parameter correlated-$\chi^2$ fits with a 
single-exponential ansatz on the diagonal elements of the rotated 
correlation matrix to obtain a preliminary determination of the finite-volume 
spectra. These preliminary energies are used in Fig.~\ref{f:i1box} and with 
Eq.~\ref{e:ov} to obtain a qualitative picture of the spectrum and nature 
of the states. 

%As a further check of the GEVP systematics, we have performed simultaneous 
%fully correlated fits to the entire (real) matrix $\hat{C}_{ij}(t)$ to account 
%for any non-zero off-diagonal elements. Here we fit $\hat{C}_{ij}(t)$ to the ansatz
%\begin{align}
%	\hat{C}_{ij}(t) = \sum_{n}A_{ni}A_{nj}\mathrm{e}^{-E_n t}. 
%\end{align}
%These fits are consistent with the single exponential fits to the diagonal 
%elements discussed above. 

For our final analysis we employ 
a different approach which exploits the similarity (and correlation) between 
two-pion and single-pion correlation functions. As in 
Ref.~\cite{Helmes:2015gla} but here generalized to arbitrary momenta, for an 
optimized two-pion operator with pion momenta $\boldsymbol{d}_1$ and $\boldsymbol{d}_2$ ($\mathcal{O}_{\boldsymbol{d}_1,\boldsymbol{d}_2}$), we define the ratio
\begin{align}\label{e:rat}
	R(t) = \frac{\langle \mathcal{O}_{\boldsymbol{d}_1,\boldsymbol{d}_2} (t)
	\bar{\mathcal{O}}_{\boldsymbol{d}_1,\boldsymbol{d}_2}(0) \rangle}{
		\langle \mathcal{O}_{\boldsymbol{d}_1}(t) 
		\bar{\mathcal{O}}_{\boldsymbol{d}_1}(0)\rangle
		\langle \mathcal{O}_{\boldsymbol{d}_2}(t) 
	\bar{\mathcal{O}}_{\boldsymbol{d}_2}(0)\rangle}
\end{align}
which is constructed on each bootstrap sample and fit in a fully correlated manner to the ansatz 
$R(t) = A\mathrm{e}^{-\Delta Et}$. The energy shift $\Delta E$ is used to 
reconstruct the desired energy via 
\begin{align}
	a_tE = a_t\Delta E + \sqrt{(a_tm_{\pi})^2 + \left(\frac{2\pi a_s}{\xi L}\right)^2\boldsymbol{d}_1^2} + \sqrt{(a_tm_{\pi})^2 + \left(\frac{2\pi a_s}{\xi L}\right)^2\boldsymbol{d}_2^2}, 
\end{align}
where $m_{\pi}$ is obtained from the single-pion fits.  
In the $I=1$ channel, these two-hadron states mix with the $\rho$-meson. For 
such mixed states these ratio fits are still beneficial, but exhibit an 
increased amount of excited-state contamination, which will be discussed 
shortly.

Several sources of systematic error in this procedure must be addressed. First, 
the fitting range $\left[\tmin, \tmax\right]$ is varied, in particular $\tmin$. Second, systematic errors due to the small but 
non-zero off-diagonal elements of $\hat{C}_{ij}(t)$ must be assessed. 
\begin{figure}
	\includegraphics[width=\textwidth]{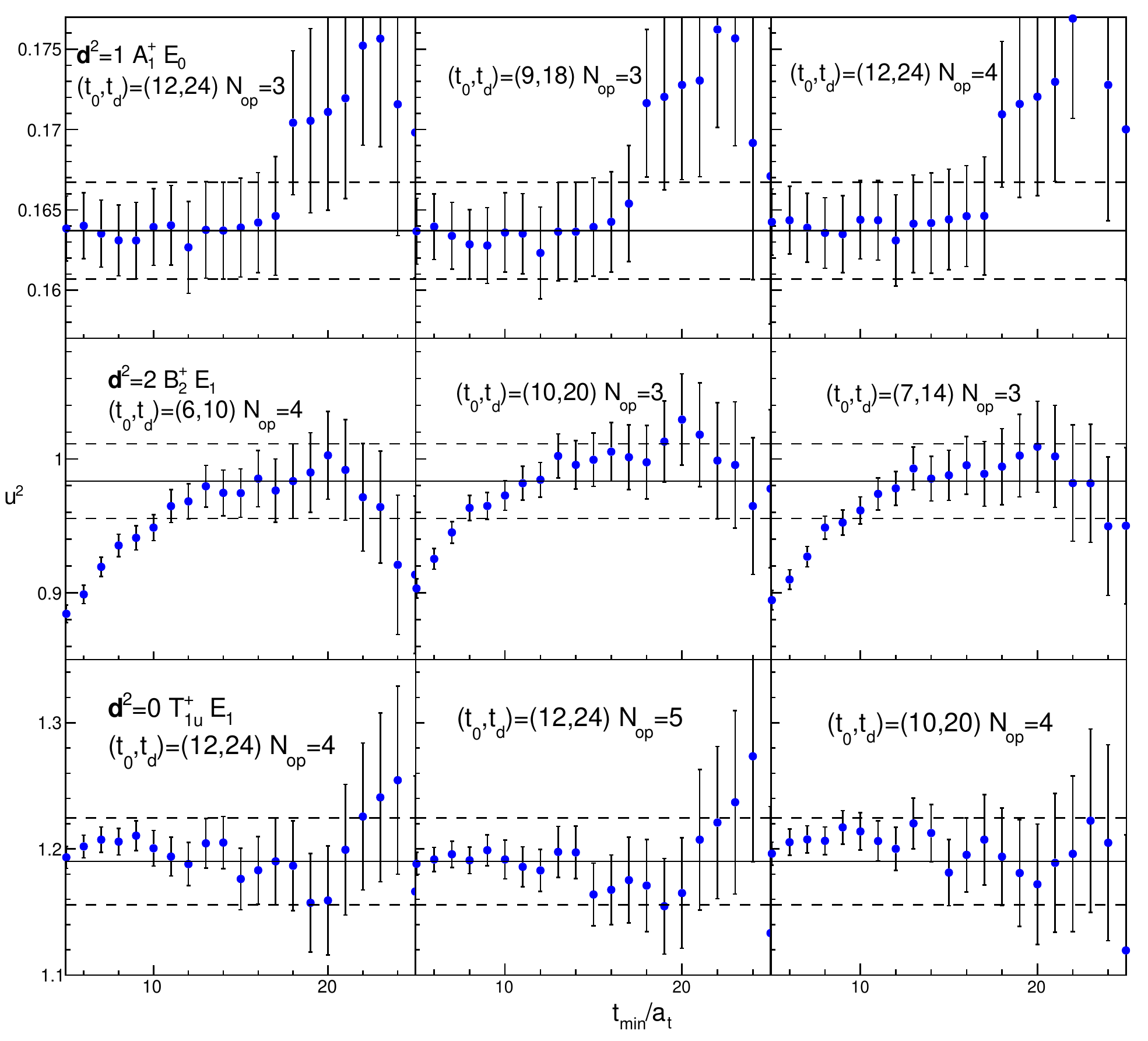}
	\caption{\label{f:exp_tmin}Variation of $\tmin$, $(t_0,t_d)$, and the 
	number of operators included in the GEVP for three representative energy 
levels in $I=1$. Each row corresponds to three different GEVP's for a single 
energy level. We have chosen a representative sample of three energy levels 
consisting of a ground state below the resonance region (top), a first excited 
state near the resonance energy (middle) and a first excited state 
somewhat above the resonance energy (bottom). The dimensionless 
center-of-mass momentum $u^2$ (defined in Eq.~\ref{e:kin}) is shown, as it determines the scattering phase shift.}
\end{figure}
To this end, we not only vary the fitting range $\left[\tmin, \tmax\right]$
but also $(t_0,t_d)$ and the operators included in the GEVP. The variation 
of these systematics for a selection of energy levels is shown in 
Fig.~\ref{f:exp_tmin}. There the dimensionless center-of-mass momentum $u^2$ 
is shown, which is defined in Eq.~\ref{e:kin}.   

Generally, systematic effects due to $\tmin$ are the largest and must be 
treated with care. To this end we fix $(t_0,t_d) = (12a_t,24a_t)$ and choose
$\tmin$ conservatively.
As minimum requirements we demand that the chosen $\tmin$ gives a suitable 
correlated $\chi^2/d.o.f. < 1.7$ and that 
\begin{align}\label{e:tmin}
	\Delta E_{\mathrm{fit}}(\tmin) - \Delta E_{\mathrm{fit}}(\tmin - \delta_t) < \sigma(\tmin)
\end{align}
where $\sigma(\tmin)$ is the bootstrap error on $\Delta E_{\mathrm{fit}}(\tmin)$ and 
$\delta_t = 4a_t$. 

While these ratio fits have the advantage of directly determining the energy 
shifts, their excited-state contamination may have a non-standard form. This 
can be seen by examining the leading excited-state corrections for the 
ratio directly 
\begin{align}
	\lim_{t\rightarrow \infty} R(t) = A\mathrm{e}^{-\Delta E t} \left[ 
		1 + B_{\boldsymbol{d}_1,\boldsymbol{d}_2}
		\mathrm{e}^{-\Delta E_{\boldsymbol{d}_1,\boldsymbol{d}_2} t} -
		B_{\boldsymbol{d}_1}\mathrm{e}^{-\Delta E_{\boldsymbol{d}_1} t} 
	- B_{\boldsymbol{d}_2}\mathrm{e}^{-\Delta E_{\boldsymbol{d}_2} t} \right], 
\end{align}
where $\Delta E_{\boldsymbol{d}_1,\boldsymbol{d}_2}$ is the energy gap from
the two-pion correlator in the numerator 
and $B_{\boldsymbol{d}_1,\boldsymbol{d}_2}$ the relevant interpolator-dependent 
prefactor, while $B_{\boldsymbol{d}_{1,2}}$ and 
$\Delta E_{\boldsymbol{d}_{1,2}}$ are the analogous 
quantities for each of the single-pion correlators in the denominator.
If the first two excited states in the numerator effectively consist of 
one pion in the ground state and the other in an excited state, the 
overall excited-state contamination in $R(t)$ will be very small.
However, in general the excited-state contamination from the denominator 
enters with different sign, 
possibly causing a non-monotonically decreasing `bump'-type behavior 
in $\tmin$-plots. Such 
bumps must be taken into account when choosing fit ranges for the 
strongly-interacting $I=1$ states. 
Apart from Fig.~\ref{f:exp_tmin}, $\tmin$-plots for ratio fits performed to 
all correlation functions used in the phase shift analysis are shown in 
App.~\ref{s:app1} and App.~\ref{s:app2}. Although bumps are evident for 
some levels, we choose conservative fit ranges in those cases to 
ensure systematic effects from excited states are smaller than the 
statistical error. 

\subsection{Scattering phase shifts}\label{s:phase}

After discussing the procedure for extracting finite-volume energies, we 
now turn to using them to calculate elastic scattering phase shifts.
The relation between finite-volume energy spectra and infinite-volume 
elastic scattering amplitudes is derived in Ref.~\cite{Luscher:1990ux} and generalized to 
non-zero total momentum in Ref.~\cite{Rummukainen:1995vs}. A useful summary of 
the method for several different situations may be found in 
Ref.~\cite{Gockeler:2012yj}, while generalizations to asymmetric spatial 
volumes~\cite{Feng:2004ua}, multiple coupled two-particle 
channels~\cite{He:2005ey} and three-particle 
scattering~\cite{Polejaeva:2012ut,Hansen:2014eka,Hansen:2015zga} have been developed. 

For scattering between two identical particles of mass $m$, we denote by $E$ 
 the energy measured in the lattice (`lab') frame in a particular 
 irrep with total momentum $\boldsymbol{P} = \frac{2\pi\boldsymbol{d}}{L}$. We 
 define the kinematical variables  
 \begin{align}\label{e:kin}
	E_{\cm} &= \sqrt{E^2 - \boldsymbol{P}^2}, \qquad \gamma=\frac{E}{E_{\cm}}, 
	\qquad \boldsymbol{q}_{\cm}^2 = \frac{1}{4}E_{\cm}^2 - m^2, \qquad 
	%\\\nonumber
u^2 = \frac{L^2\boldsymbol{q}^2_{\cm}}{(2\pi)^2}. % \qquad \boldsymbol{s} = . %\left( 1 + \frac{m_1^2 - m_2^2}{E_{\cm}^2}\right)\boldsymbol{d}.
\end{align}

Up to exponentially suppressed finite-volume effects, the elastic scattering 
matrix is related to the finite-volume energy spectra via the well-known 
quantization condition, which is a matrix equation of the form 
\begin{align}
	\mathrm{det} \left[ 1 + F^{(\boldsymbol{d}, \gamma,u)} (S-1)\right] = 0 
\end{align}
where $S$ is the infinite-volume scattering matrix and the determinant is taken over the indices $(J,m_{J},L,\sigma)$ 
corresponding to total angular momentum, its projection along some axis, 
orbital angular momentum, and spin, respectively. 
Note that the matrix $F$ in general mixes different partial waves. 

For elastic 
scattering between identical spin-zero particles $\sigma=0$ and $J=L$. In this case 
the matrix $F$ is given by 
\begin{align}
	F^{(\boldsymbol{d},\gamma,u)}_{L'm_{L'};Lm_L} &= \frac{1}{2}\left( 
\delta_{L'L}\delta_{m_{L'}m_L} + W_{L'm_{L'};Lm_L}\right), \\
W_{L'm_{L'};Lm_L} &= \frac{2i}{\pi\gamma u^{\ell+1}}Z_{\ell m}(\boldsymbol{d},
		\gamma, u^2) \int d^2\Omega Y^{*}_{L'm_{L'}}(\Omega) Y_{\ell m}^{*}(\Omega)
		Y_{Lm_L}(\Omega), 
\end{align}
where $\ell$ and $m$ are summed over and we have introduced the L\"{u}scher 
zeta functions $Z_{\ell m}(\boldsymbol{d}, \gamma, u^2)$. We use a 
representation of the zeta functions given in App. A of 
Ref.~\cite{Gockeler:2012yj} for their numerical evaluation, which is 
consistent with an independent implementation based on an alternative representation discussed in 
Ref.~\cite{Fahy:2014jxa}.

While we have expressed $F$ in the $Lm$ basis, it is more convenient to 
express the relation in terms of finite-volume irreps, as both $F$ and $S$ 
become block diagonal, facilitating the evaluation of the determinant. 
\begin{table}
	\centering
	\begin{tabular}{|c|c|c|c|}
		\hline
		$\ell$ & $\boldsymbol{d}_{\mathrm{ref}}$ & irrep & $q_{\cm}^{2\ell + 1}\cot \delta_{\ell}$ \\
		\hline 
		0     & $(0,0,0)$ & $A_{1g}$ & $w^{(0)}_{00}$ \\
		\hline
		& $(0,0,n)$ & $A_{1}$ & $w^{(0)}_{00}$ \\
		\hline
		& $(0,n,n)$ & $A_{1}$ & $w^{(0)}_{00}$ \\
		\hline
		& $(n,n,n)$ & $A_{1}$ & $w^{(0)}_{00}$ \\
		\hline 
		\hline
		1     &  $(0,0,0)$ & $T_{1u}$ & $w^{(1)}_{00}$  \\
		\hline
		&  $(0,0,n)$ & $A_{1}$ & $w^{(1)}_{00} + \frac{2}{\sqrt{5}}w^{(1)}_{20}$  \\
		&            & $E$ & $w^{(1)}_{00} - \frac{1}{\sqrt{5}}w^{(1)}_{20}$  \\
		\hline
		&  $(0,n,n)$ & $A_{1}$ & $w^{(1)}_{00} + \frac{1}{2\sqrt{5}}w^{(1)}_{20} - 
		i\sqrt{\frac{6}{5}}w^{(1)}_{21} - \sqrt{\frac{3}{10}}w^{(1)}_{22}$ \\	
		&            & $B_{1}$ & $w^{(1)}_{00} - \frac{1}{\sqrt{5}}w^{(1)}_{20} + 
		\sqrt{\frac{6}{5}}w^{(1)}_{22}$ \\
		&            & $B_2$   & $w^{(1)}_{00} + \frac{1}{2\sqrt{5}}w^{(1)}_{20} + 
		i\sqrt{\frac{6}{5}}w^{(1)}_{21} -\sqrt{\frac{3}{10}}w^{(1)}_{22}$ \\
		\hline
		&  $(n,n,n)$ & $A_{1}$ & $w^{(1)}_{00} + 2i\sqrt{\frac{6}{5}}w^{(1)}_{22}$ \\
		&            & $E$     & $w^{(1)}_{00} -  i\sqrt{\frac{6}{5}}w^{(1)}_{22}$ \\
		\hline
		%\hline
		%2     &  $(0,0,0)$ & $E_{g}$  & $w^{(2)}_{00} + \frac{6}{7}w^{(2)}_{40}$    \\
		%\hline
		%&  $(0,0,n)$ & $B_{1}$  & $w^{(2)}_{00} - \frac{\sqrt{5}}{7}w^{(2)}_{20} + 
		%\frac{1}{7}w^{(2)}_{40} + \sqrt{\frac{10}{7}}w^{(2)}_{44}$    \\
		%&           & $E$  & $w^{(2)}_{00} + \frac{\sqrt{5}}{7}w^{(2)}_{20} - 
		%\frac{4}{7}w^{(2)}_{40}$    \\
		%\hline
	\end{tabular}
	\caption{\label{t:phase} Expressions for the scattering phase shifts in each 
	irreducible representation for both the $\ell=0$ and $\ell=1$ partial waves in terms of the quantities defined in Eq.~\ref{e:w}.}
\end{table}
After performing this block diagonalization and neglecting the contribution of higher partial waves, the relationship between the scattering phase shifts and 
\begin{align}\label{e:w}
	w^{(\ell)}_{lm} =  \left(\frac{2\pi}{L}\right)^{2\ell+1} \frac{u^{2\ell-l}}{\gamma\pi^{3/2}} Z_{lm}(\boldsymbol{d},\gamma, u^2)
\end{align}
for each irrep is shown in Tab.~\ref{t:phase}. 
One advantage of employing expressions relating the real part of the inverse 
scattering amplitude to the 
$w_{l m}^{(\ell)}$ is that the analyticity of 
$q_{\cm}^{2\ell+1}\cot \delta_{\ell}$ near threshold is explicit. For weakly 
interacting 
channels such as the $I=2$ $A_{1g}^{+}$, this enables a smooth 
behavior between positive and negative $q_{\cm}^2$. 
As we treat identical-particle scattering, 
particle-exchange symmetry prevents mixing between successive partial waves
in moving frames. 
Neglecting the remaining partial wave mixing amounts to neglecting 
the $I=1$, $\ell = 3$ 
and $I=2$, $\ell =2$ partial waves. 
%For the determination of the $I=2$, $\ell=2$
%partial wave, we choose irreps in which the $\ell=0$ partial wave does not 
%contribute and thus neglect the $I=2$, $\ell=4$ partial wave in these cases. 
 
\section{Results}\label{s:res}

This section contains our results for elastic scattering phase shifts. We 
neglect exponential finite-volume corrections and, as discussed in 
Sec.~\ref{s:phase}, treat only the lowest partial wave which contributes to 
each lattice irrep. Our results are interpreted in 
terms of the effective range expansion, which provides the correct 
threshold behavior of the scattering amplitude while also accommodating 
resonances. Finite-volume energy levels near or above the inelastic threshold
$E_{\cm}/m_{\pi}=4$ are 
not described by the elastic L\"{u}scher formulae of Sec.~\ref{s:phase} and 
thus not used. 

\subsection{$I=1$}\label{s:res1}

\begin{figure}
	\includegraphics[width=\textwidth]{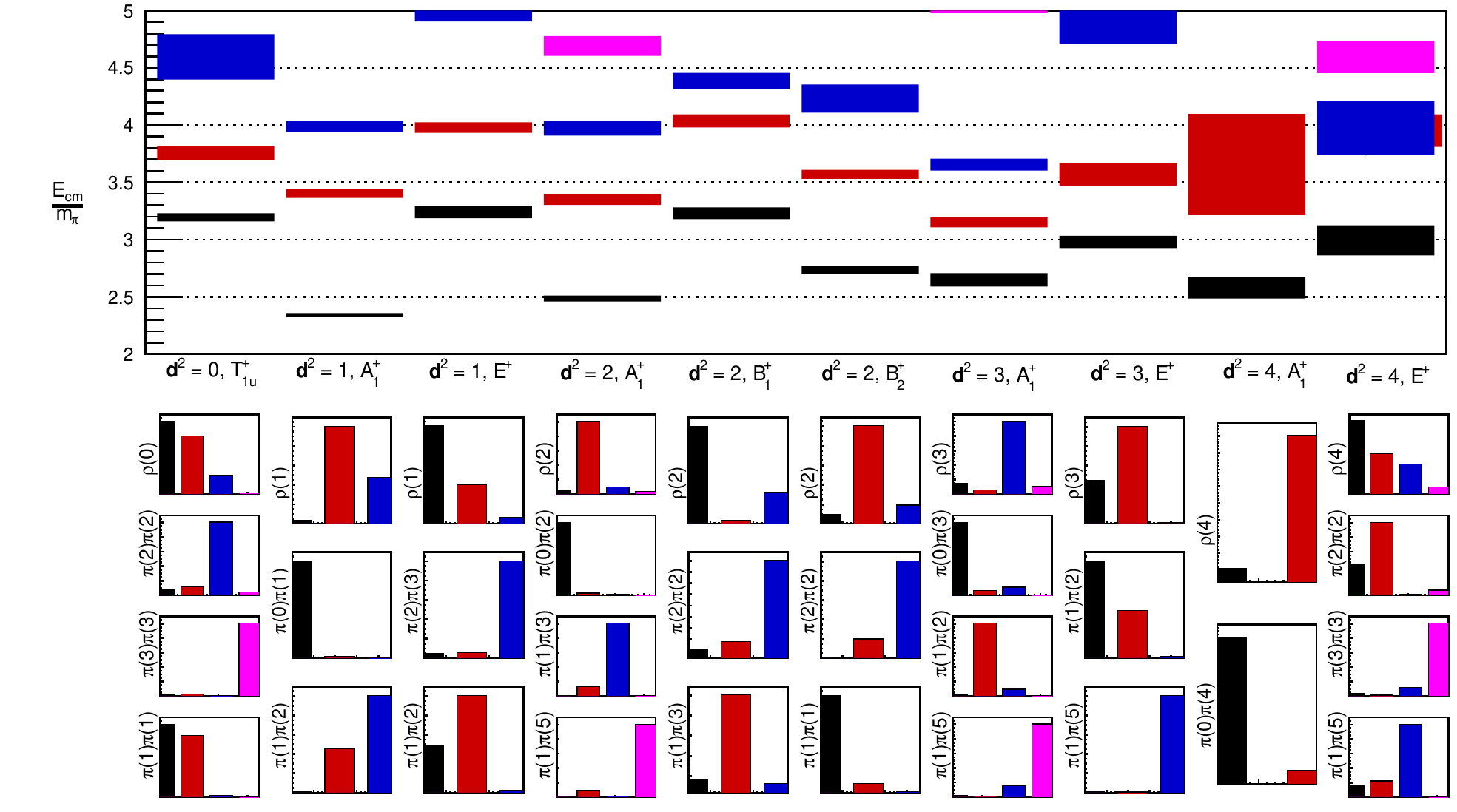}
	\caption{\label{f:i1box}(color online) $I=1$ center-of-mass energies (upper panel) for 
		each irrep together with the overlaps of each interpolator. 
	Each column (across both the upper and lower panels) corresponds to a 
single irrep and the colors are consistent between the energy levels and 
overlap plots. States significantly below the resonance mass 
(located at $E_{\cm}/m_{\pi}\approx 3.4$) 
have significant overlap with two-pion operators only, while those near the 
resonance region overlap with both two-pion and single-$\rho$ interpolators.}
\end{figure}

The $I=1$, $\ell=1$ partial wave contains the $\rho$-resonance. Not only is 
this evident in the scattering phase shifts, but it is also  
suggested by examining the overlaps of interpolating operators onto finite-volume Hamiltonian eigenstates. Specifically, we estimate 
$Z_{in} = |\langle 0 | \hat{\mathcal{O}}_i | n \rangle|^2$ by forming the ratio 
\begin{align}\label{e:ov}
Z_{in}(t) = \left|\frac{\sum_{j} C_{ij}(t)v_{nj}(t_0,t_d)}{\mathrm{e}^{-\frac{E_n}{2}t}\sqrt{\hat{C}_{nn}(t)}}\right|^2, 
\end{align}
where $E_n$ is the fitted energy, 
and taking $t=20a_t$. For each interpolating operator the overlaps onto the 
Hamiltonian eigenstates are plotted in Fig.~\ref{f:i1box} together with the  
energies extracted from single-exponential fits. Center-of-mass energies 
are shown in that figure to facilitate comparison between channels with 
different total momenta. 

As expected, local $\rho$-meson interpolating operators have 
significant overlap with energy eigenstates near the resonance mass 
$E_{\cm}/m_{\pi} \approx 3.4$, where mixing with two-pion operators can be 
observed. However, only two-pion interpolating operators have significant 
overlap with energy eigenstates outside this resonance region. For  
states which have significant overlap onto two-pion interpolators only, the 
ratio fits described previously have very little excited state contamination. 
Clearly, there are a number of states near or above the four-pion threshold 
$E_{\cm}/m_{\pi} = 4$. While these states can be extracted with suitable 
statistical precision, their interpretation in terms of infinite-volume 
scattering amplitudes is unknown. 
\begin{table}
	\centering
	\begin{tabular}{|c|c|c|c|c|c|c|c|c|}
\hline
$\boldsymbol{d}^{2}$ &  irrep  &   level    & $(\boldsymbol{d}_{1}^2, \boldsymbol{d}_{2}^2)$ & $\tmin/a_t$ & $\chi^2$ & $a_t\Delta E$ &  $E_{\cm}/m_{\pi}$  &   $\left(q_{\cm}/m_{\pi}\right)^{3}\cot\delta_{1}$ \\
\hline
$0$   &    $T_{1u}^{+}$   & 0  & $(1, 1)$ & 19 & 1.18 & -0.01214(91) &   3.206(25) & 2.05(29) \\ 
       &            & 1  & $(1, 1)$ & 17 & 0.84 & 0.0086(15) &   3.734(40) & -5.1(1.1) \\ 
\hline
\hline
$1$   &    $A_{1}^{+}$   & 0  & $(0, 1)$ & 14 & 1.22 & -0.00106(18) &   2.3168(61) & 7.9(1.5) \\ 
       &            & 1  & $(1, 2)$ & 17 & 0.99 & -0.01429(89) &   3.373(27) & 0.34(17) \\ 
\hline
       &    $E^{+}$   & 0  & $(1, 2)$ & 19 & 0.9 & -0.0196(11) &   3.226(34) & 1.28(21) \\ 
\hline
\hline
$2$   &    $A_{1}^{+}$   & 0  & $(0, 2)$ & 19 & 0.84 & -0.00225(42) &   2.486(15) & 4.8(1.1) \\ 
       &            & 1  & $(1, 3)$ & 19 & 1.2 & -0.0216(12) &   3.325(36) & -1.05(12) \\ 
\hline
       &    $B_{1}^{+}$   & 0  & $(1, 3)$ & 19 & 0.83 & -0.0248(12) &   3.232(37) & 0.90(19) \\ 
\hline
       &    $B_{2}^{+}$   & 0  & $(1, 1)$ & 22 & 1.0 & -0.00351(72) &   2.749(24) & 3.8(1.1) \\ 
       &            & 1  & $(1, 1)$ & 18 & 1.07 & 0.0210(11) &   3.495(34) & -1.08(18) \\ 
\hline
\hline
$3$   &    $A_{1}^{+}$   & 0  & $(0, 3)$ & 15 & 1.33 & -0.00199(63) &   2.650(22) & 6.6(2.6) \\ 
       &            & 1  & $(1, 2)$ & 15 & 1.39 & -0.00091(56) &   3.132(22) & 2.3(3.4) \\ 
       &            & 2  & $(1, 5)$ & 20 & 1.67 & -0.0330(27) &   3.498(85) & -2.10(39) \\ 
\hline
       &    $E^{+}$   & 0  & $(1, 2)$ & 19 & 1.07 & -0.0060(11) &   2.966(38) & 3.22(98) \\ 
       &            & 1  & $(1, 2)$ & 17 & 1.07 & 0.0129(20) &   3.570(62) & -3.61(48) \\ 
\hline
\hline
$4$   &    $A_{1}^{+}$   & 0  & $(0, 4)$ & 13 & 0.99 & -0.00269(92) &   2.751(35) & 5.2(2.5) \\ 
       &            & 1  & $(0, 4)$ & 18 & 1.06 & 0.0154(23) &   3.382(79) & -1.78(40) \\ 
\hline
       &    $E^{+}$   & 0  & $(0, 4)$ & 16 & 1.03 & 0.0107(16) &   3.225(58) & 1.89(67) \\ 
       &            & 1  & $(0, 4)$ & 18 & 1.31 & 0.0296(32) &   3.84(10) & -0.1(2.3) \\ 
\hline
\end{tabular}

	\caption{\label{t:i1dat}Results for the center-of-mass energies and 
	scattering phase shifts in $I=1$. For each total momentum 
	($\boldsymbol{d}^2$), lattice irrep and energy level, the two single-pion 
	correlation functions used in the ratio fits are denoted by $(\boldsymbol{d}^2_1,\boldsymbol{d}^2_2)$. The minimum time included in the fit range, the 
	correlated-$\chi^2$, fitted energy shift, reconstructed center-of-mass energy,
	and scattering phase shift are also given for each energy level. 
}
\end{table}

Numerical results for our final analysis using ratio fits are listed in Tab.~\ref{t:i1dat}, 
where $(q_{\cm}/m_{\pi})^3\cot{\delta_{1}}$ is obtained by applying the formulae of 
Tab.~\ref{t:phase}.  
\begin{figure}
	\includegraphics[width=0.49\textwidth]{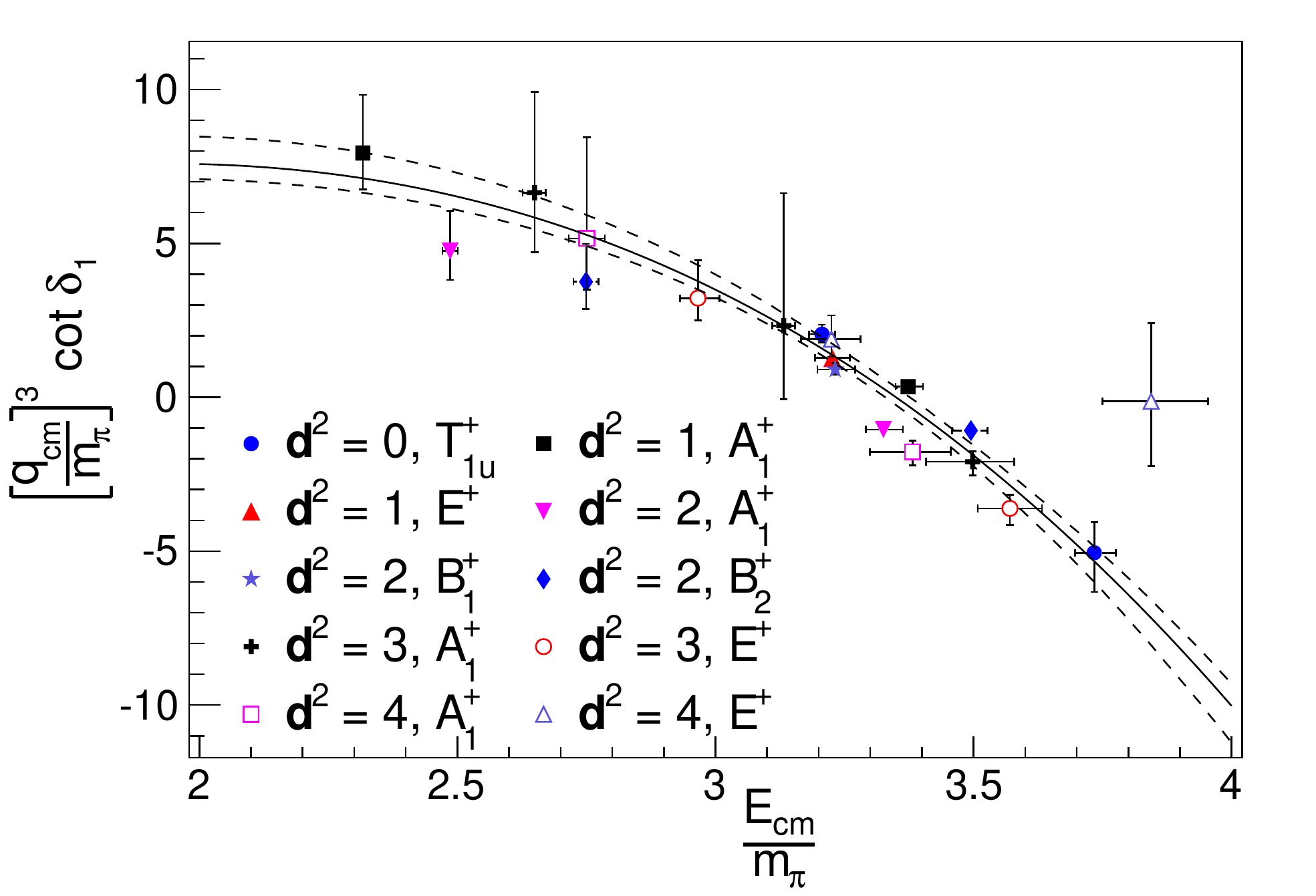}
	\includegraphics[width=0.49\textwidth]{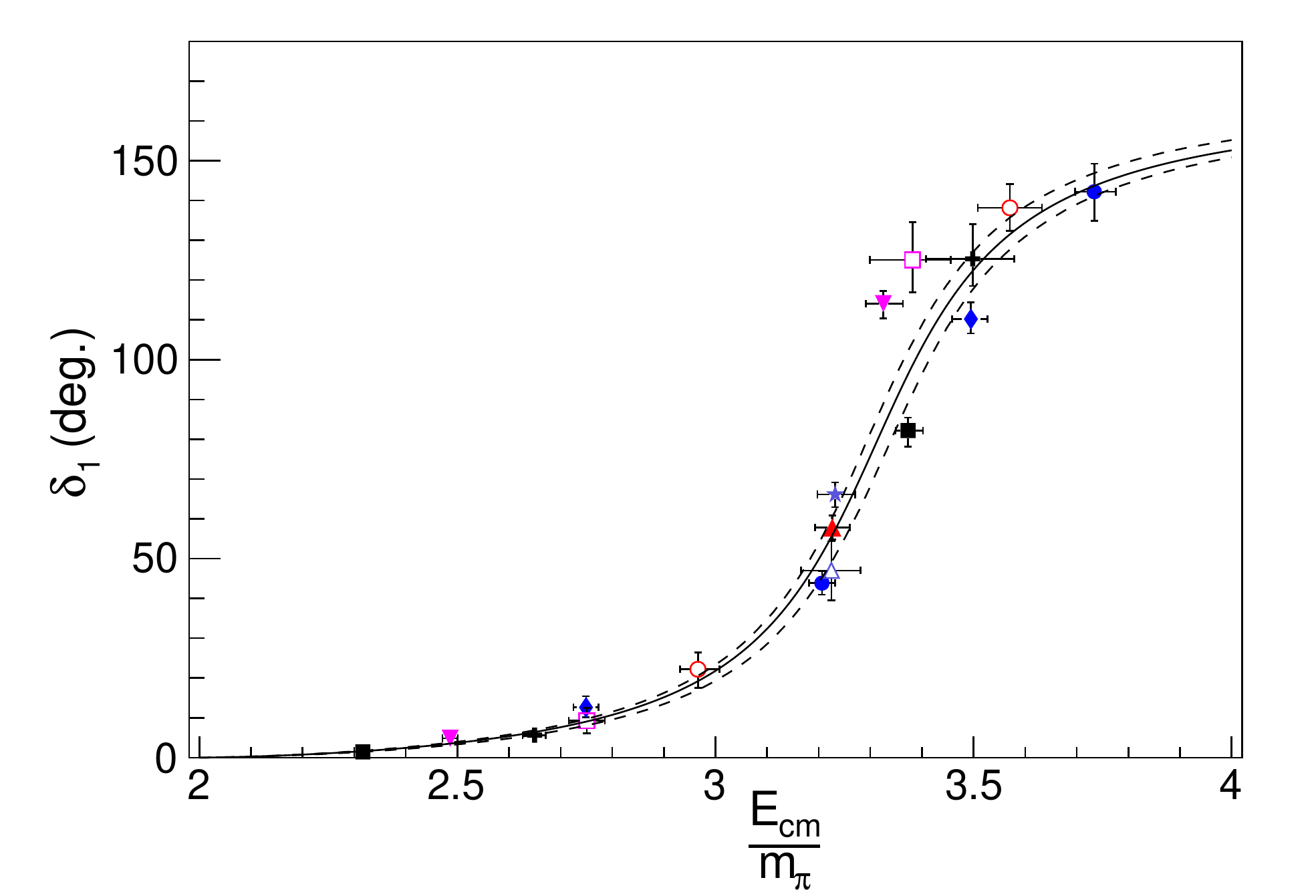}
	
	\includegraphics[width=0.49\textwidth]{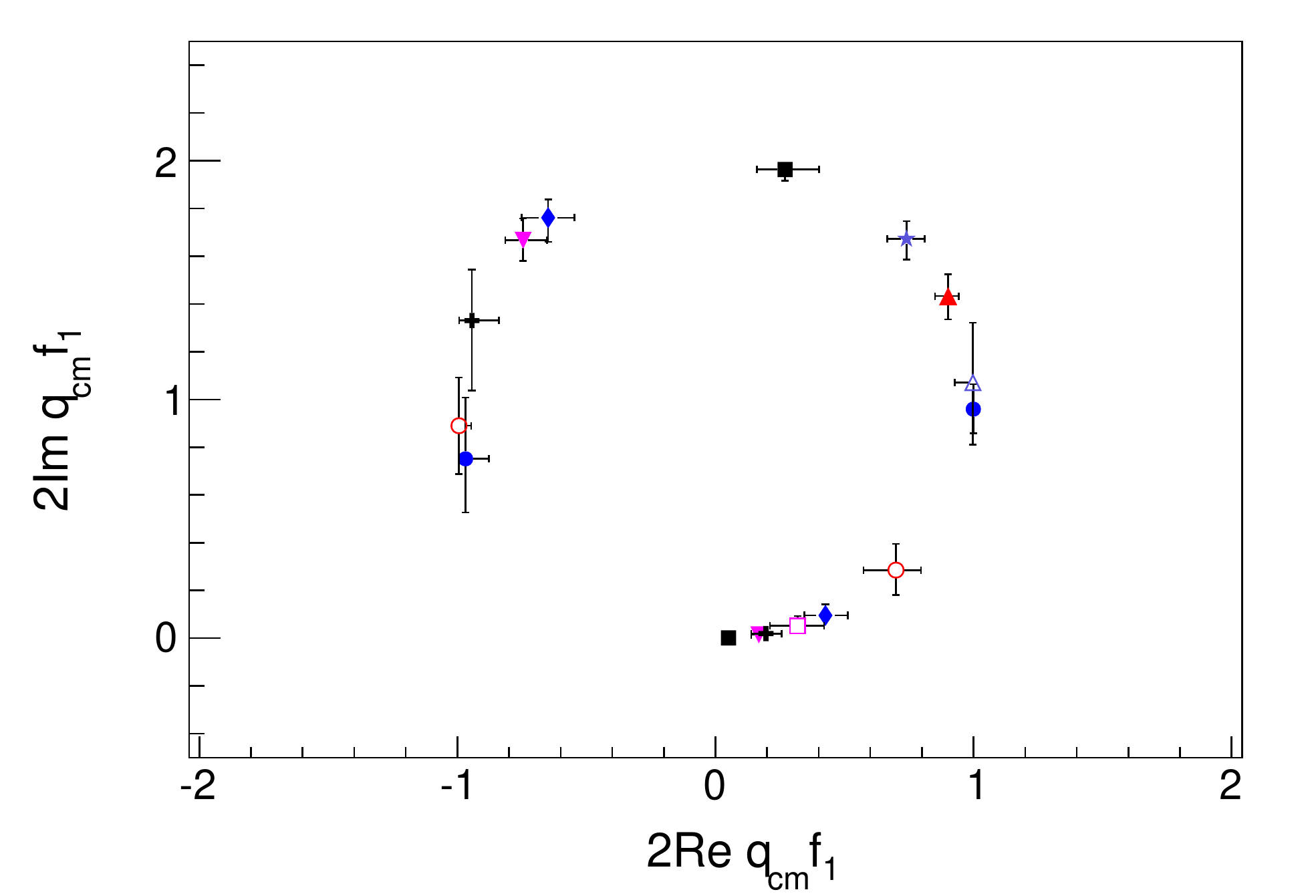}
	\includegraphics[width=0.49\textwidth]{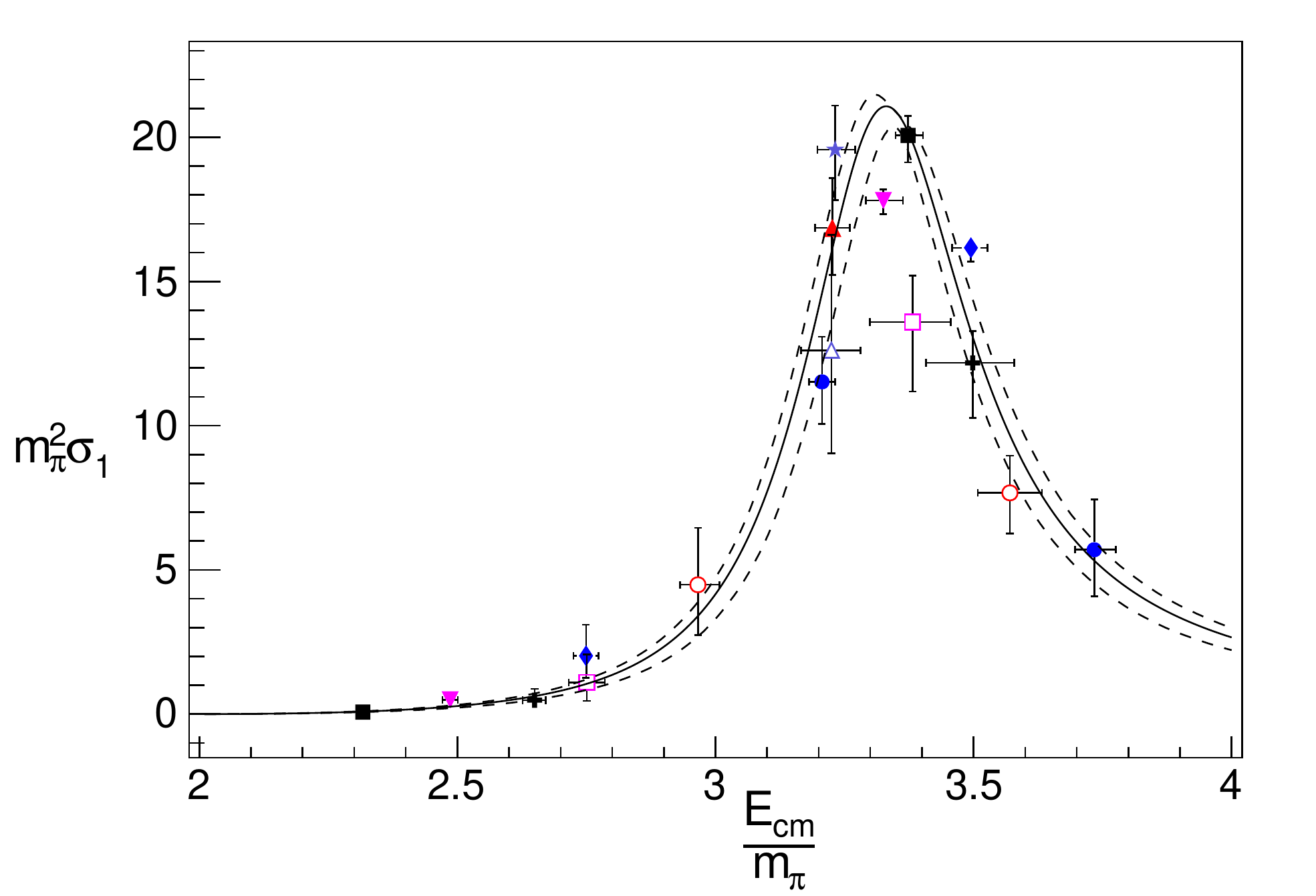}
	\caption{\label{f:i1scat} The real part of the inverse scattering 
		amplitude (top left), phase shift (top right), Argand plot (lower left), and partial wave amplitude (lower right) 
	 for the $I=1$, $\ell =1$ partial wave. The dotted lines indicate a 
 fit to data in the upper left plot.  
 Points from the first excited state in the $\boldsymbol{d}^2=3$, $A_{1}^{+}$ and $\boldsymbol{d}^2=4$, $E^{+}$ channels are omitted from the 
 phase shift, Argand, and amplitude plots due to their large errors.}
\end{figure}
This particular quantity is the real part of the inverse scattering 
amplitude and is thus analytic in the complex momentum plane 
near the two-pion threshold $E_{\cm}/m_{\pi} = 2$, making it a natural choice
for fits of the amplitude's energy dependence. 

For this resonant $\ell=1$ partial wave, $(q_{\cm}/m_{\pi})^{3}\cot \delta_{1}$ can be 
described by the Breit-Wigner parametrization 
\begin{align}\label{e:bw}
	\left(\frac{q_{\cm}}{m_{\pi}}\right)^{3}\cot \delta_1 = 
	\left(\frac{m^2_{\rho}}{m^2_{\pi}} - \frac{E^2_{\cm}}{m_{\pi}^2}\right)
	\frac{6\pi E_{\cm}}{g_{\rho\pi\pi}^2m_{\pi}}
\end{align}
which also has the correct threshold behavior dictated by the effective range expansion. 
A two-parameter (fully-correlated) $\chi^2$-fit to Eq.~\ref{e:bw} is performed.
This fit must not only take into account the correlation between 
different data points, but also the correlation between $E_{\cm}/m_{\pi}$ 
and $	\left(q_{\cm}/m_{\pi}\right)^{3}\cot \delta_1 $ for each data 
point. In order to do this, we employ the correlated-$\chi^2$ which is 
the maximum likelihood estimator for the distribution of the residuals $d_i$
\begin{align}
	\chi^2 &= \sum_{i,j} d_i \, \mathrm{Cov}^{-1}(i,j) \,d_j, 
	\\\nonumber d_i(m_{\rho}/m_{\pi}, g_{\rho\pi\pi}) &= 	\left[\left(\frac{q_{\cm}}{m_{\pi}}\right)^{3}\cot \delta_1\right]_i - \left[\left(\frac{m^2_{\rho}}{m^2_{\pi}} - \frac{E^2_{\cm}}{m_{\pi}^2}\right)
	\frac{6\pi E_{\cm}}{g_{\rho\pi\pi}^2m_{\pi}}\right]_i.
\end{align}
As with the other fits in this work, the bootstrap estimator is used to 
estimate the covariance between the $\{d_i\}$. However, each $d_i$ depends 
nontrivially on the fit parameters $m_{\rho}/m_{\pi}$ and $g_{\rho\pi\pi}$ 
so the bootstrap estimate of the covariance must be recalculated on each 
call to the correlated-$\chi^2$ function. In other words, on each bootstrap sample, each call to the correlated-$\chi^2$ employs all bootstrap samples to 
estimate the covariance. While this method may seem cumbersome, it ensures that 
all correlations among the data are taken into account.  

The results of this fit are 
\begin{align}\label{e:rhofit}
	\frac{m_{\rho}}{m_{\pi}} = 3.350(24), \qquad g_{\rho\pi\pi} = 5.99(26) , \qquad \chi^2/d.o.f. = 1.04. 
\end{align}
While other resonance parametrizations have been applied  
(in e.g. Ref.~\cite{Dudek:2012xn}) which maintain unitarity above the resonance 
region, given the proximity of the four-pion threshold such parametrizations 
seem poorly motivated here. However, we test the dependence of these resonance 
parameters on the Breit-Wigner fit form by employing a non-relativistic 
ansatz to $\left(q_{\cm}/m_{\pi}\right)^{3}\cot \delta_1$
\begin{align}
	\tan \delta_1 = \frac{\Gamma/2}{m_{\rho} - E_{\cm}} + A, 
	\quad \Gamma = \frac{g_{\rho\pi\pi}^2}{48\pi m_{\rho}^2}(m_{\rho}^2 - 4m_{\pi}^2)^{3/2},
\end{align}
where $\Gamma$ is an energy-independent width and $A$ parametrizes a 
slowly-varying background. This three-parameter fit gives 
\begin{align}\label{e:rhofit2}
	\frac{m_{\rho}}{m_{\pi}} = 3.352(23), \qquad g_{\rho\pi\pi} = 5.84(34) , \qquad A = -0.160(26) , \qquad \chi^2/d.o.f. = 1.03. 
\end{align}

A 
summary of our data as well as the fit of Eq.~\ref{e:rhofit} are shown in Fig.~\ref{f:i1scat}. 
Several different representations of the data are shown in that figure. First,
the $	(q_{\cm}/m_{\pi})^{3}\cot \delta_1 $ data points are 
shown with the corresponding fit to them. Then, $\delta_1$ is shown (in 
$[0,\pi]$) with the fit to $(q_{\cm}/m_{\pi})^{3}\cot \delta_1 $. The rapid variation of the phase shift is clear in this plot. Further 
evidence of this rapid variation is seen in an Argand plot showing the real 
and imaginary parts of the partial wave amplitude (following the conventions 
of Ref.~\cite{taylor2012scattering}) 
\begin{align}
	q_{\cm} f_{1} = \mathrm{e}^{i\delta_1}\sin \delta_1.
\end{align}
Finally, a plot of the 
partial wave cross section 
\begin{align}
	m_{\pi}^2 \sigma_1 = 12\pi m_{\pi}^2 \frac{\sin^2 \delta_1}{q_{\cm}^2}
\end{align}
shows a clear enhancement due to the resonance. 

Due to the singular nature of the L\"{u}scher zeta functions at 
non-interacting energies, the distribution of bootstrap samples of the 
quantities shown in Fig.~\ref{f:i1scat} can show significant asymmetry. In that 
figure we therefore display asymmetric $1\sigma$ bootstrap error bars. 
Displaying the points in this manner indicates the level of asymmetry but
ignores the correlation between the horizontal and vertical error bars.

\subsection{$I=2$}

\begin{table}
	\centering
	\begin{tabular}{|c|c|c|c|c|c|c|c|c|}
\hline
$\boldsymbol{d}^{2}$ &  irrep  &   level    & $(\boldsymbol{d}_{1}^2, \boldsymbol{d}_{2}^2)$ & $\tmin/a_{t}$ & $\chi^2$ & $a_{t}\Delta E$ &  $(q_{\cm}/m_{\pi})^2$  &   $\left(q_{\cm}/m_{\pi}\right)\cot\delta_{0}$ \\
\hline
$0$   &    $A_{1g}^{+}$   & 0  & $(0, 0)$ & 9 & 1.4 & 0.00082(17) &   0.0210(42) & -16.5(3.2) \\ 
       &            & 1  & $(1, 1)$ & 10 & 1.16 & 0.00519(63) &   2.324(40) & -7.9(1.1) \\ 
\hline
\hline
$1$   &    $A_{1}^{+}$   & 0  & $(0, 1)$ & 14 & 0.97 & 0.00170(35) &   0.439(12) & -10.8(2.1) \\ 
       &            & 1  & $(1, 2)$ & 12 & 1.07 & 0.0075(11) &   2.939(66) & -5.6(1.1) \\ 
\hline
\hline
$2$   &    $A_{1}^{+}$   & 0  & $(0, 2)$ & 11 & 1.24 & 0.00133(26) &   0.693(12) & -10.3(2.2) \\ 
       &            & 1  & $(1, 1)$ & 10 & 1.14 & 0.00191(23) &   1.130(16) & -7.81(84) \\ 
\hline
\hline
$3$   &    $A_{1}^{+}$   & 0  & $(0, 3)$ & 10 & 1.43 & 0.00158(48) &   0.922(24) & -6.9(2.9) \\ 
       &            & 1  & $(1, 2)$ & 10 & 1.05 & 0.00447(44) &   1.732(31) & -8.01(72) \\ 
\hline
\hline
$4$   &    $A_{1}^{+}$   & 0  & $(1, 1)$ & 10 & 1.13 & 0.00089(33) &   0.029(14) & -7.3(3.9) \\ 
       &            & 1  & $(0, 4)$ & 14 & 1.27 & 0.0057(19) &   1.137(67) & -5.7(4.1) \\ 
       &            & 2  & $(2, 2)$ & 12 & 1.23 & 0.00017(68) &   2.131(51) & -23(31) \\ 
\hline
\end{tabular}

	\caption{\label{t:i2dat}The same as Tab.~\ref{t:i1dat} but $I=2$ data for the $\ell=0$ partial wave.}
\end{table}
The $I=2$ channel is weakly interacting and thus a good test 
of the stochastic LapH method. As in the $I=1$ case, we examine the real part 
of the inverse scattering amplitude, which is analytic near the 
two-pion threshold. Our fitted energies and 
resultant phase shifts for the $\ell =0$ partial wave 
are shown in Tab.~\ref{t:i2dat}. 
\begin{figure}
	\includegraphics[width=0.49\textwidth]{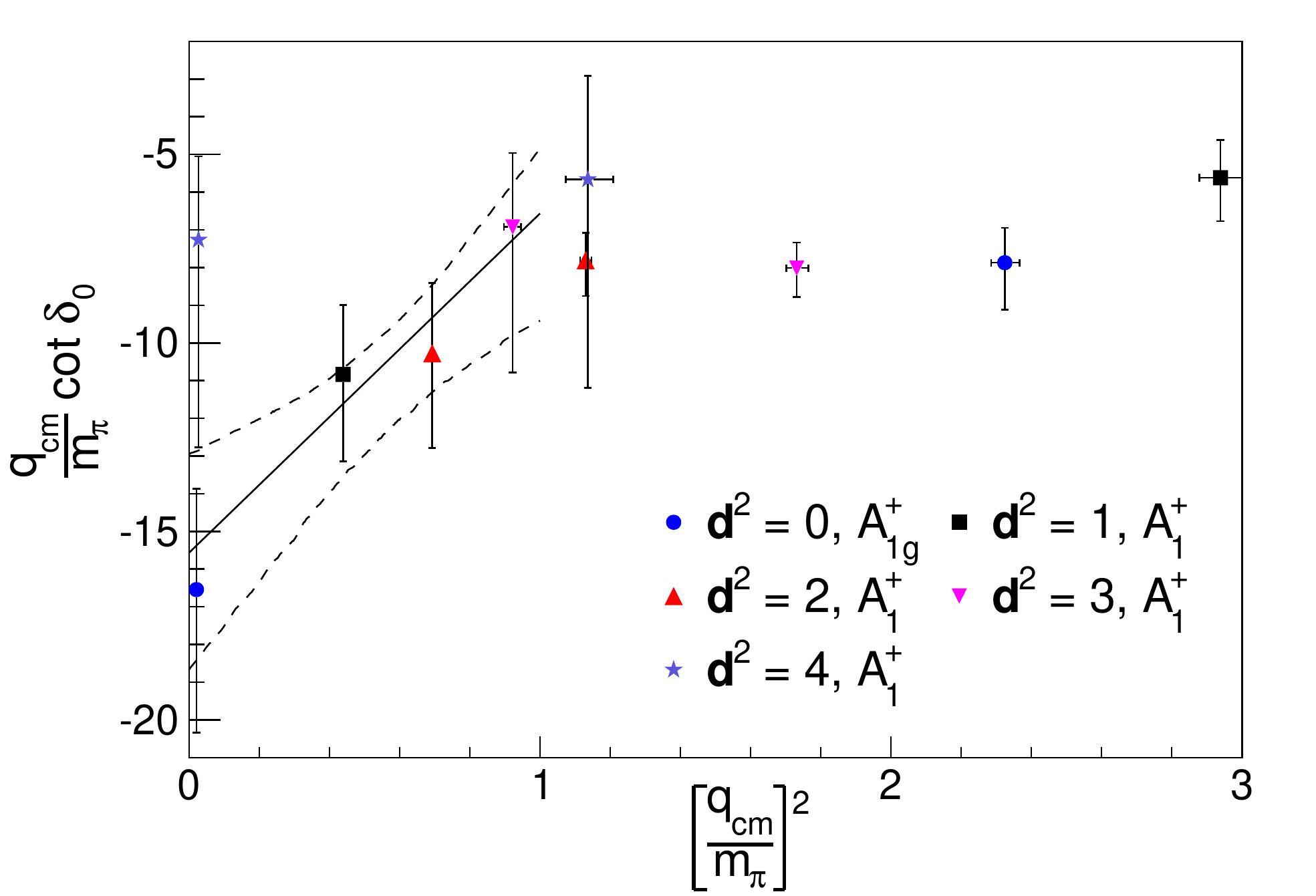}
	\includegraphics[width=0.49\textwidth]{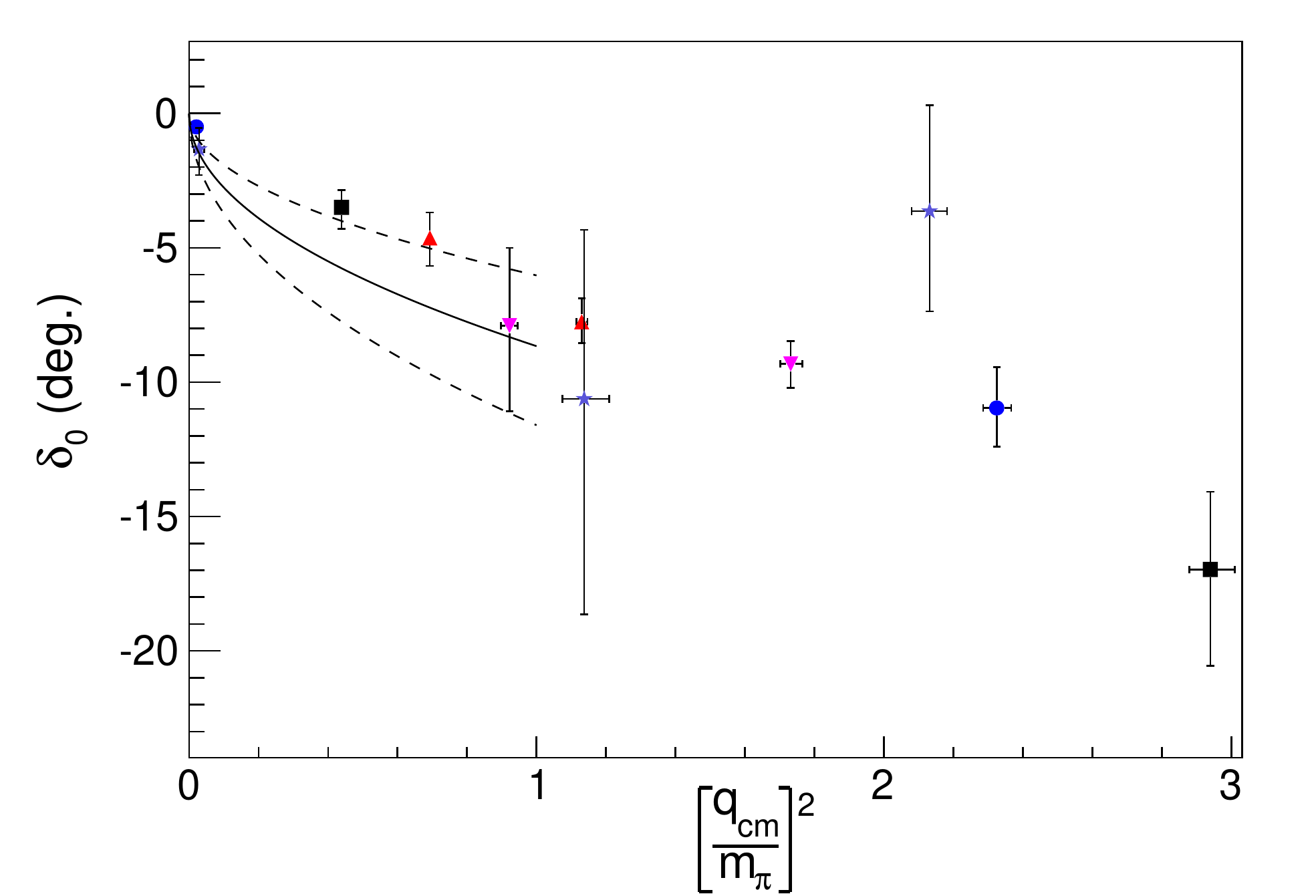}
	\caption{\label{f:i2scat} The real part of the inverse scattering amplitude 
		(left) and the scattering phase shift (right) for the $I=2$, $\ell=0$ 
		partial wave.}
\end{figure}
The weakly interacting nature of this channel motivates its description by the lowest few terms of the effective range expansion
\begin{align}\label{e:er}
	\left(\frac{q_{\cm}}{m_{\pi}}\right)\cot \delta_0 = \frac{1}{m_{\pi}a_{0}} + 
	\frac{1}{2}(m_{\pi}r)\left( \frac{q_{\cm}}{m_{\pi}}\right)^2. 
\end{align}
This parametrization is expected to be valid for momenta below the 
$t$-channel cut $q_{\cm} \ll m_{\pi}$~\cite{Beane:2011sc}. 

Our results are collected in Fig.~\ref{f:i2scat}. 
Due to the smaller number of finite-volume irreps in which the $\ell =0$ 
partial wave appears, there are only four points in this low-momentum region.
A two-parameter fit to the effective range ansatz of Eq.~\ref{e:er} yields 
\begin{align}\label{e:i2fit}
	m_{\pi} a^{I=2}_{0} = -0.064(12) ,\qquad m_{\pi} r = 18.1(8.4), \qquad \chi^2/d.o.f. = 0.19. 
\end{align}
The small number of points in this channel suggests that not much can be 
gained by adding the next term in the effective range expansion which contains 
the shape parameter. The scattering length is determined with about $20\%$ 
precision and is consistent with the (continuum) $\chi$PT extrapolation of 
(e.g.) Ref.~\cite{Helmes:2015gla} but the pion mass used in this work is 
lighter than those employed there.

\section{Conclusions}\label{s:concl}

The elastic $I=1$ and $I=2$ $\pi-\pi$ scattering phase shifts are determined from a $\nf=2+1$ dynamical lattice QCD simulation in a 
large spatial volume with a light pion mass. In particular, the stochastic 
estimation scheme employed here performs efficiently, and determines the 
correlation functions with sufficient precision to extract the finite-volume energies and 
scattering phase shifts. 
This suggests that larger volumes and lighter pions are possible due to the 
favorable scaling of the stochastic LapH method. 

\begin{figure}
	\includegraphics[width=0.49\textwidth]{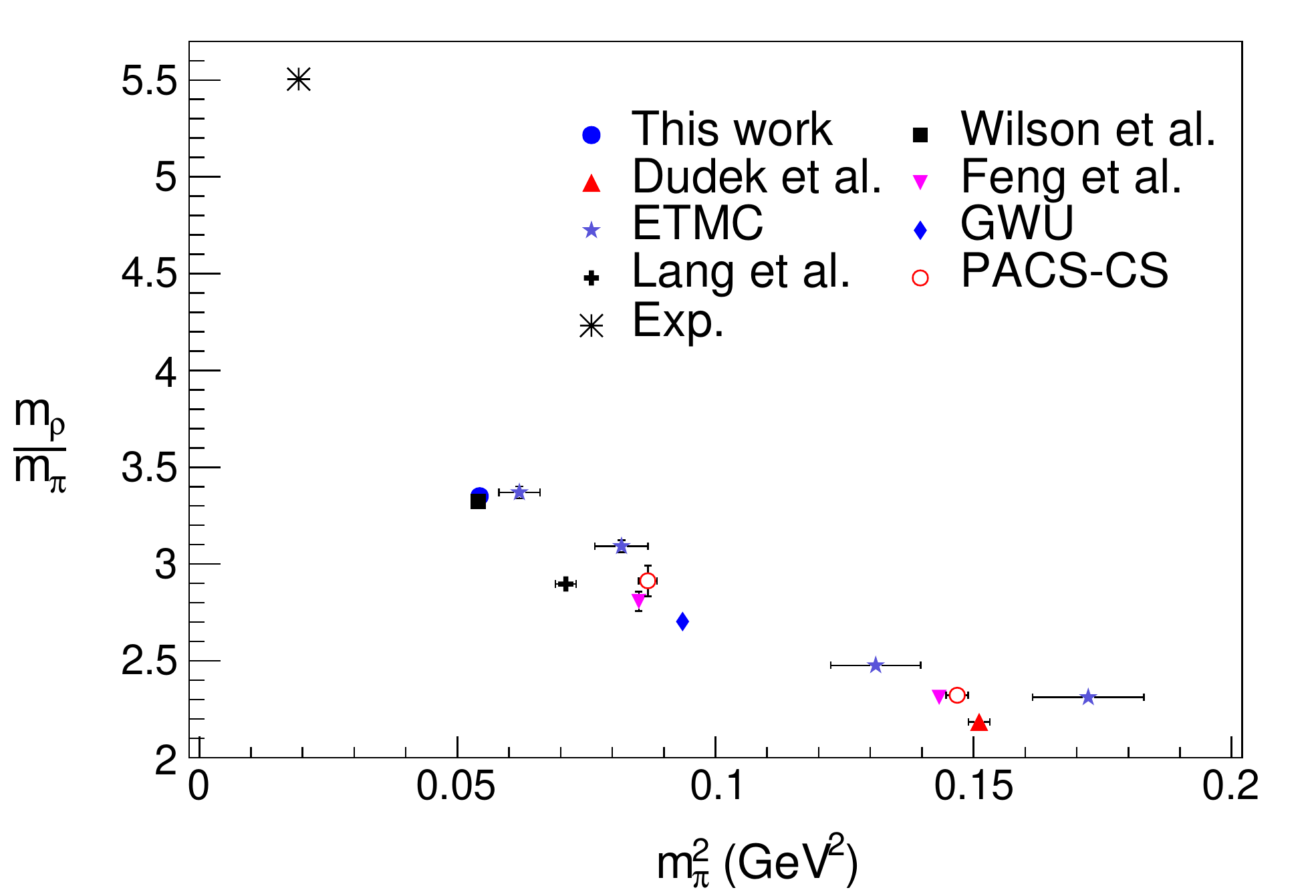}
	\includegraphics[width=0.49\textwidth]{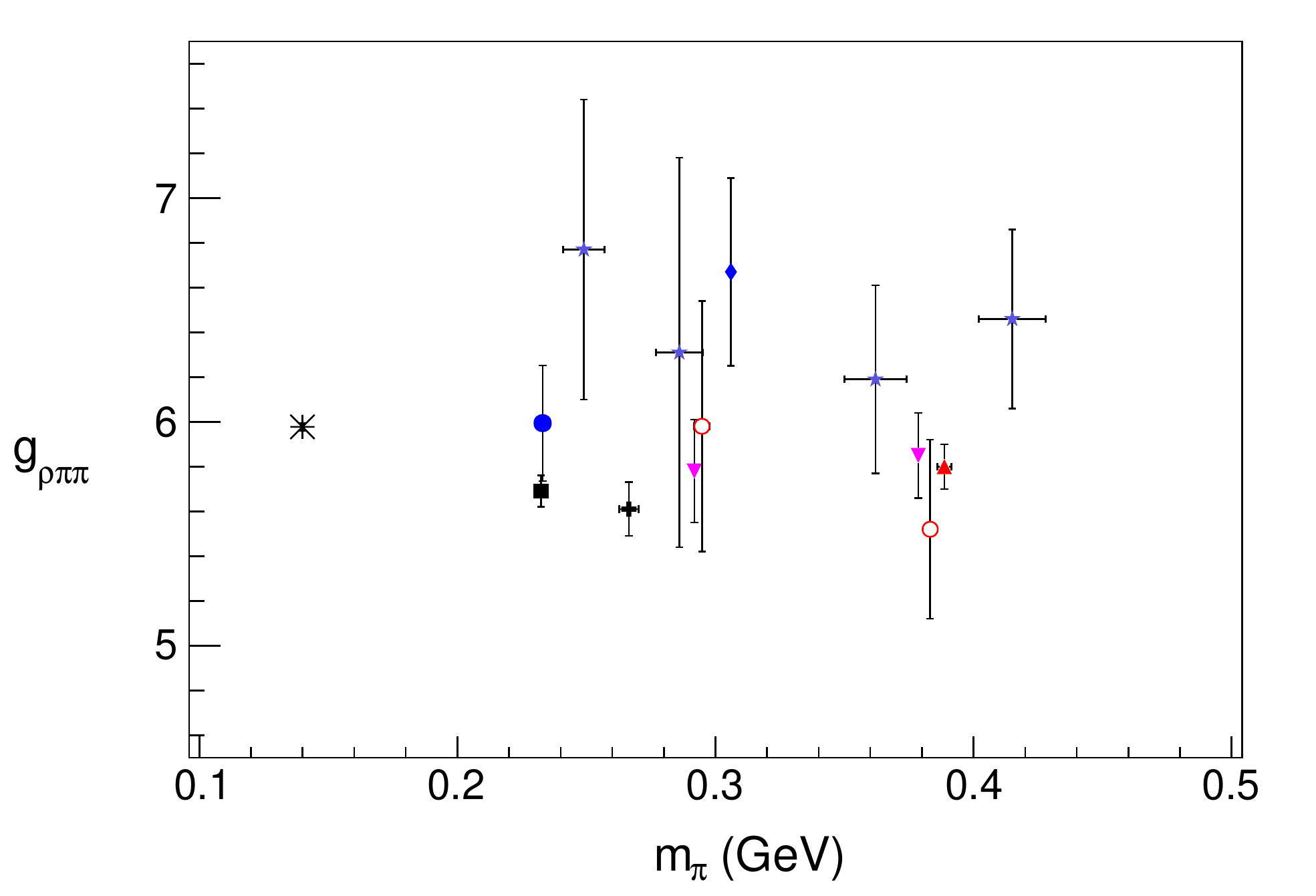}
	\caption{\label{f:rho_sum}Summary of recent published work for $m_{\rho}$ and 
	$g_{\rho\pi\pi}$. The legend denotes Ref.~\cite{Dudek:2012xn} `Dudek et al.',
  Ref.~\cite{Feng:2010es} `ETMC', Ref.~\cite{Lang:2011mn} `Lang, et al.', 
	Ref.~\cite{Wilson:2015dqa} `Wilson et al.', Ref.~\cite{Feng:2014gba} `Feng, et al.', Ref.~\cite{Pelissier:2012pi} `GWU', and Ref.~\cite{Aoki:2011yj} 
`PACS-CS'.}
\end{figure}

After extracting finite-volume energy levels, the L\"{u}scher method is employed to 
calculate elastic scattering phase shifts. 
The $I=1$, $\ell=1$ partial wave is well described by a Breit-Wigner form and 
exhibits rapid phase motion indicative of a resonance. 
Our main results are Fig.~\ref{f:i1scat} and Eq.~\ref{e:rhofit}. 
We have compiled recent published calculations of the $\rho$-resonance in 
Fig.~\ref{f:rho_sum} 
indicating that this calculation (together with Ref.~\cite{Wilson:2015dqa}) is 
the closest to the physical quark masses achieved so far. Fig.~\ref{f:rho_sum} 
compares $m_{\rho}/m_{\pi}$ to reduce scale uncertainties, as none of the 
results are extrapolated to the continuum limit. 
The results for 
the mass are generally in good agreement, but $g_{\rho\pi\pi}$ is known 
with considerably less precision.   

Due to our light quark
masses, the lowest inelastic threshold (due to four pions) is close to the 
resonance region limiting
the applicability of the elastic L\"{u}scher formulae. Hopefully, existing 
work on extending the L\"{u}scher formulae to three-particle 
scattering~\cite{Polejaeva:2012ut,Hansen:2014eka,Hansen:2015zga} can 
be adapted to treat these thresholds in the future. Of course, the problem 
 worsens as the quark masses are lowered to their physical values as 
experimentally $m_{\rho} > 4m_{\pi}$. Once this threshold 
can be treated quantitatively its effect may be small, as the 
experimental branching fraction for $\rho\rightarrow4\pi$ is below the percent
level. 

We have less points below inelastic threshold for the $I=2$, $\ell=0$ partial 
wave, as there are fewer lattice irreps in which it appears. Still,
our data below the $t$-channel cut $q_{\cm} = m_{\pi}$ is well-described by the 
first two terms in the effective range expansion and provides a determination of the scattering length to about 20\%. Our results for $I=2$
are shown in Fig.~\ref{f:i2scat} and Eq.~\ref{e:i2fit}.  Calculations of the $I=2$ $s$-wave scattering length\footnote{For a recent review of 
these calculations see Ref.~\cite{Helmes:2015gla} and the references quoted therein.} 
are considerably more advanced than in $I=1$, so a single-ensemble result is not fit for direct comparison.
However, the $\approx 20\%$ error on $a_{0}$ is somewhat remarkable given our stochastic estimation of the all-to-all quark propagators
and the precise calculation of small energy shifts required to obtain a signal.

As mentioned in the introduction, Ref.~\cite{Wilson:2015dqa} appeared 
during the preparation of this manuscript which uses the full 
distillation method to treat the required all-to-all propagators and 
 can be viewed as the maximal dilution limit of our approach. 
We compare results in Tab.~\ref{t:comp} for a selection of published 
\begin{table}
\centering
\begin{tabular}{|c|c|c|c|c|c|c|c|c|}
	\hline
		Ref. & $N_{D}$ & $a_{t}m_{\pi}$ & $T_{1u}^{+}$ $E_0$ & 
	$T_{1u}^{+}$ $E_1$ & $a_tm_{\rho}$ & $g_{\rho\pi\pi}$ \\
		\hline
		This work & 2304 & 0.03939(19) & 0.12625(94) & 0.1470(16) &
		  0.13190(87) & 5.99(26) \\ 
		\hline
		Ref.~\cite{Wilson:2015dqa} & 393216 & 0.03928(18) & 0.12488(40) 
		&  0.14534(52) & 0.13175(35) &  5.688(70)\\ 
		\hline
\end{tabular}
\caption{\label{t:comp}Comparison of the number of Dirac matrix inversions per
configuration $N_D$, the pion mass, the first two energies in the $I=1$, 
$\boldsymbol{d}^2=0$, $T_{1u}^{+}$ channel, and the $\rho$ resonance parameters 
between this work and Ref.~\cite{Wilson:2015dqa}. We see that while the 
precision on the pion mass is comparable, the distillation method is 
2-3 times more precise for the other quantities while requiring about 
170 times more Dirac matrix inversions.}
\end{table}
numbers as well as the required number of Dirac matrix inversions per 
configuration. Although $a_tm_{\rho}$ and $g_{\rho\pi\pi}$ are also obtained 
in Ref.~\cite{Wilson:2015dqa} from a Breit-Wigner ansatz, their fitting method 
constructs a correlated-$\chi^2$ directly from the finite-volume energies 
rather than $(q_{\cm}/m_{\pi})^{3}\cot \delta_1$. However, the errors on 
$m_{\pi}$ and $\xi$ (which are comparable to those on the energies) are not 
taken into account in their fit procedure. It is unclear what effect this has 
on the resultant fit parameters and their errors. Our methods for extracting 
finite-volume energies are also different from those employed in 
Ref.~\cite{Wilson:2015dqa}.

We see that the distillation results are comparable in precision for the pion, 
but have roughly half the statistical error for two-pion states, 
while requiring about 170 times more Dirac matrix inversions per configuration. The Dirac matrix inversion cost for the distillation method is significantly 
larger than the cost for the gauge generation and does 
not include the (sizeable) cost of constructing correlation functions from the 
sources and solutions, which also scales poorly with 
the volume. Even so, presumably a 170-fold increase in computational effort 
would reduce the error on our results by more than an order of 
magnitude, making them significantly more precise than Ref.~\cite{Wilson:2015dqa}.  

Overall, this first large-volume scattering calculation using stochastic LapH 
is promising for future work. 
 As it is clear 
that scattering calculations are entering a new era of increased 
statistical precision, it is important to quantify the remaining systematic 
errors. These include exponential finite-volume effects, 
the effect of higher partial waves, the presence of inelastic thresholds, and 
lattice spacing effects.
To this end, work has progressed~\cite{Bulava:2015qjz} in applying the stochastic LapH method to state-of-the-art ensembles generated by 
the Coordinate Lattice Simulations (CLS) consortium~\cite{Bruno:2014jqa}. Apart from the elastic scattering phase shifts presented here, these 
isotropic ensembles simplify the renormalization and $\mathrm{O}(a)$-improvement pattern of composite operators, enabling the determination 
of resonance matrix elements. Preliminary work on the simplest such matrix element, the timelike pion form factor, is also reported in Ref.~\cite{Bulava:2015qjz}. 
Finally, pushing to lighter pions would be desired. While this can be done 
using these CLS ensembles, the lower inelastic thresholds limit the 
applicability of the 
L\"{u}scher formula. More theoretical work is required to rigorously treat 
these thresholds.

\vskip 0.3cm

\noindent
{\bf Acknowledgements}: 
We acknowledge helpful conversations with Robert Edwards, David Wilson, and Max Hansen.
BH is supported by Science Foundation Ireland under Grant No. 11/RFP/PHY3218. CJM acknowledges support from the U.S. NSF under award PHY-1306805 and through TeraGrid/XSEDE
resources provided by TACC, SDSC, and NICS under grant number TG-MCA07S017.
The USQCD QDP++ library~\cite{Edwards:2004sx} and the Improved
BiCGStab solver in Chroma were used in developing
the software for early stages of the calculations reported here.

\appendix
\newpage
\section{$\tmin$-plots for moving pions}\label{a:pi} 
\begin{figure}[!htb]
	\includegraphics[width=0.32\textwidth]{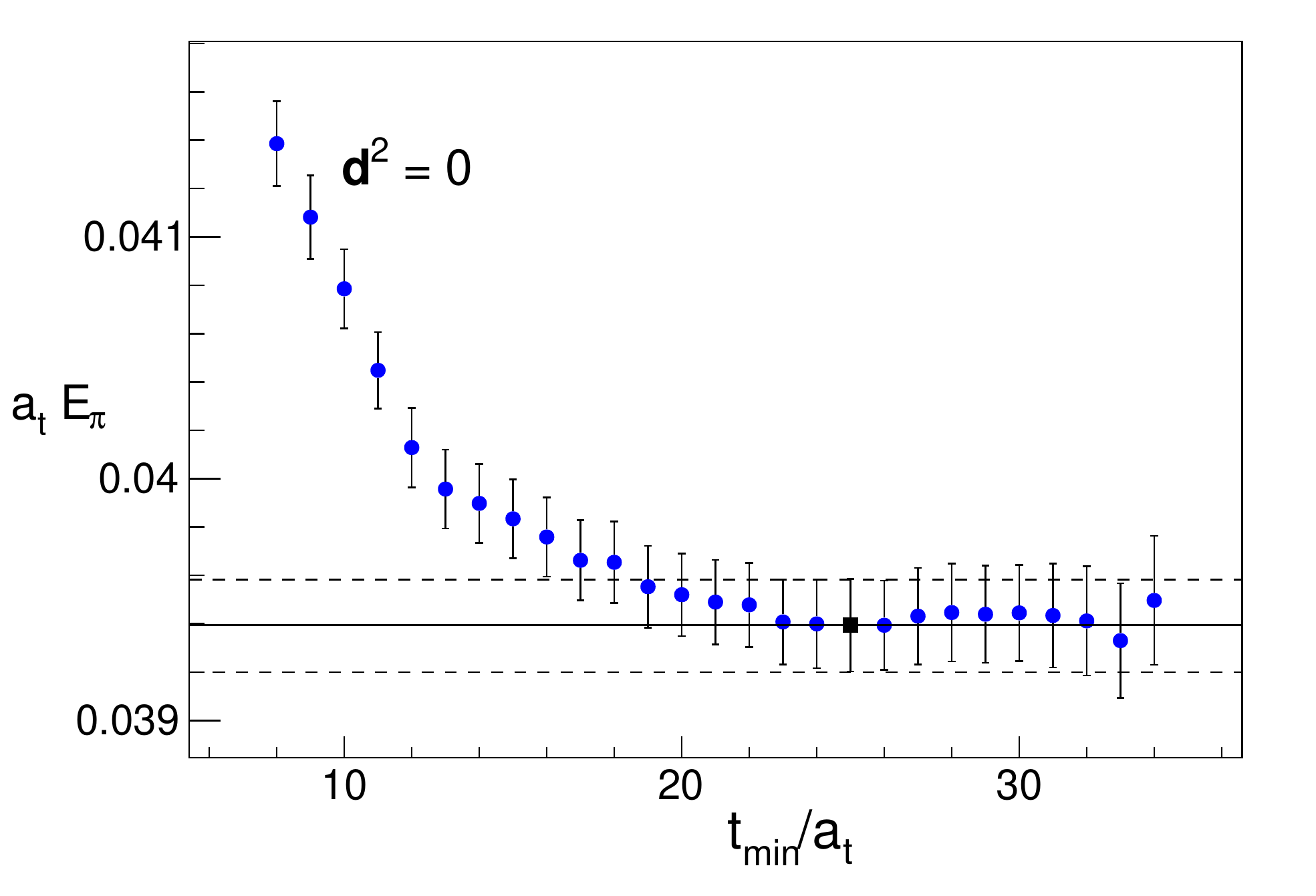}
	\includegraphics[width=0.32\textwidth]{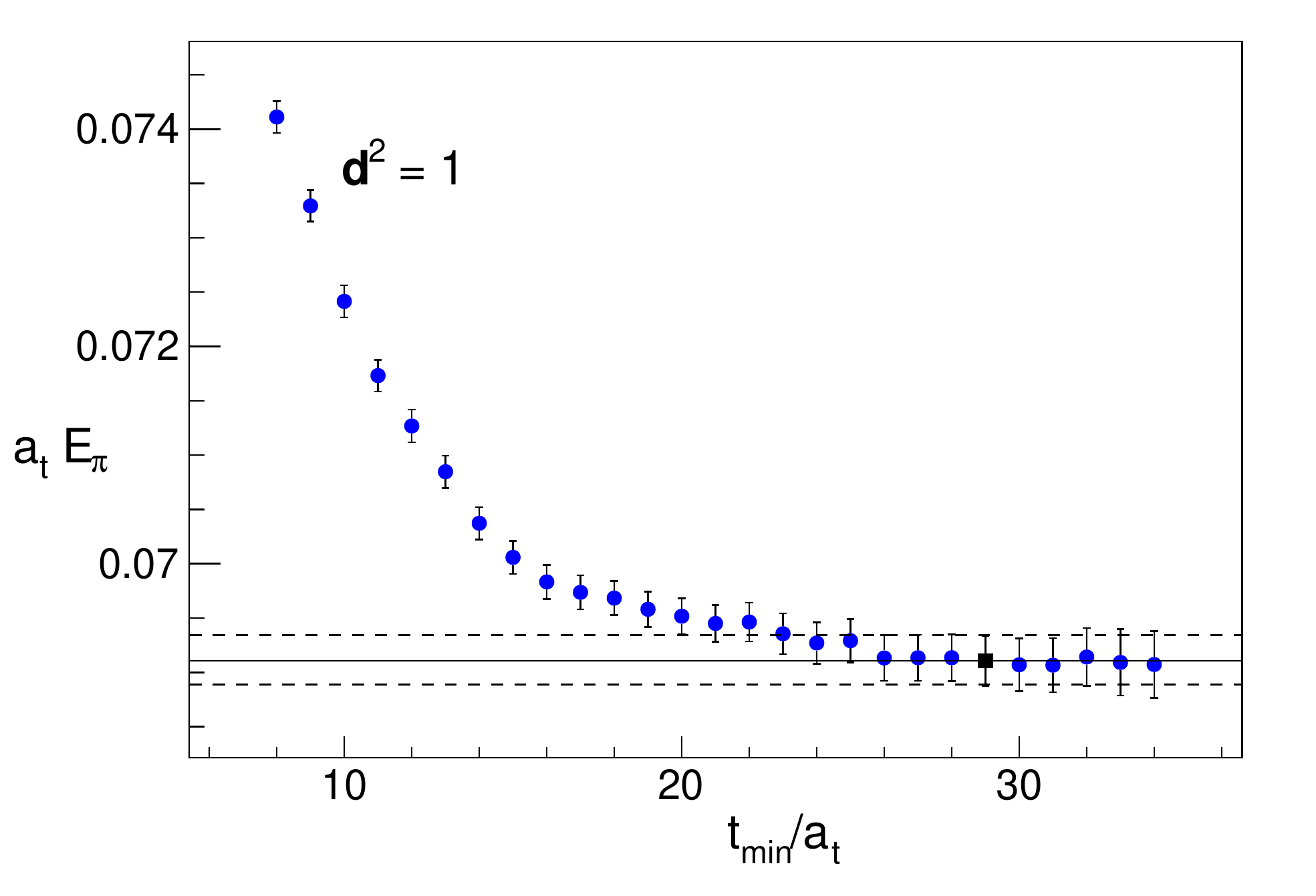}
	\includegraphics[width=0.32\textwidth]{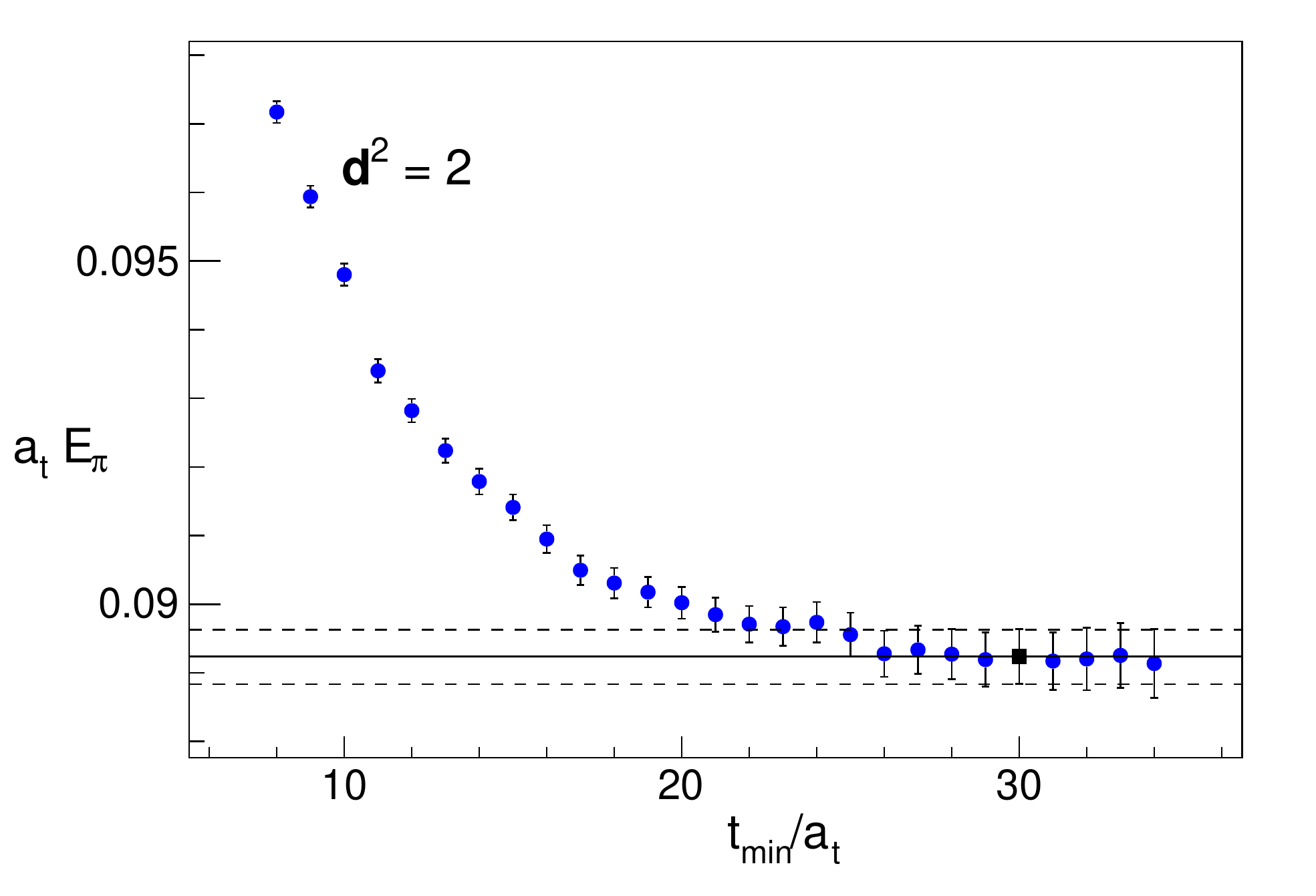}
	
	\includegraphics[width=0.32\textwidth]{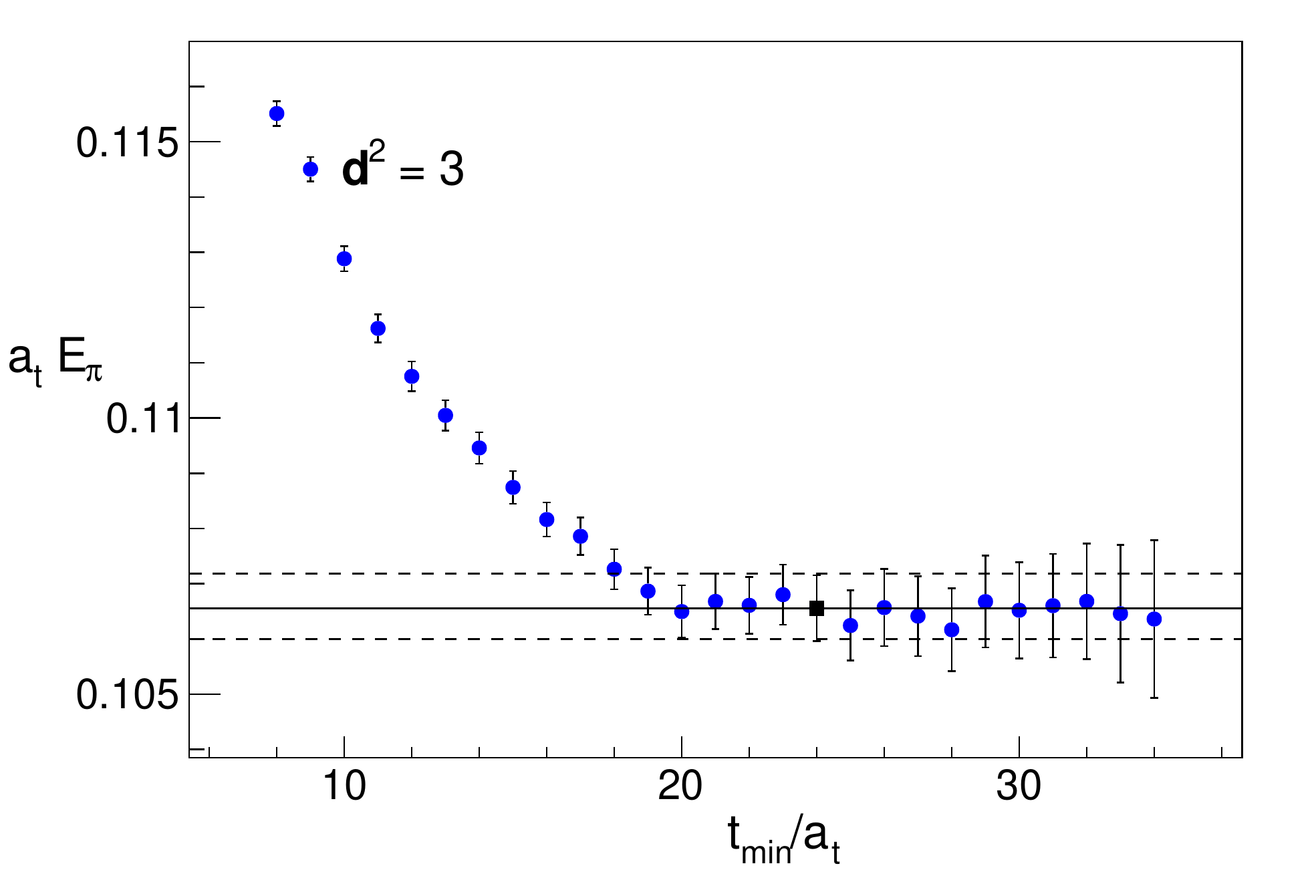}
	\includegraphics[width=0.32\textwidth]{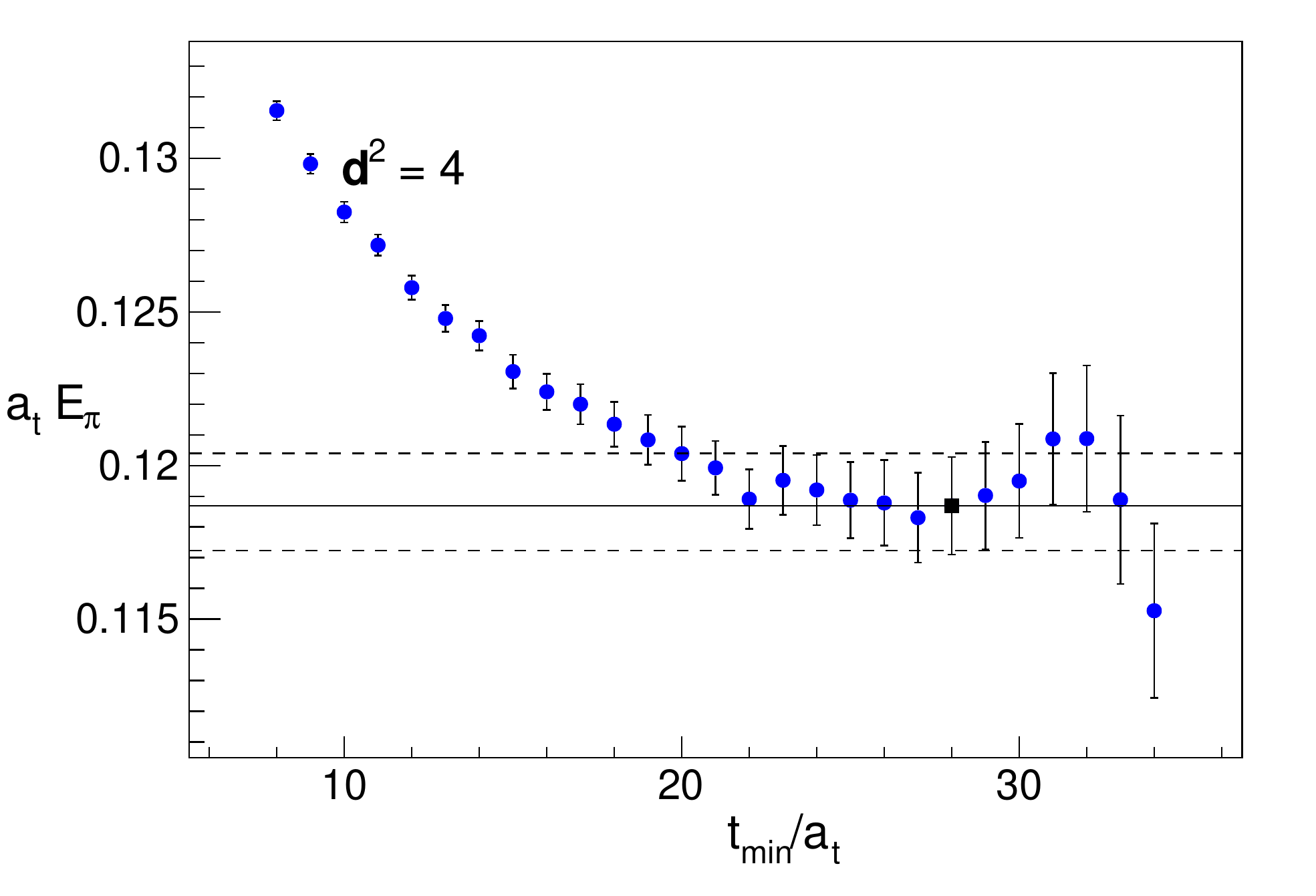}
	\includegraphics[width=0.32\textwidth]{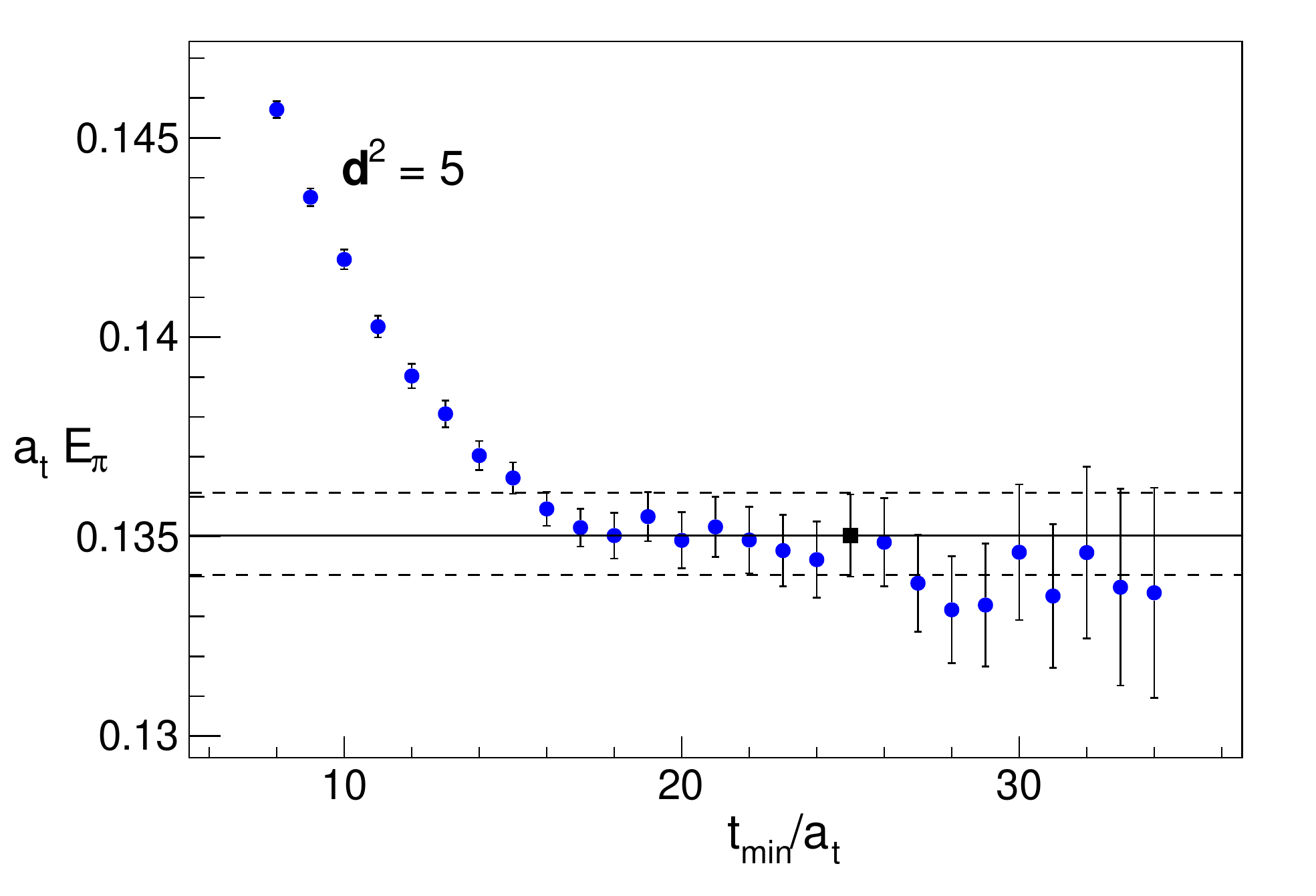}
	
	\includegraphics[width=0.32\textwidth]{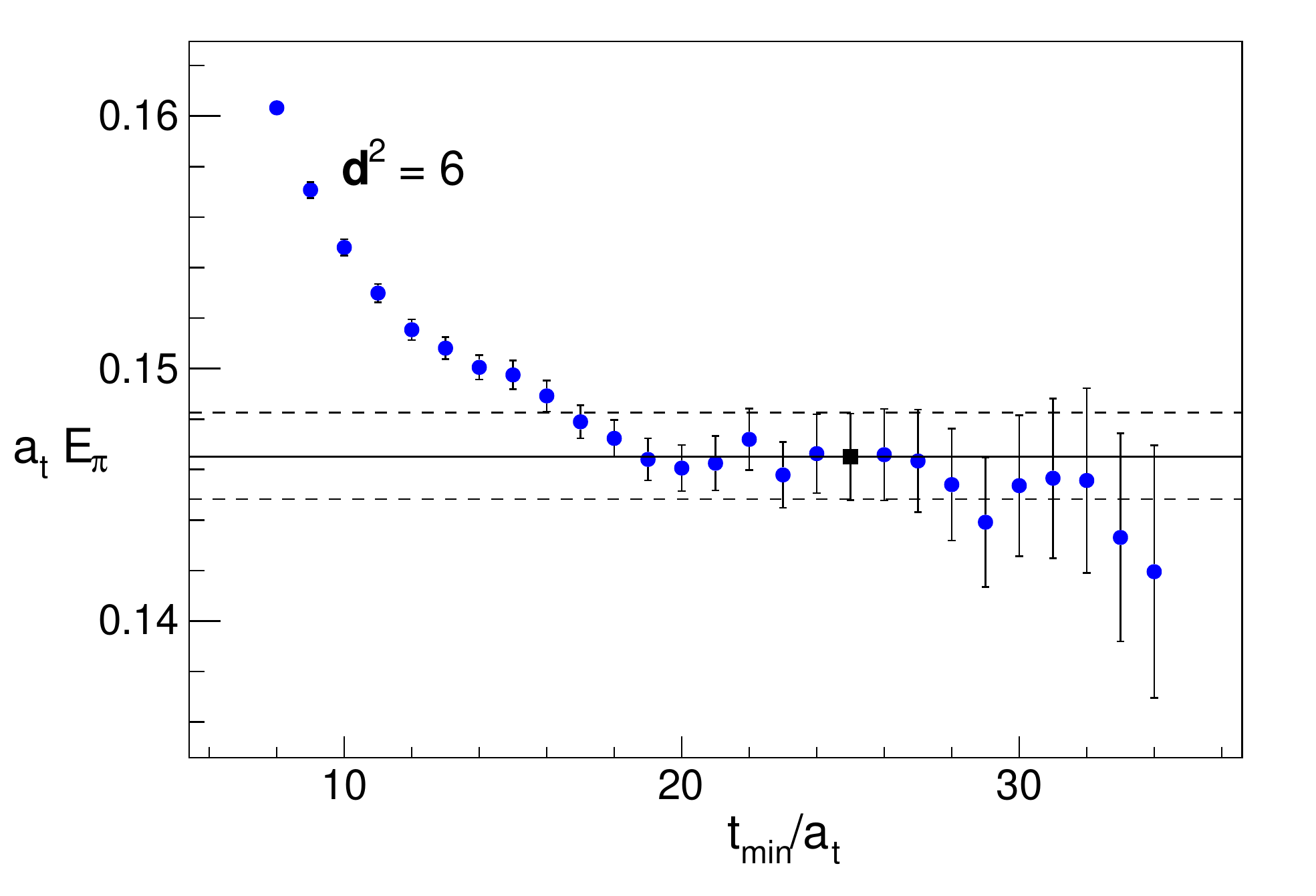}
	\includegraphics[width=0.32\textwidth]{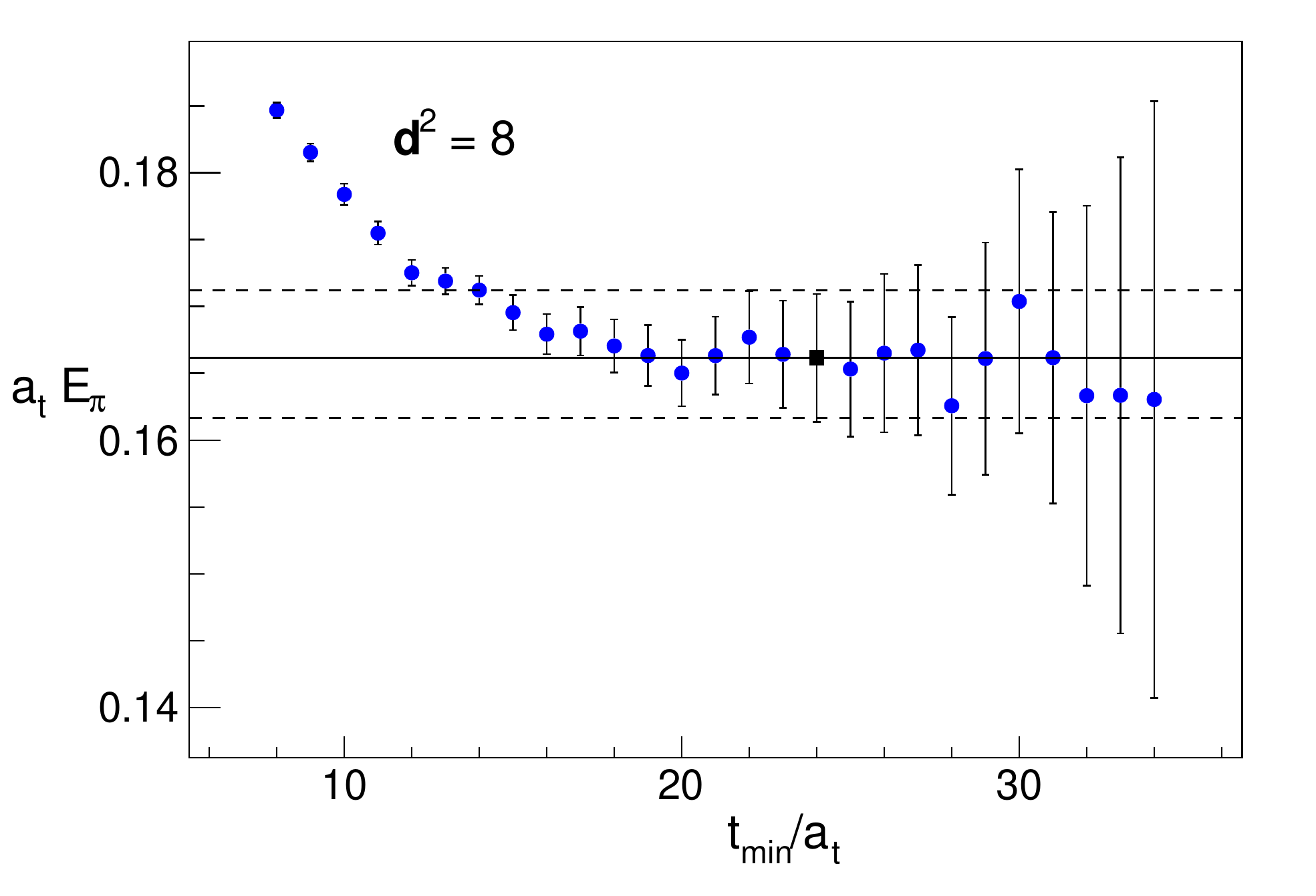}
	\includegraphics[width=0.32\textwidth]{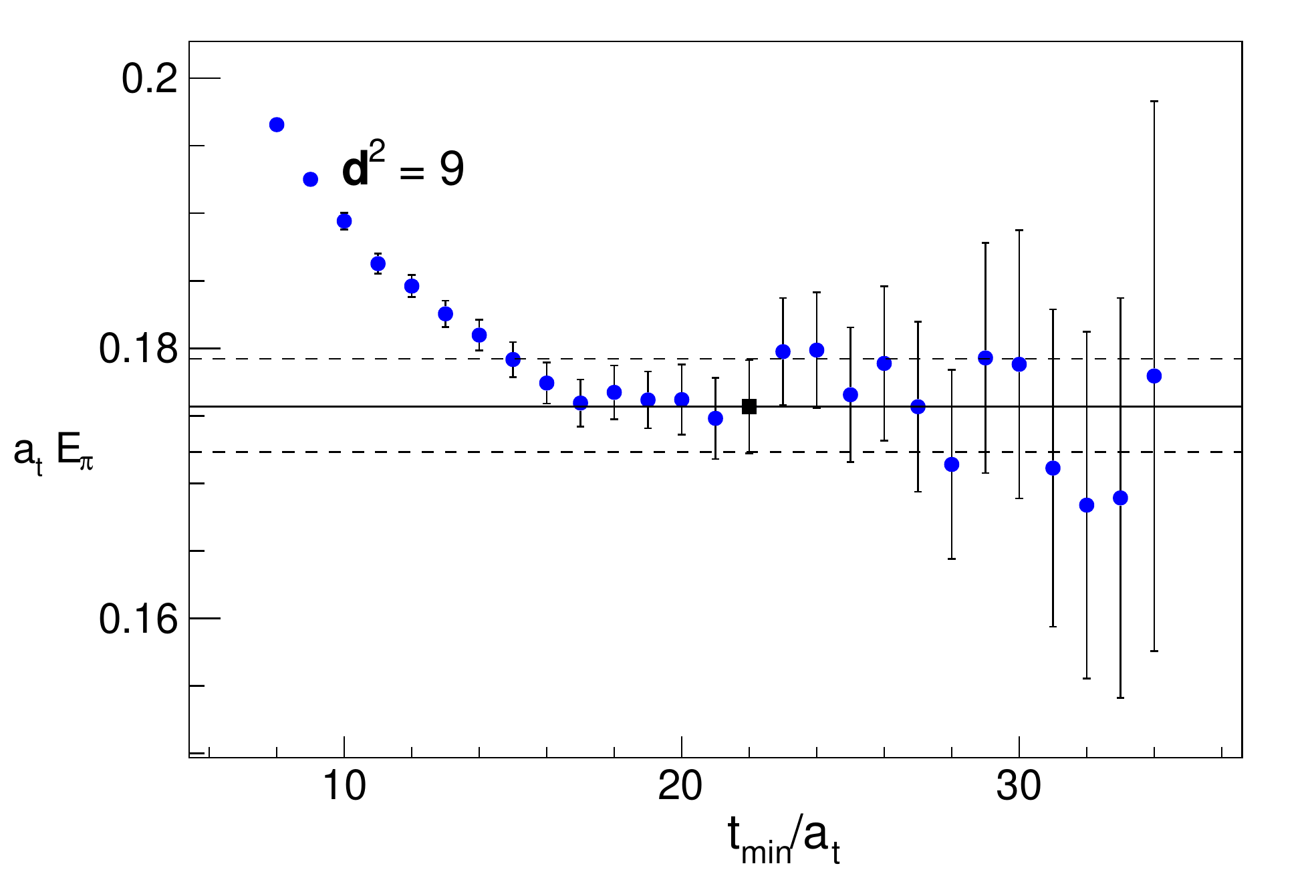}
\caption{\label{f:pi_tmins}$\tmin$-plots of single-pion energies 
 for all total momenta $0\le\boldsymbol{d}^2\le9$. The chosen
$t_{\mathrm{min}}$ is indicated by the solid line (central value) and dotted lines (1$\sigma$ errors) as well as the black square.}
\end{figure}

All $\tmin$-plots for fits to single-pion correlation functions at various momenta are shown in Fig.~\ref{f:pi_tmins}. These energies are used in Strategy 1 
discussed in Sec.~\ref{s:ens} to determine the renormalized anisotropy $\xi$. 

%\FloatBarrier
\section{$\tmin$-plots for all $I = 1$ levels}\label{s:app1}
\begin{figure}[!htp]
	\centering
\includegraphics[width=0.32\textwidth]{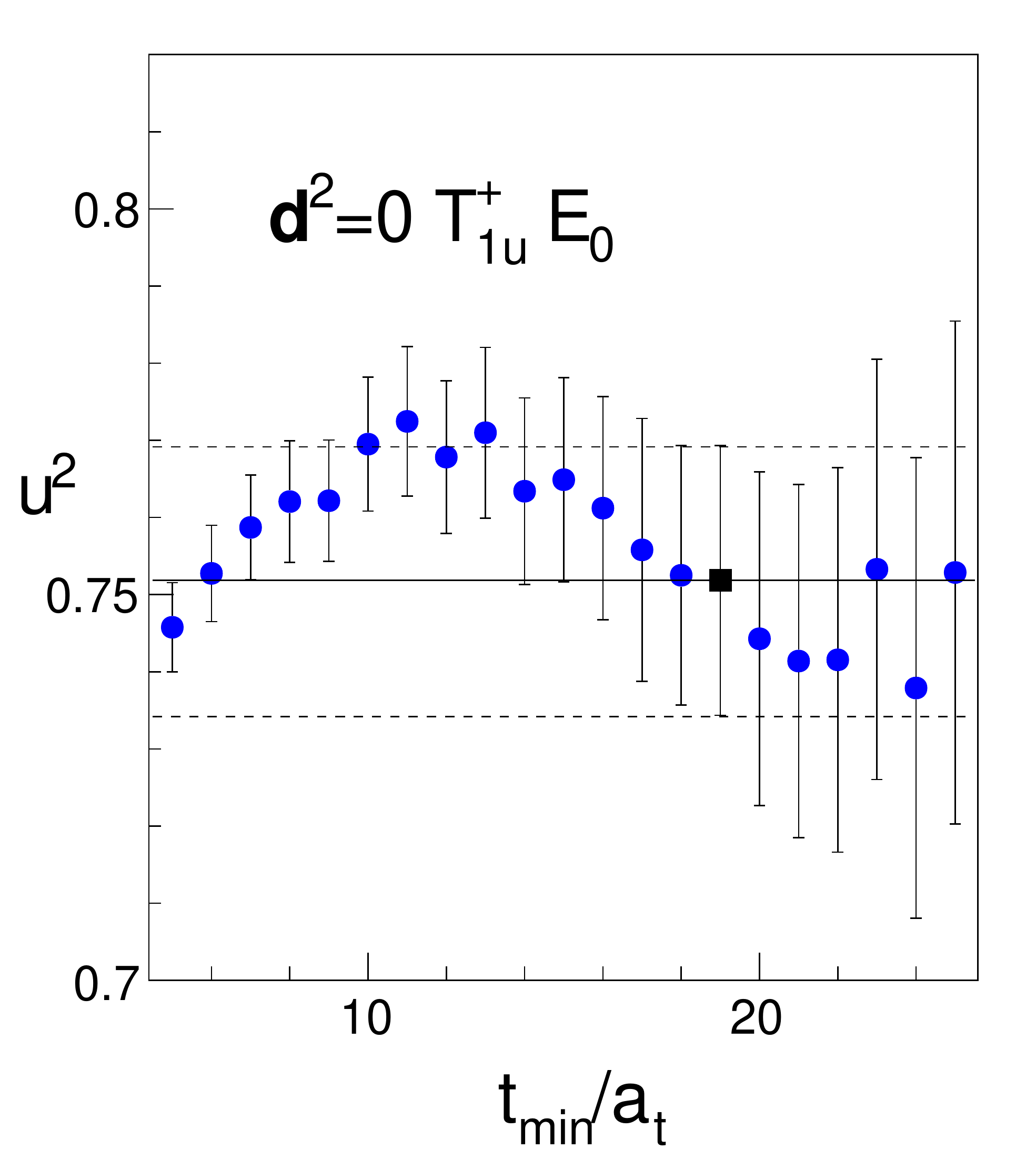}
\includegraphics[width=0.32\textwidth]{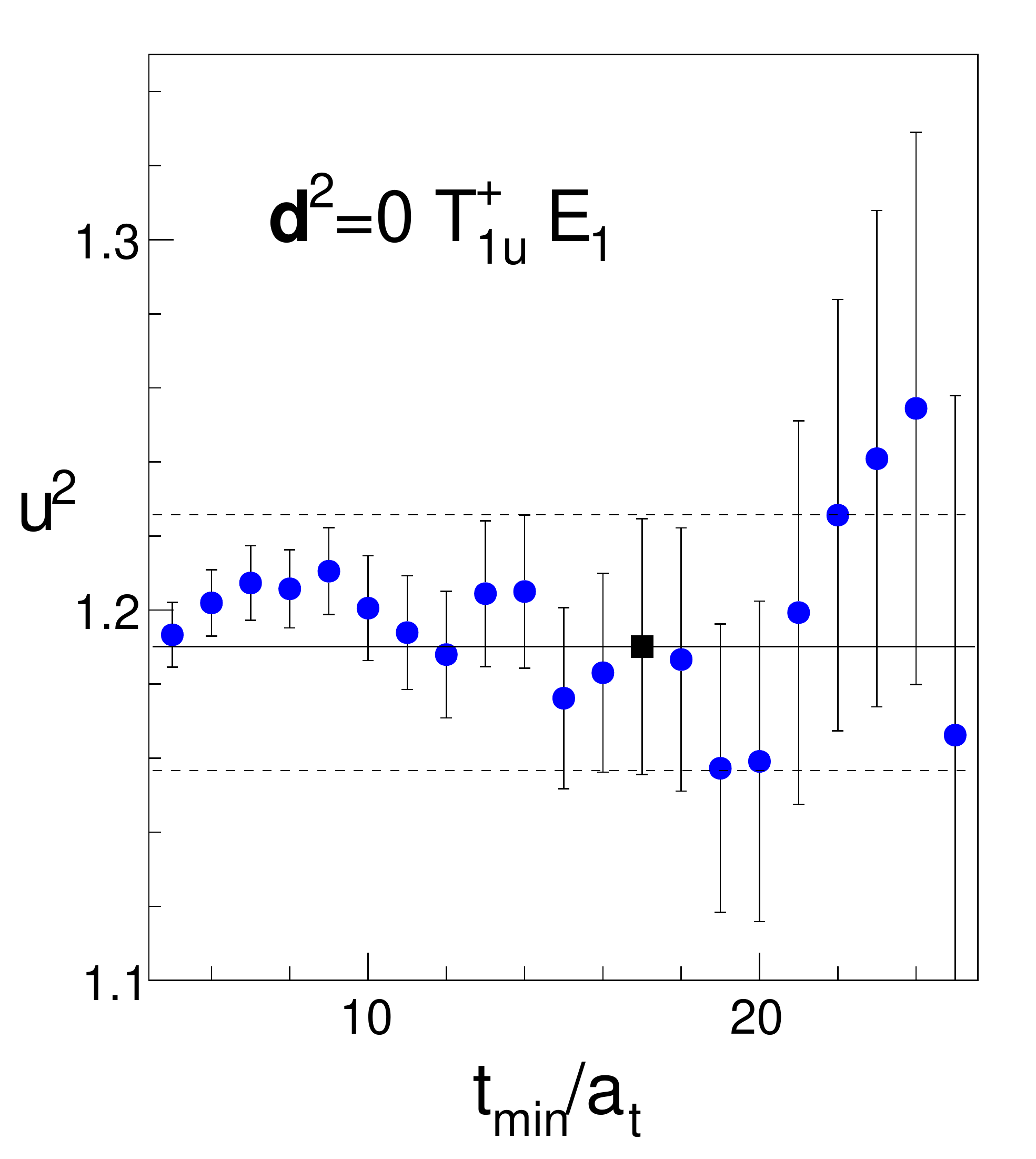}
\caption{\label{f:i1d0} $t_{\mathrm{min}}$-plots of 
the dimensionless center-of-mass momentum $u^2$ for $I=1$, 
$\boldsymbol{d}^2=0$. The chosen $\tmin$ is indicated by the black square and the lines.}
\end{figure}

\begin{figure}[!htb]
\includegraphics[width=0.32\textwidth]{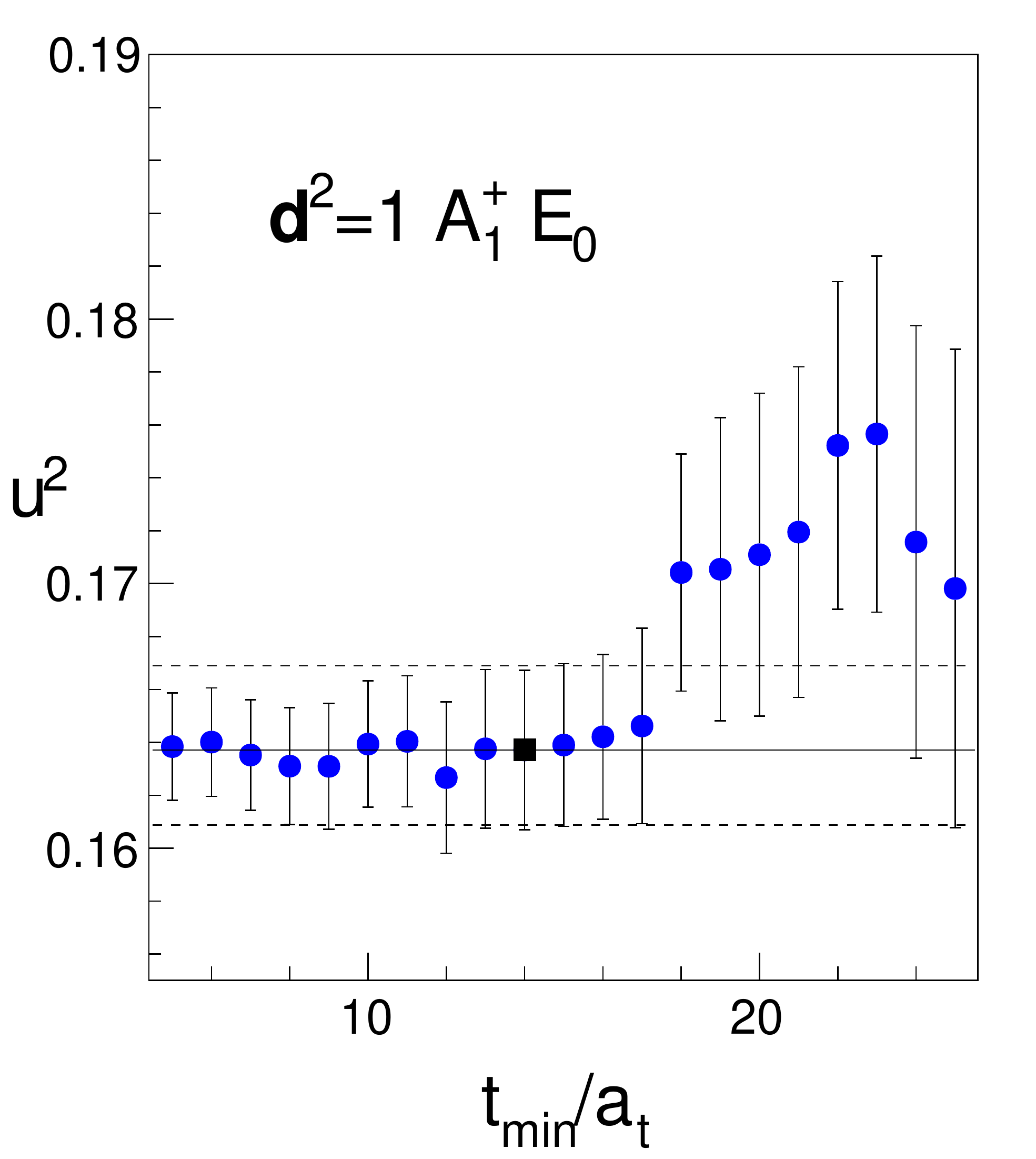}
\includegraphics[width=0.32\textwidth]{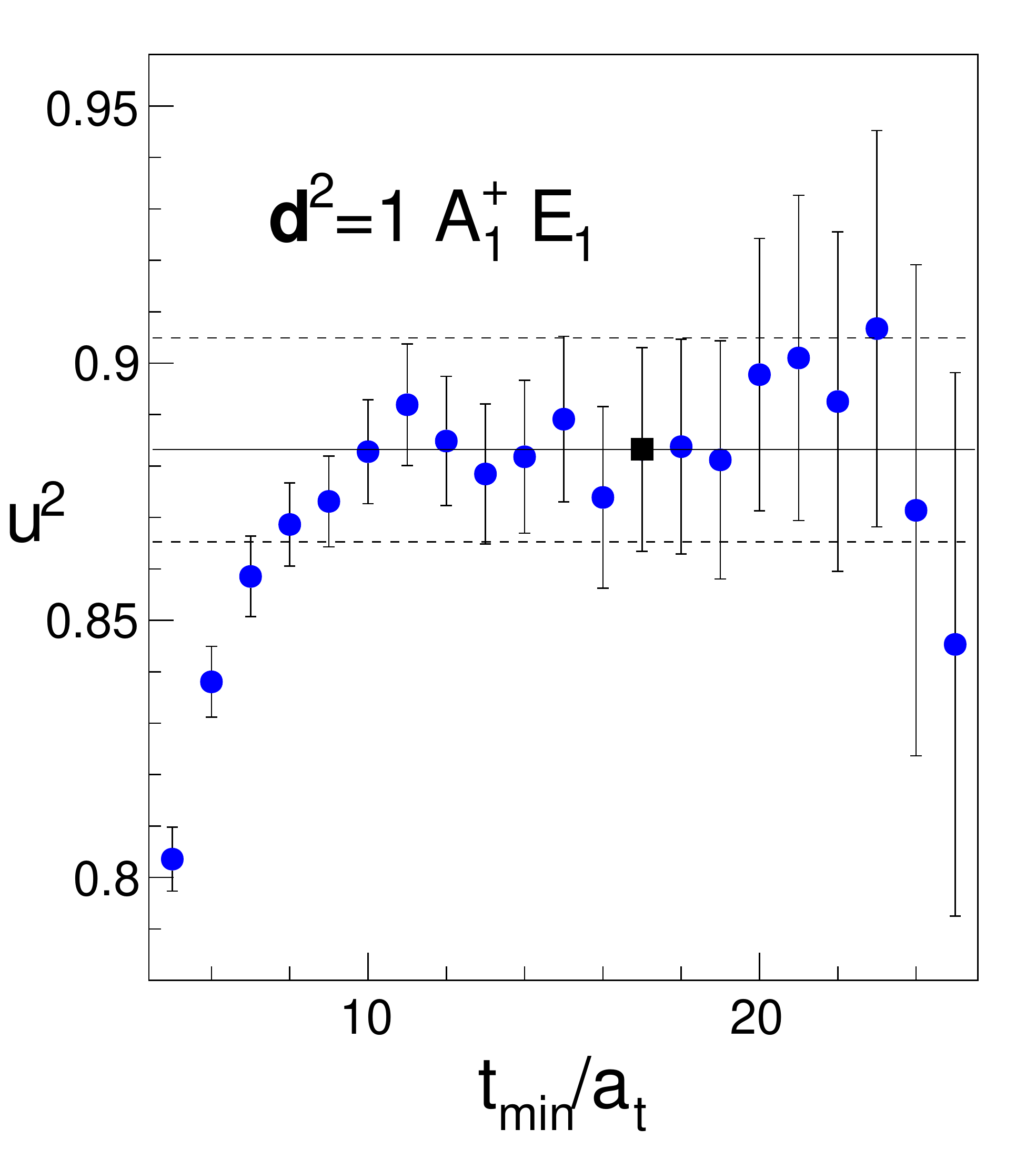}
\includegraphics[width=0.32\textwidth]{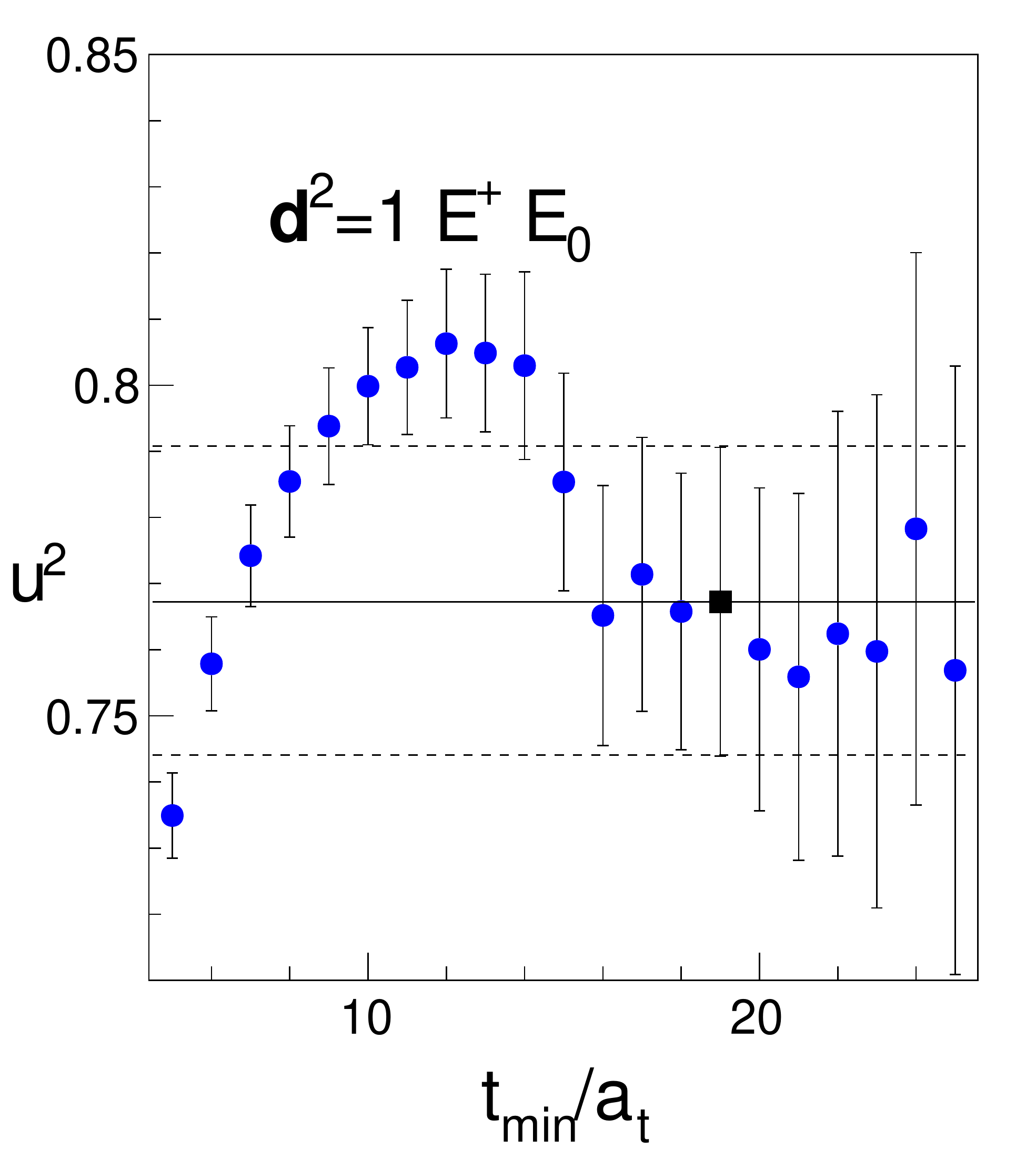}
\caption{\label{f:i1d1}Same as Fig.~\ref{f:i1d0}  but for $I=1$, 
$\boldsymbol{d}^2=1$.}
\end{figure}

\begin{figure}[!htb]
\includegraphics[width=0.32\textwidth]{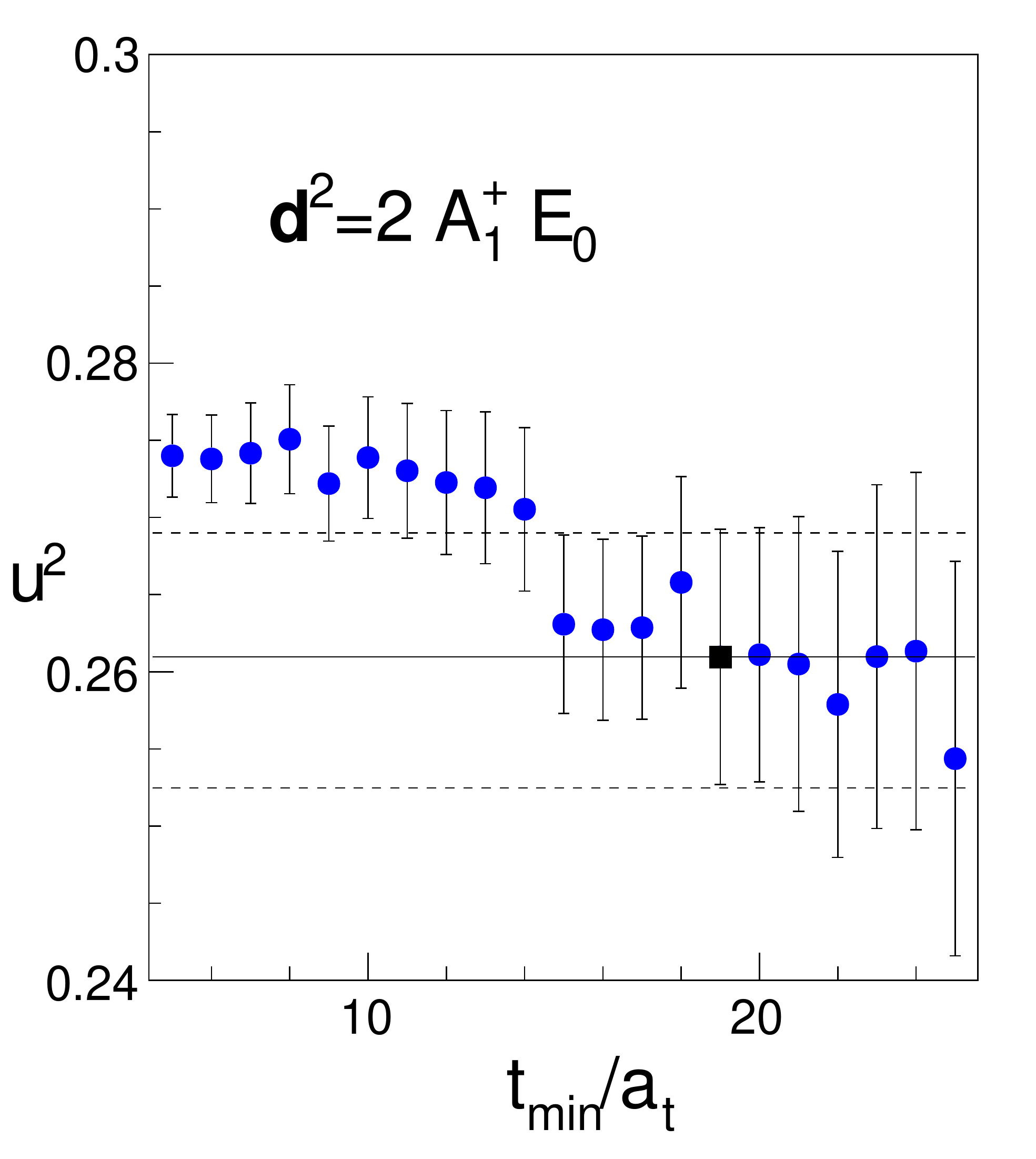}
\includegraphics[width=0.32\textwidth]{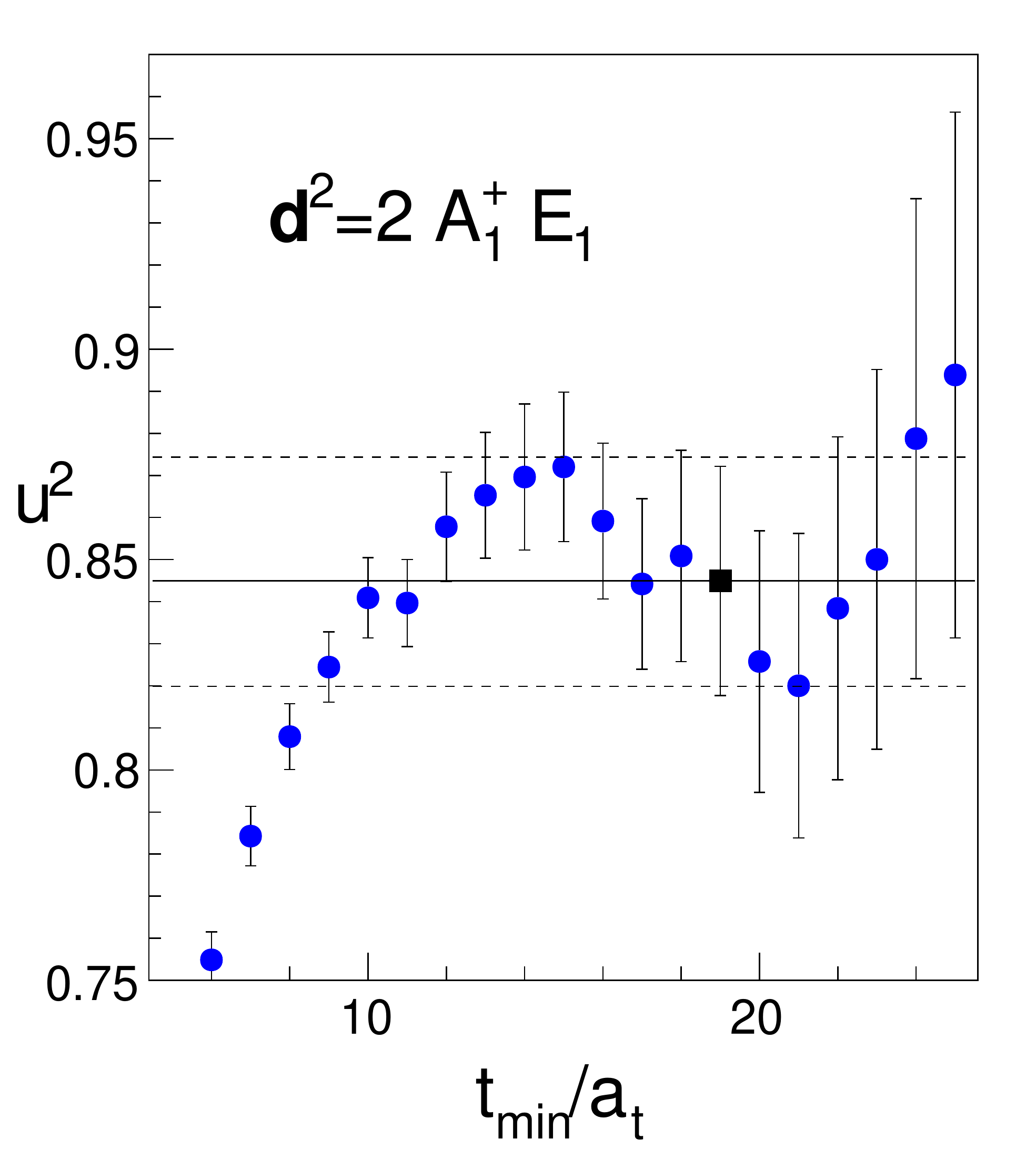}
\includegraphics[width=0.32\textwidth]{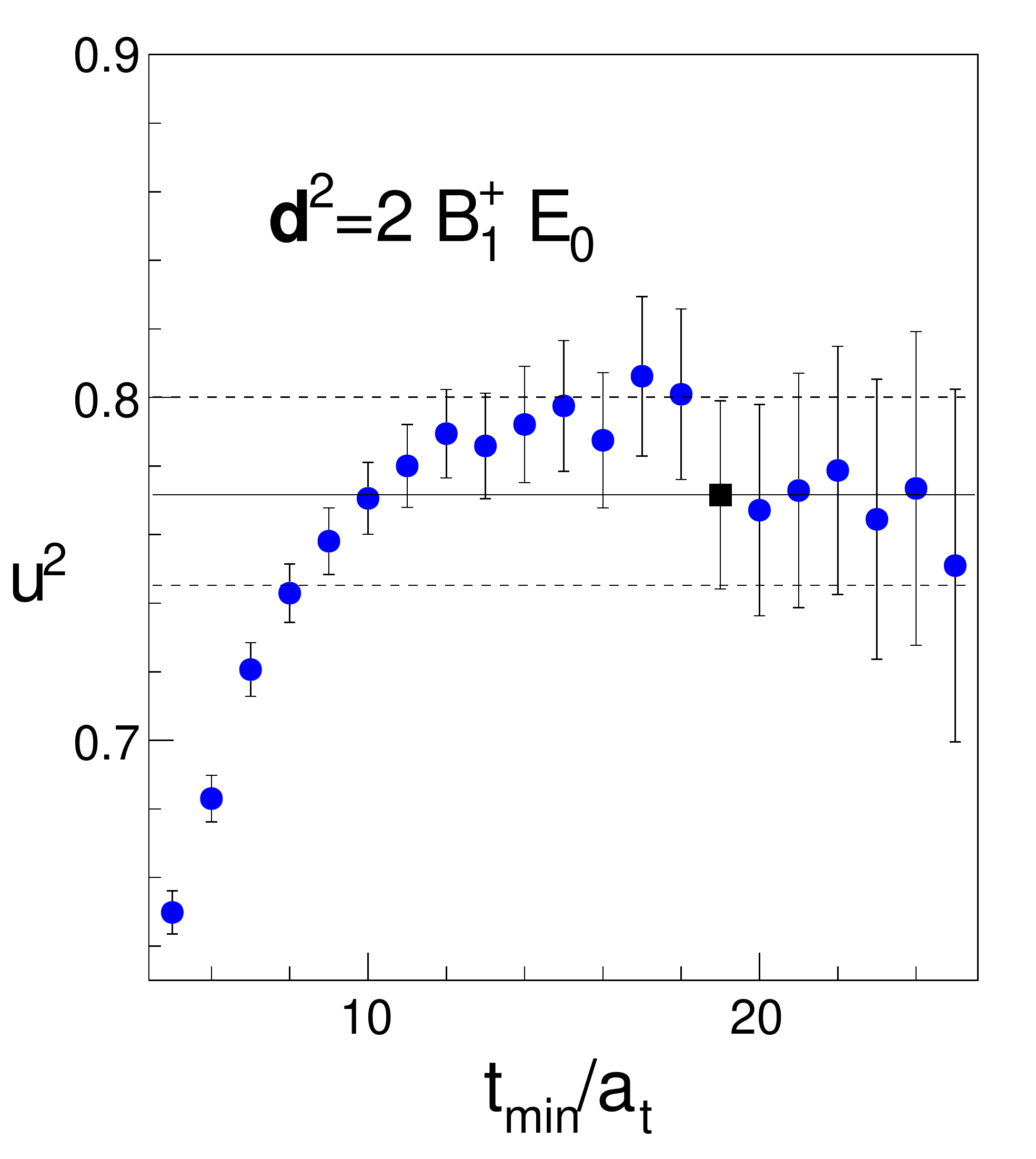}

\includegraphics[width=0.32\textwidth]{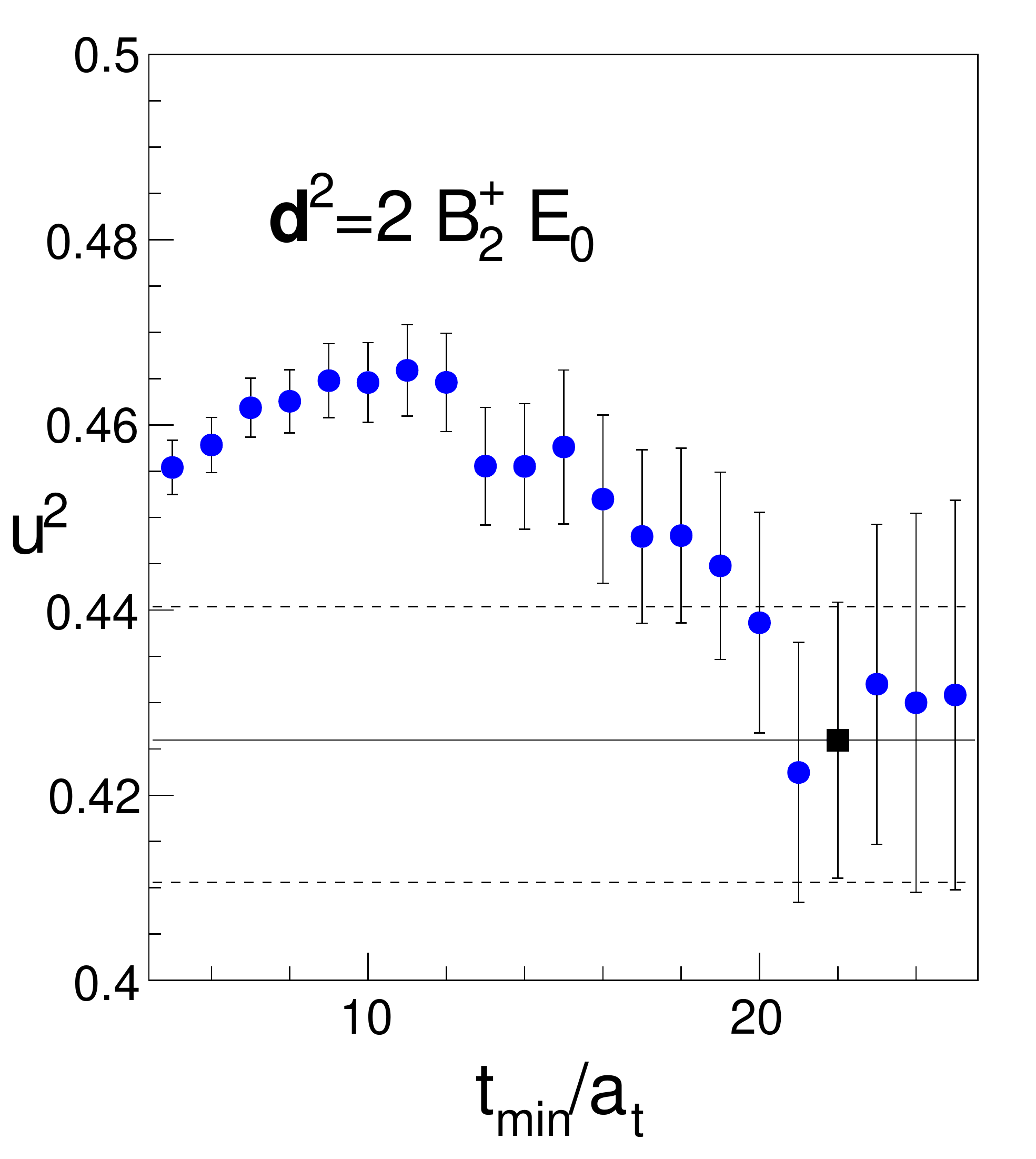}
\includegraphics[width=0.32\textwidth]{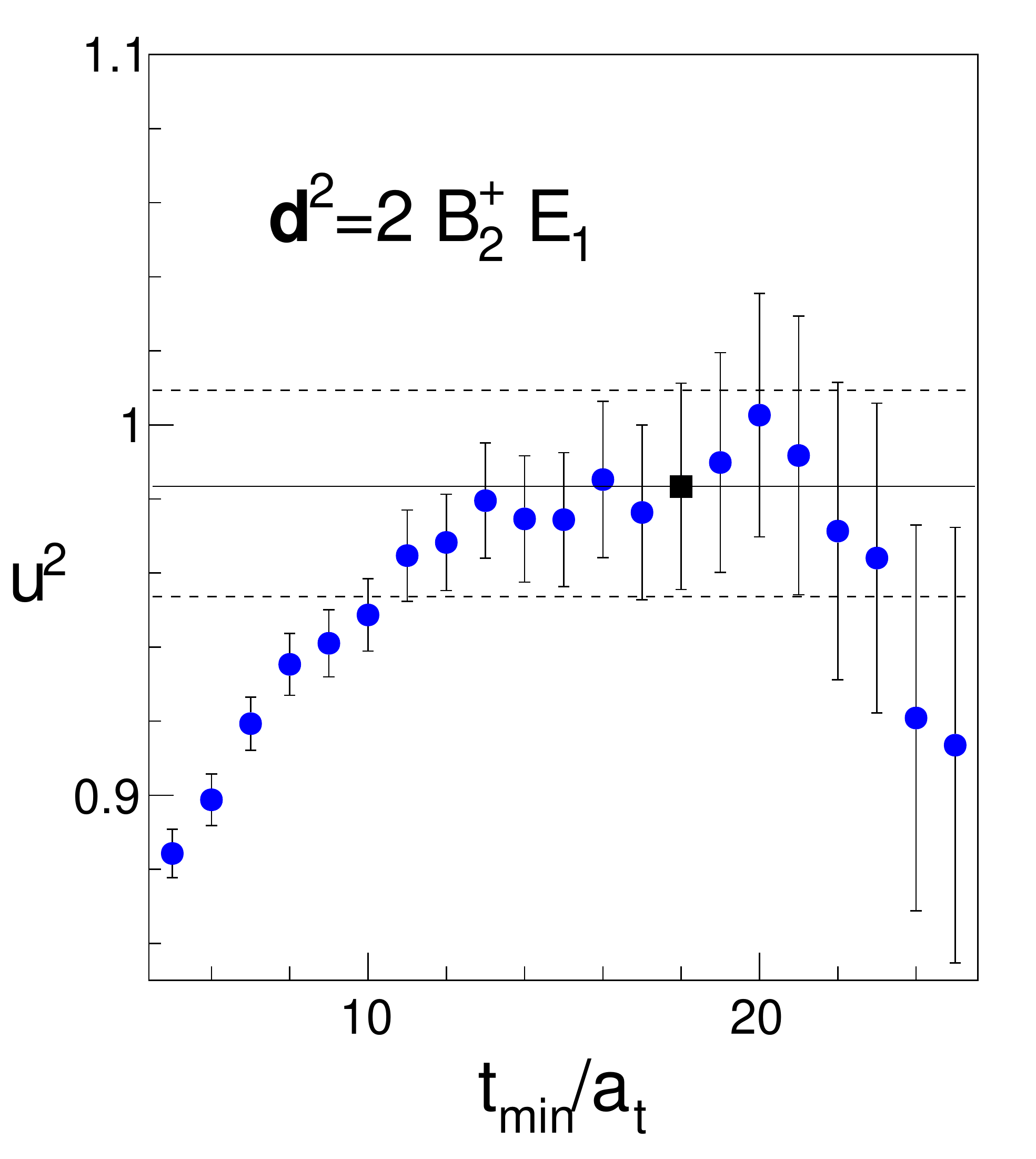}
\caption{\label{f:i1d2}Same as Fig.~\ref{f:i1d0} but for $I=1$,   
$\boldsymbol{d}^2=2$.}
\end{figure}

\begin{figure}[!htb]
\includegraphics[width=0.32\textwidth]{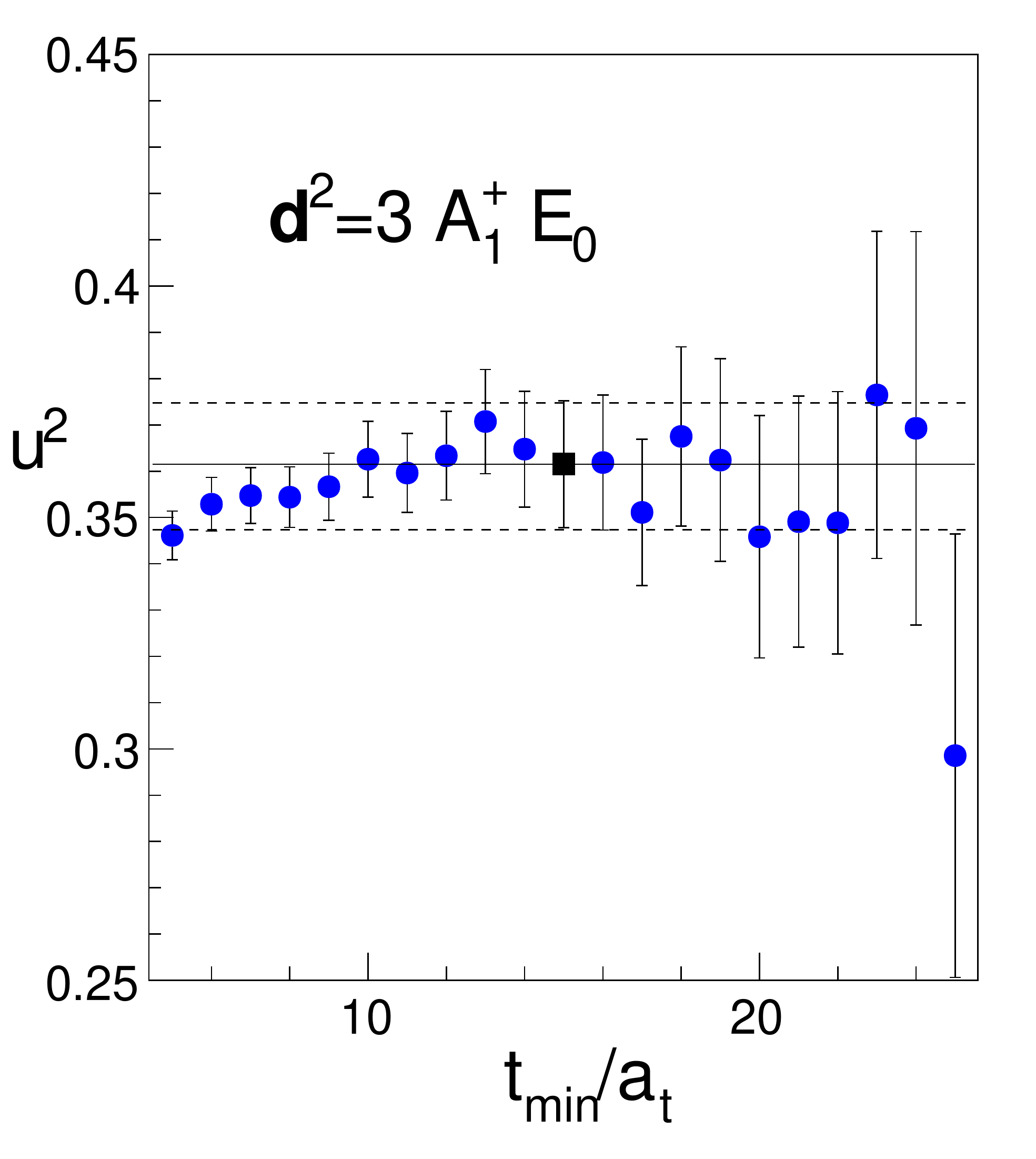}
\includegraphics[width=0.32\textwidth]{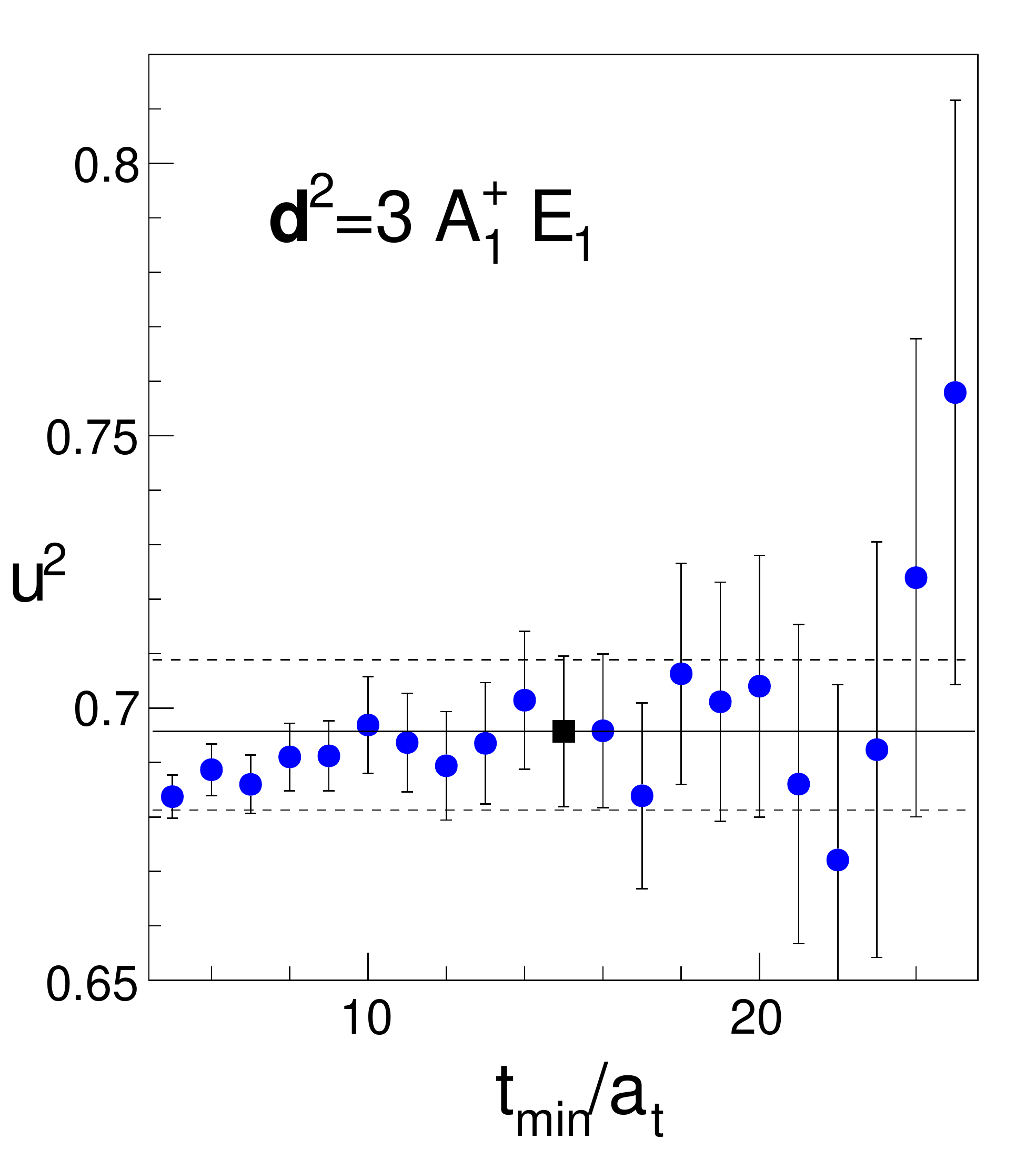}
\includegraphics[width=0.32\textwidth]{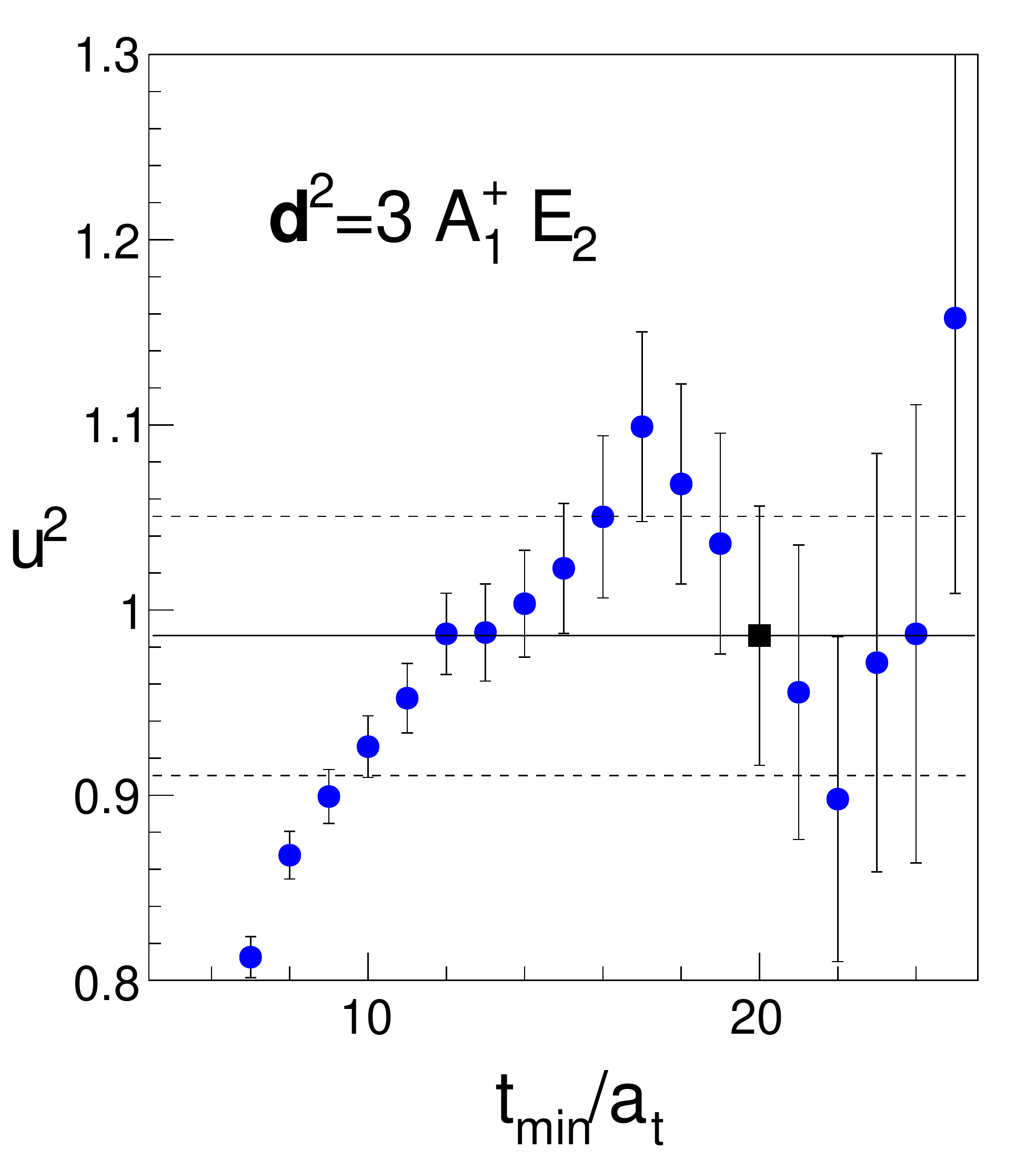}

\includegraphics[width=0.32\textwidth]{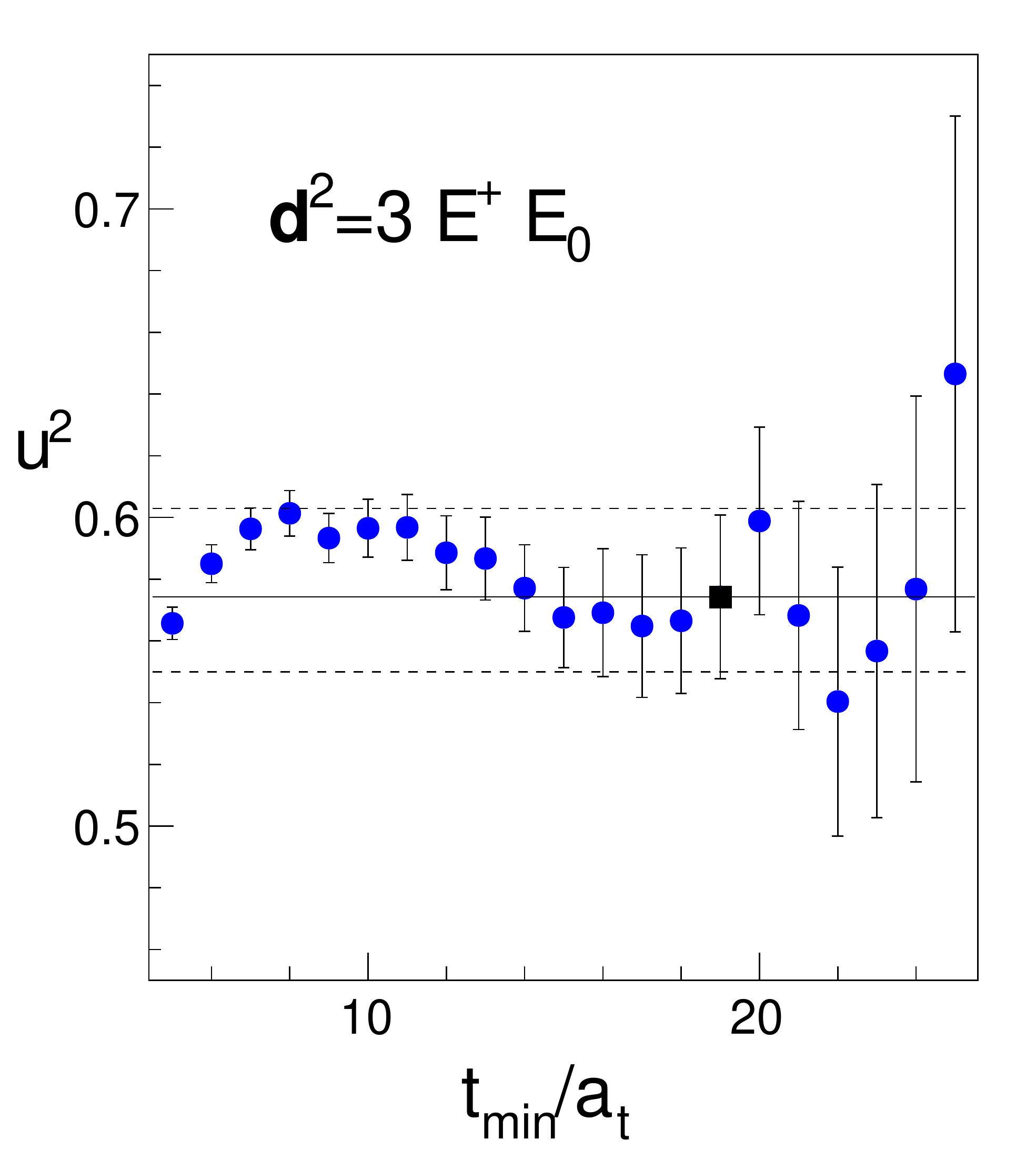}
\includegraphics[width=0.32\textwidth]{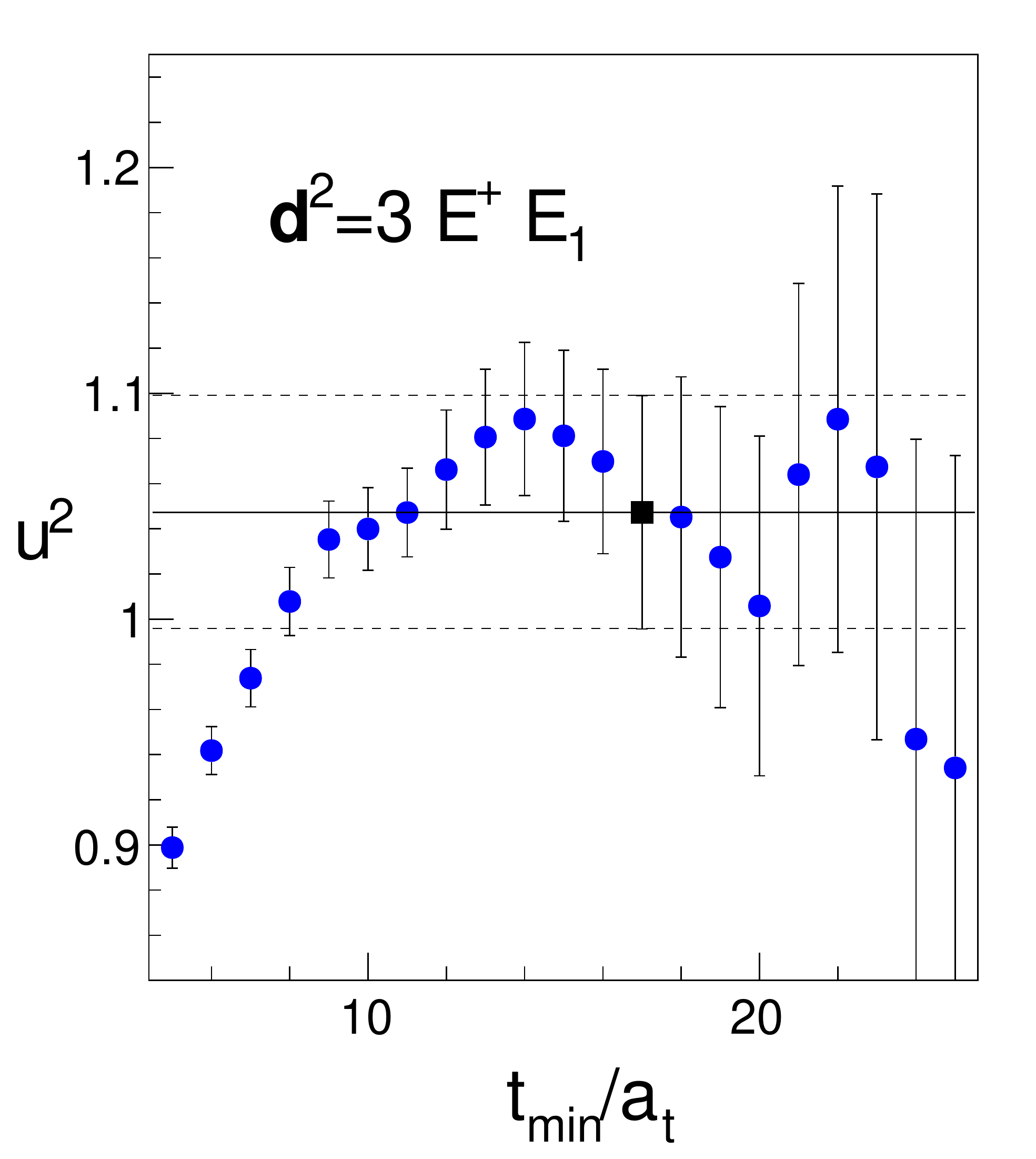}
	\caption{\label{f:i1d3}Same as Fig.~\ref{f:i1d0} for $I=1$,  $\boldsymbol{d}^2=3$.}
\end{figure}

\begin{figure}[!htb]
	\centering
\includegraphics[width=0.32\textwidth]{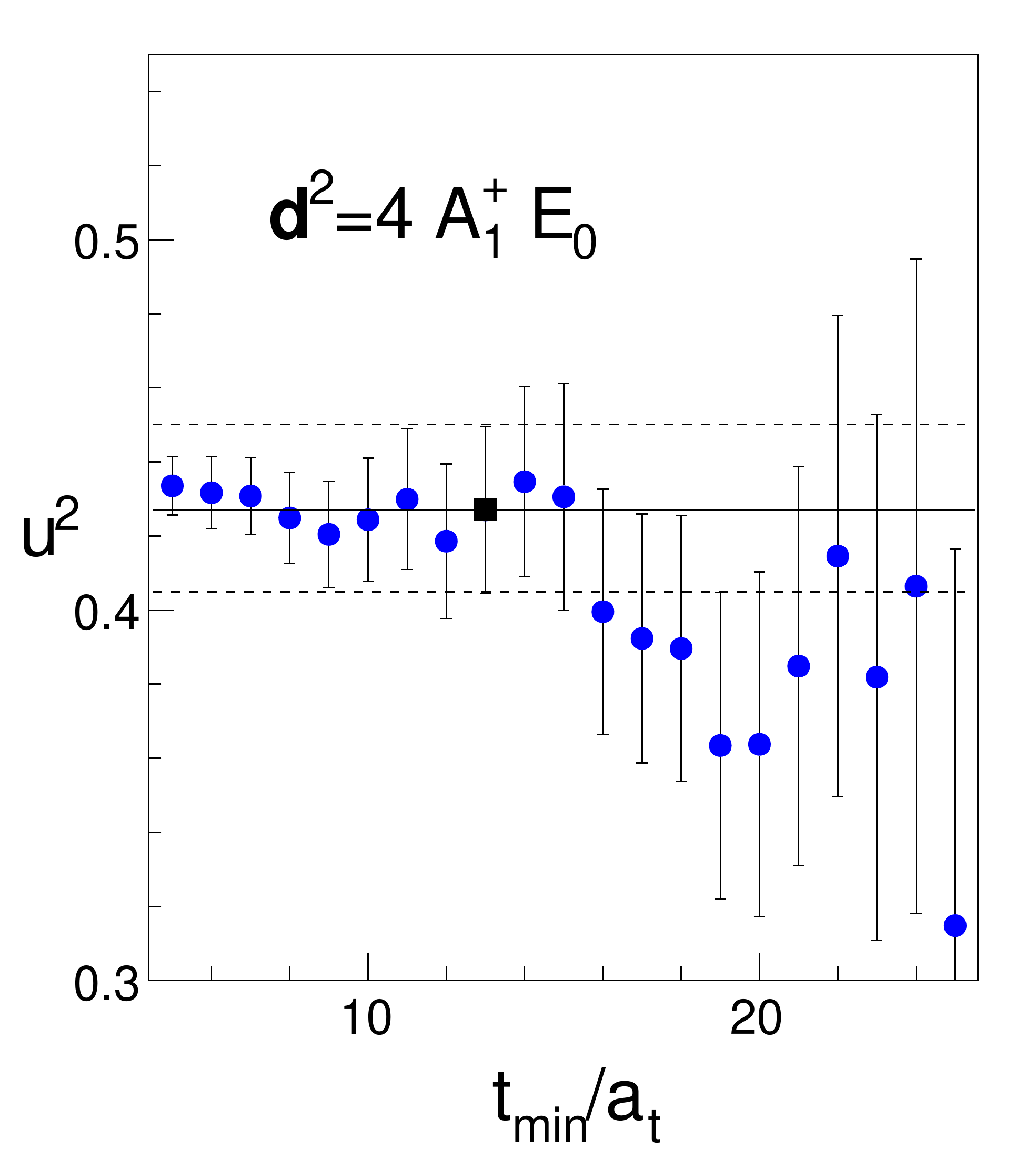}
\includegraphics[width=0.32\textwidth]{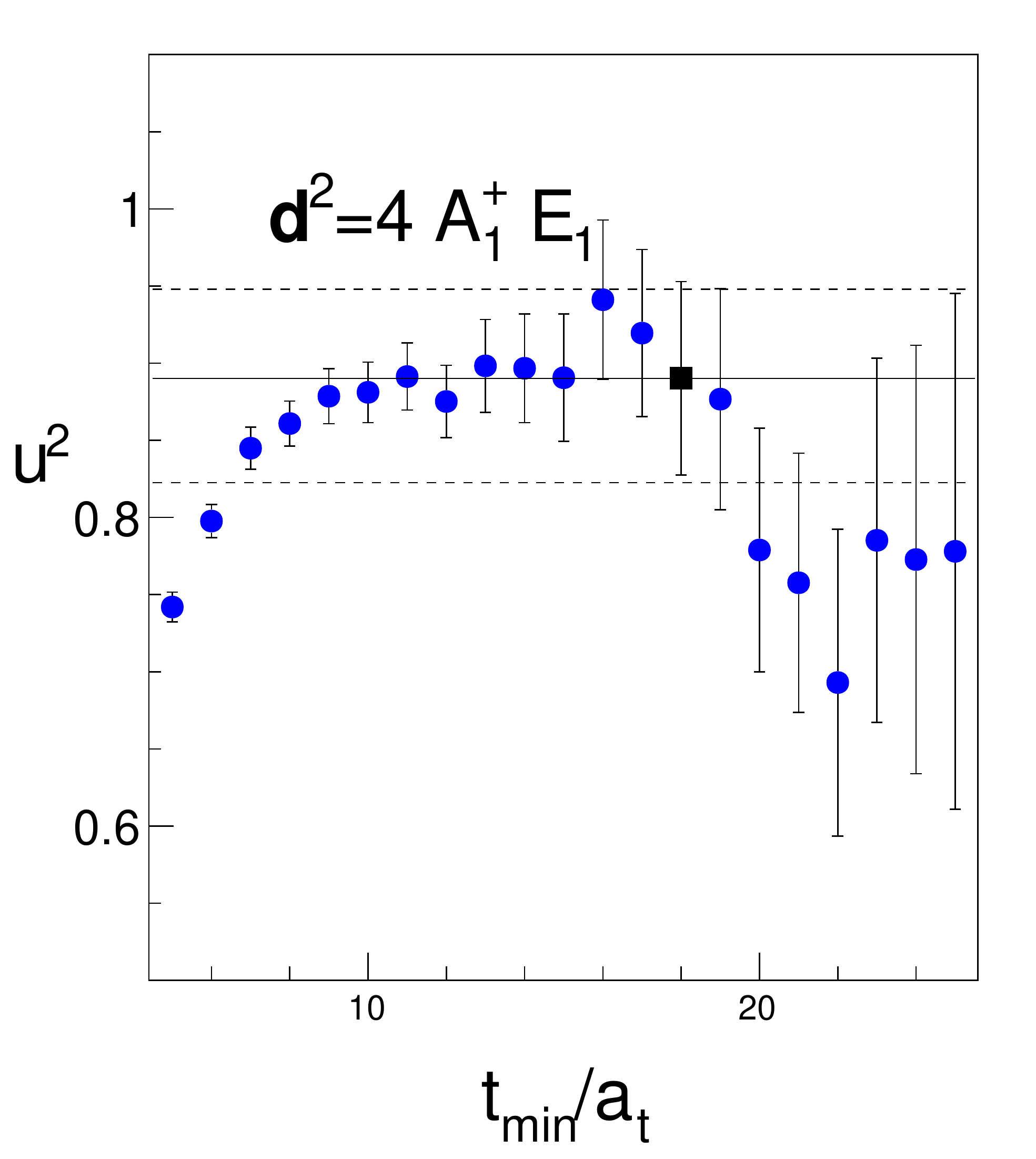}

\includegraphics[width=0.32\textwidth]{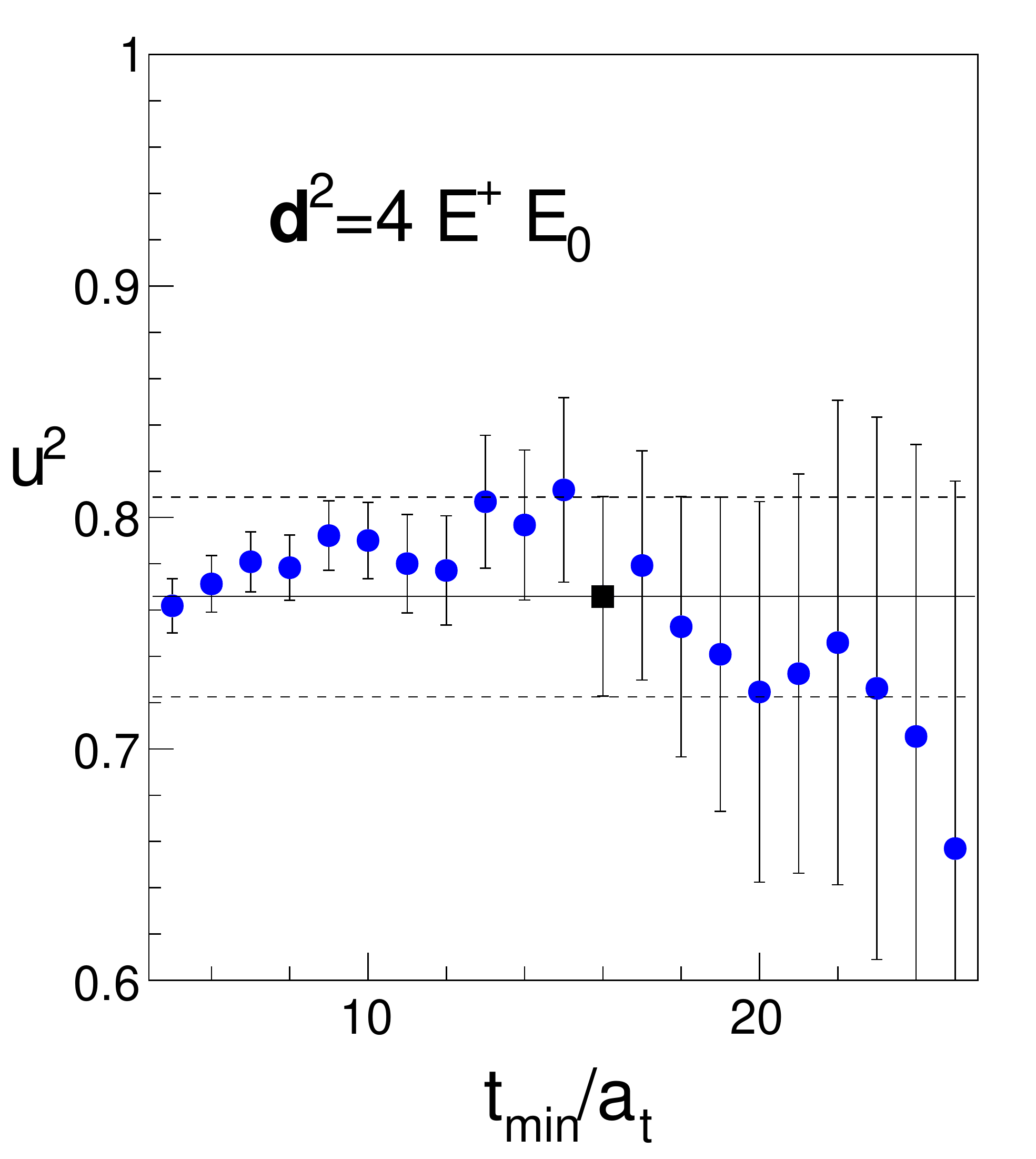}
\includegraphics[width=0.32\textwidth]{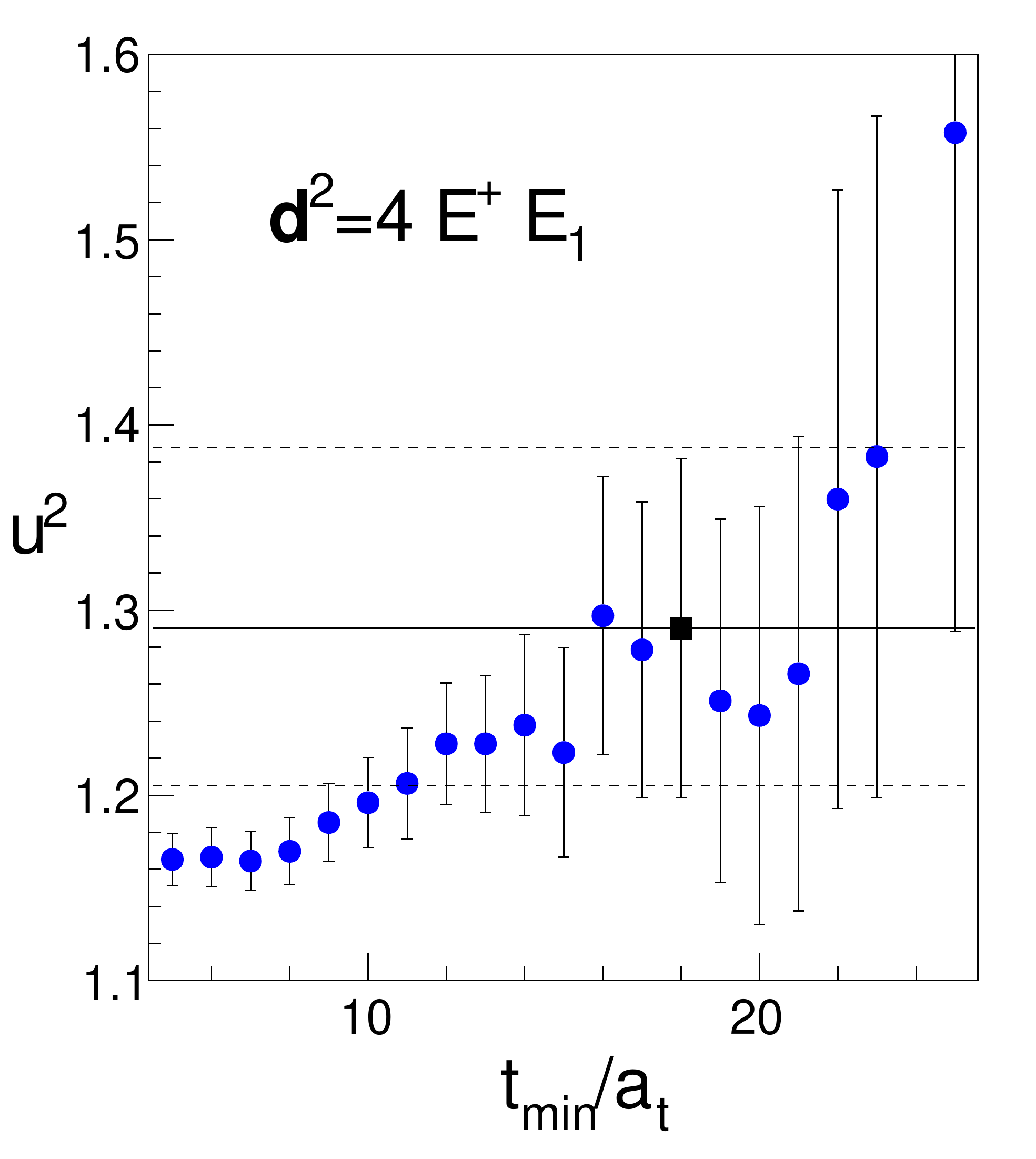}
	\caption{\label{f:i1d4}Same as Fig.~\ref{f:i1d0} but for   
	$I=1$, $\boldsymbol{d}^2=4$.}
\end{figure}

All $t_{\mathrm{min}}$-plots for finite volume energies
used in the determination of the $I=1$, $\ell=1$ elastic scattering amplitude are shown in 
Figs.~\ref{f:i1d0},~\ref{f:i1d1},~\ref{f:i1d2},~\ref{f:i1d3}, and~\ref{f:i1d4}. The ratio 
fits of Eq.~\ref{e:rat} are employed and the dimensionless center-of-mass momentum $u^2$ 
is shown.  

%\FloatBarrier
\section{$\tmin$-plots for all $I = 2$ levels}\label{s:app2}
\begin{figure}[!htb]
	\centering
\includegraphics[width=0.32\textwidth]{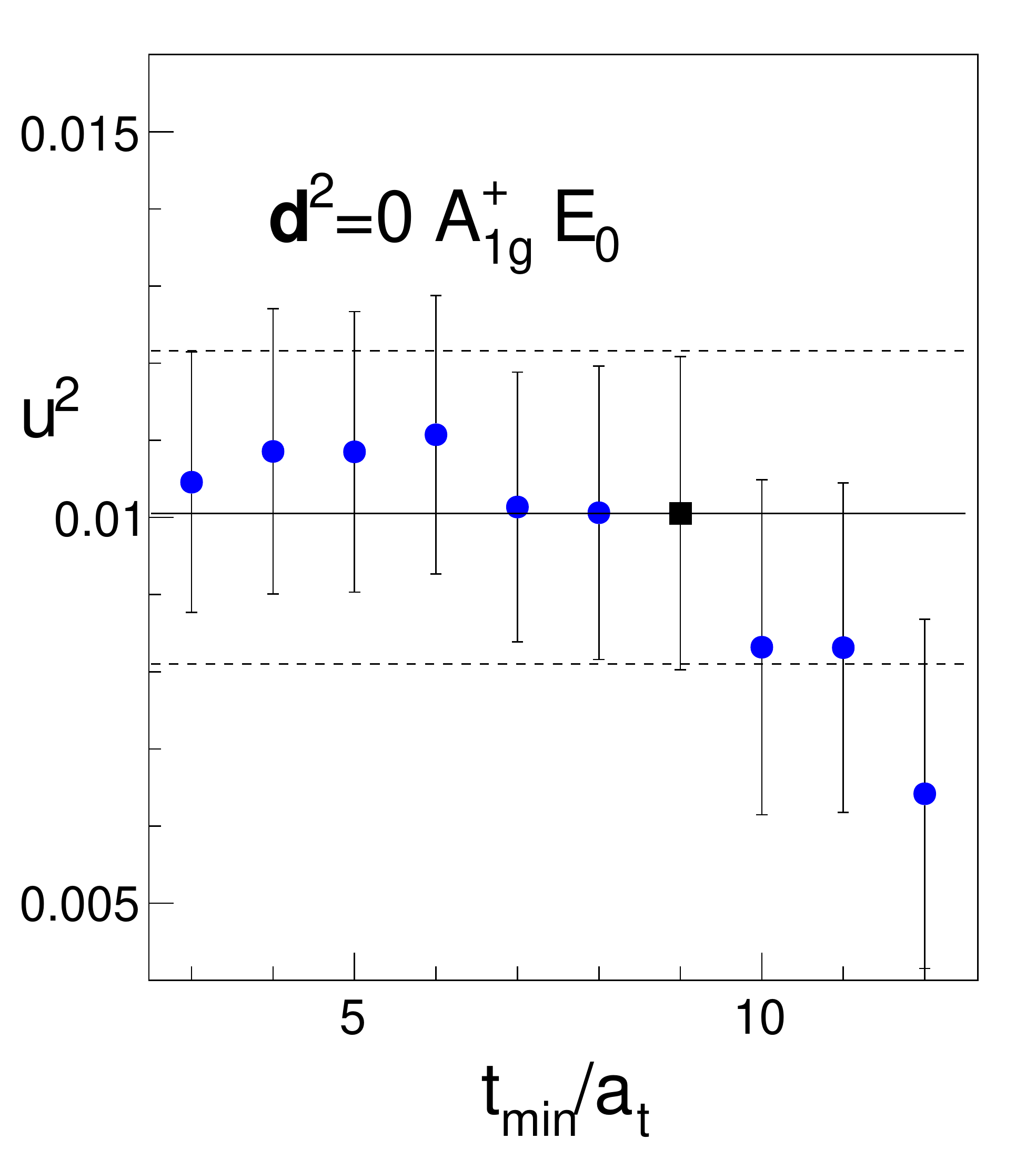}
\includegraphics[width=0.32\textwidth]{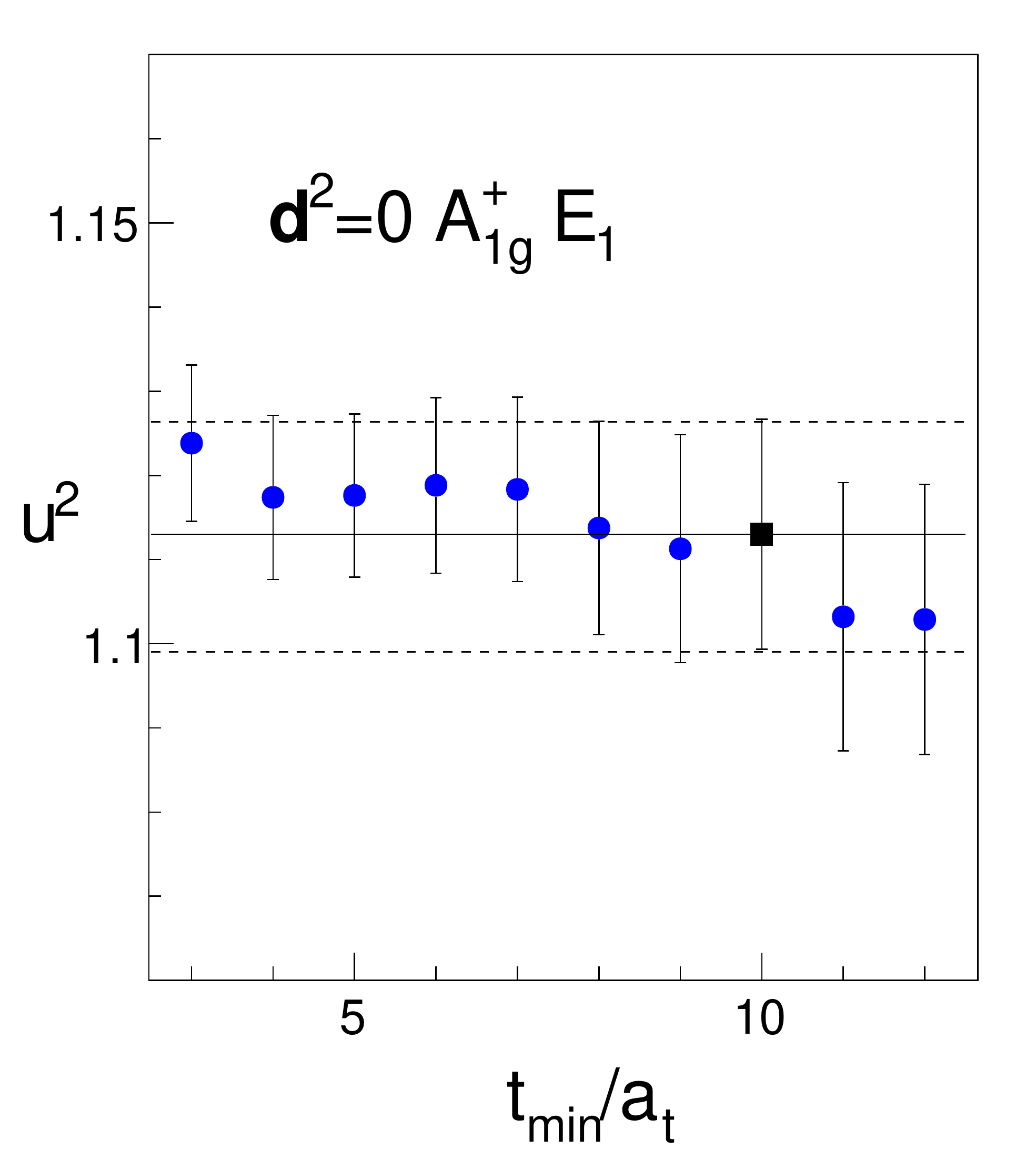}
	\caption{\label{f:i2d0}$t_{\mathrm{min}}$-plots of 
the dimensionless center-of-mass momentum $u^2$ for $I=2$, 
$\boldsymbol{d}^2=0$. The chosen $\tmin$ is indicated by the black square and the lines.}
\end{figure}

\begin{figure}[!htb]
	\centering
\includegraphics[width=0.32\textwidth]{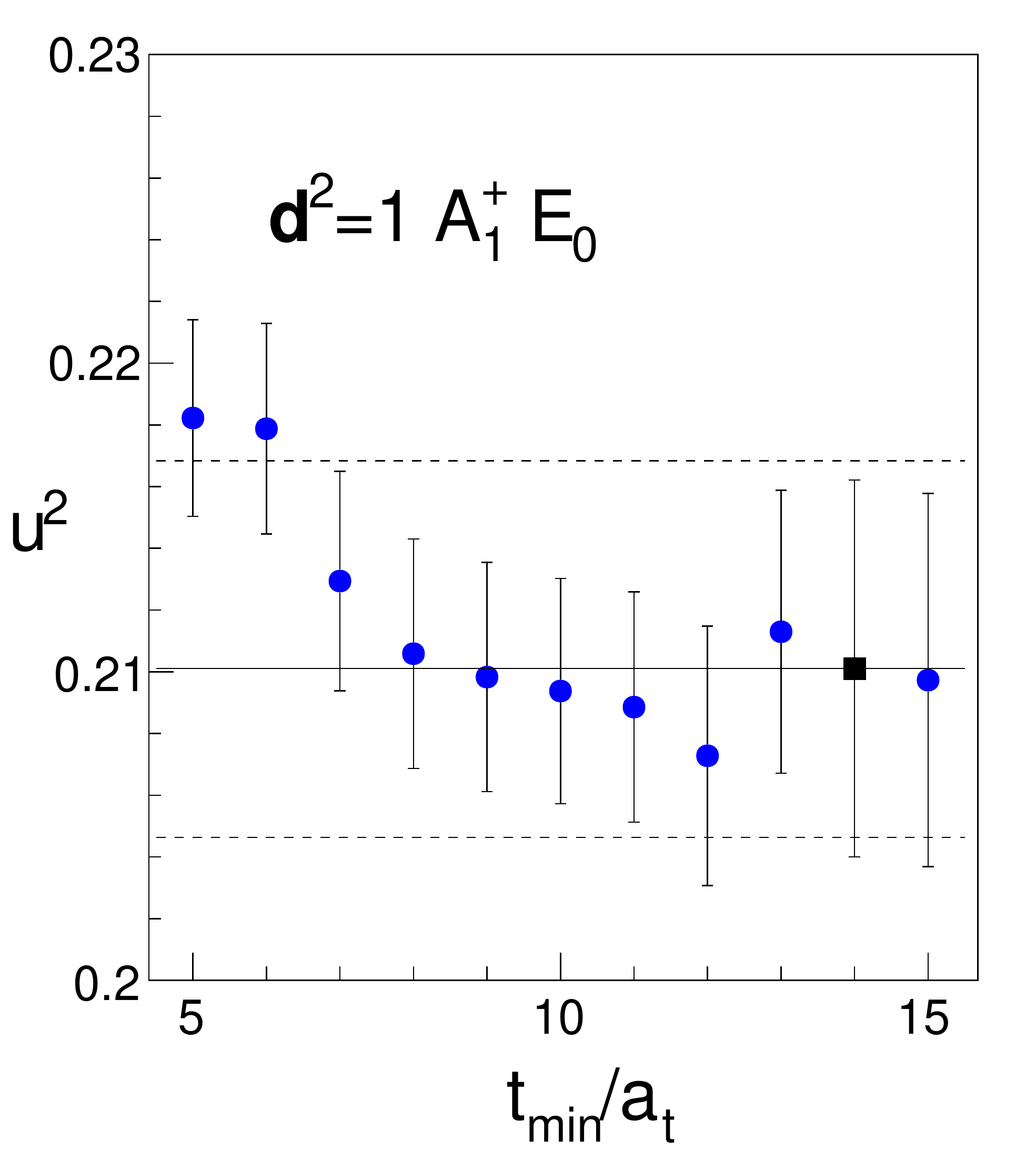}
\includegraphics[width=0.32\textwidth]{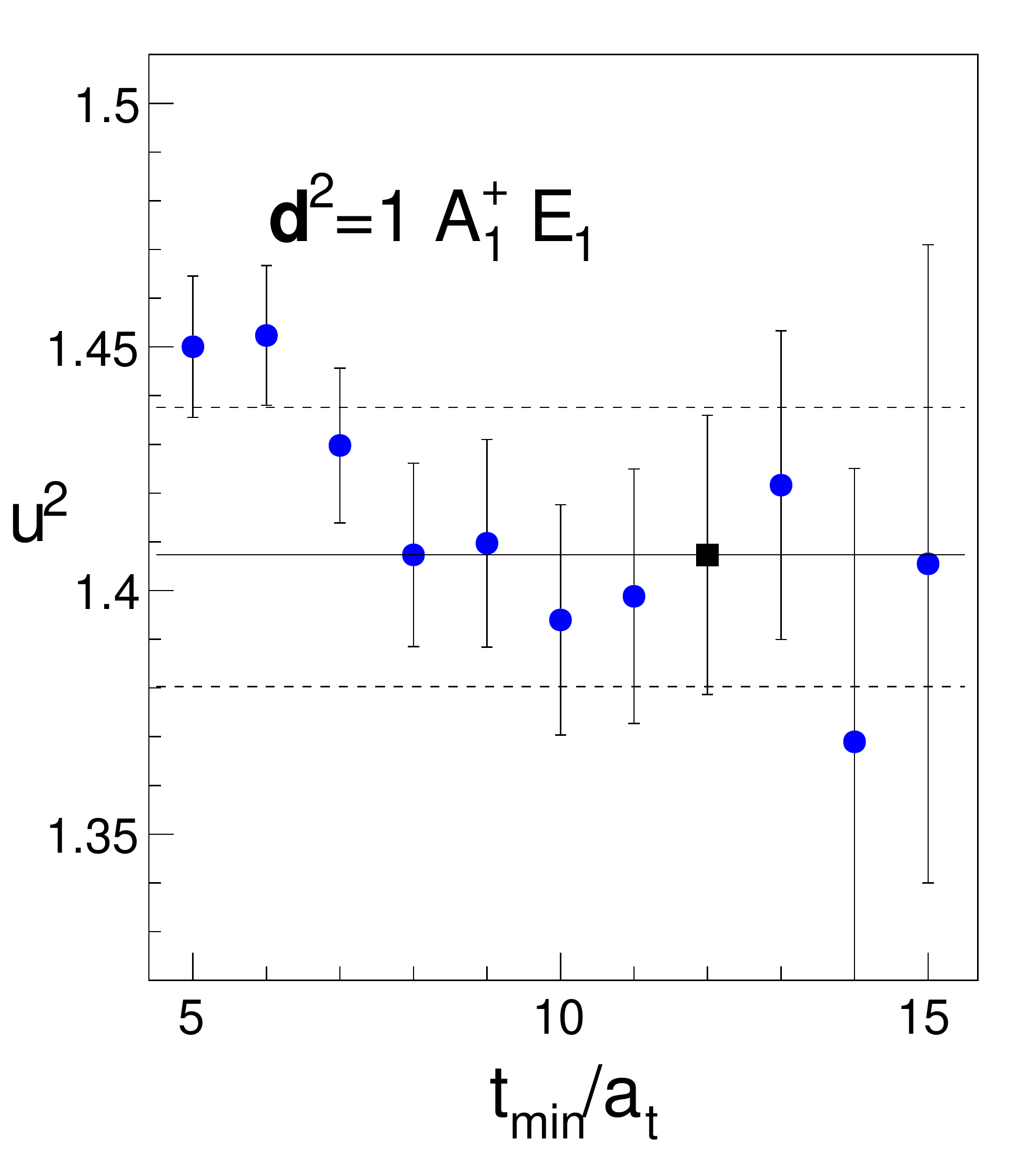}
	\caption{\label{f:i2d1}Same as Fig.~\ref{f:i2d0} but for   
$I=2$, $\boldsymbol{d}^2=1$.}
\end{figure}

\begin{figure}[!htb]
	\centering
\includegraphics[width=0.32\textwidth]{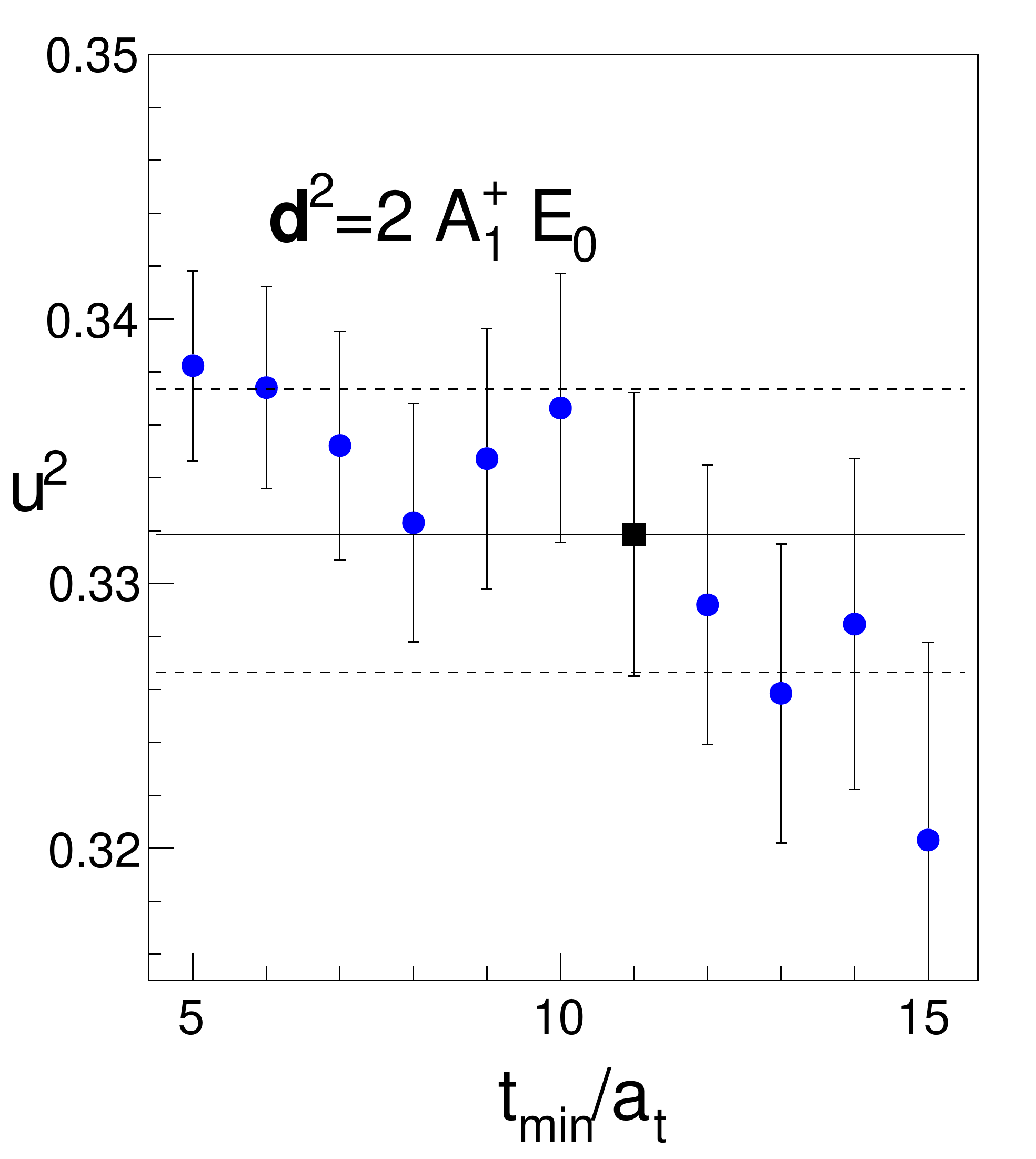}
\includegraphics[width=0.32\textwidth]{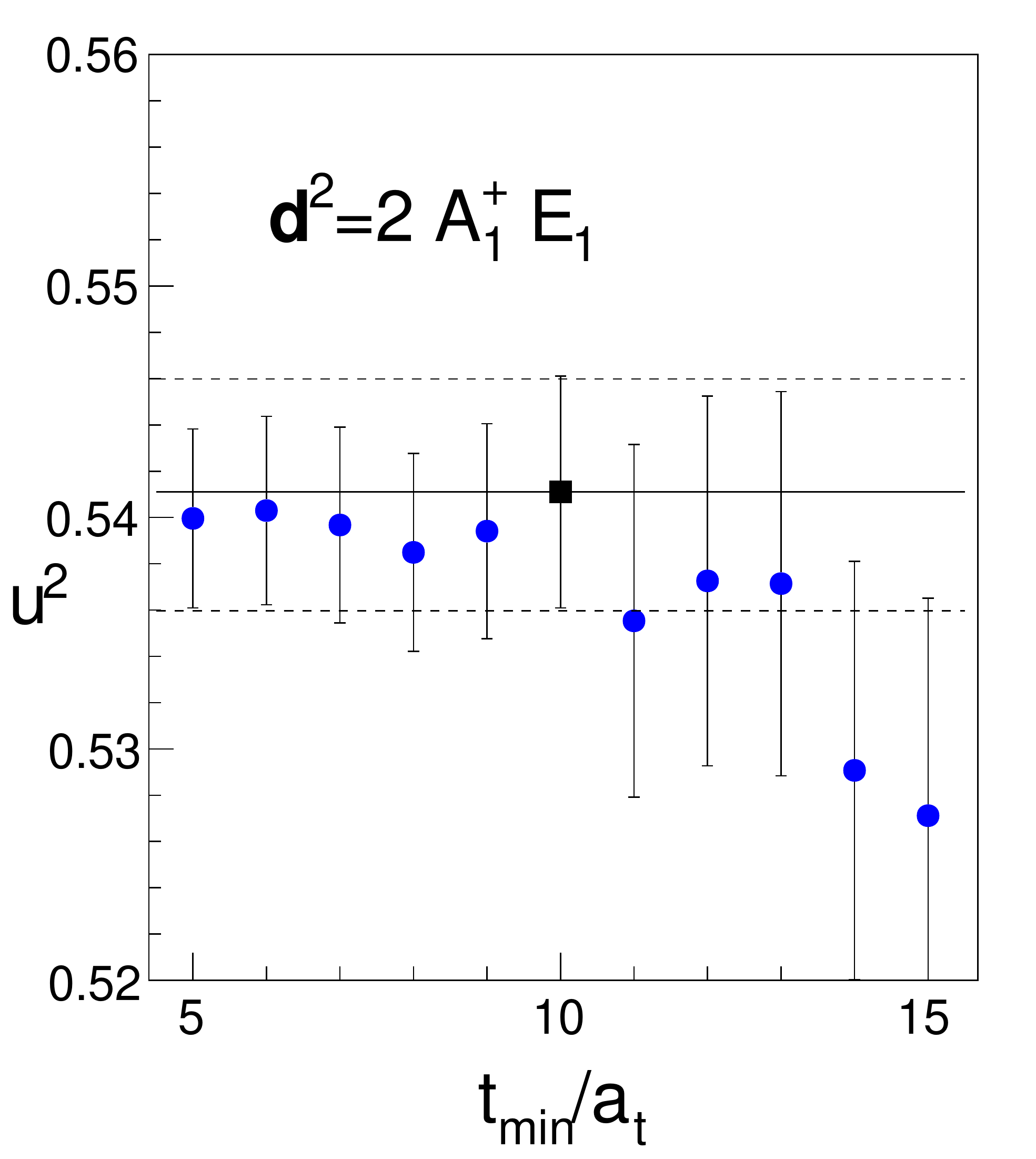}
	\caption{\label{f:i2d2}Same as Fig.~\ref{f:i2d0} but for   
$I=2$, $\boldsymbol{d}^2=2$.}
\end{figure}

\begin{figure}[!htb]
	\centering
\includegraphics[width=0.32\textwidth]{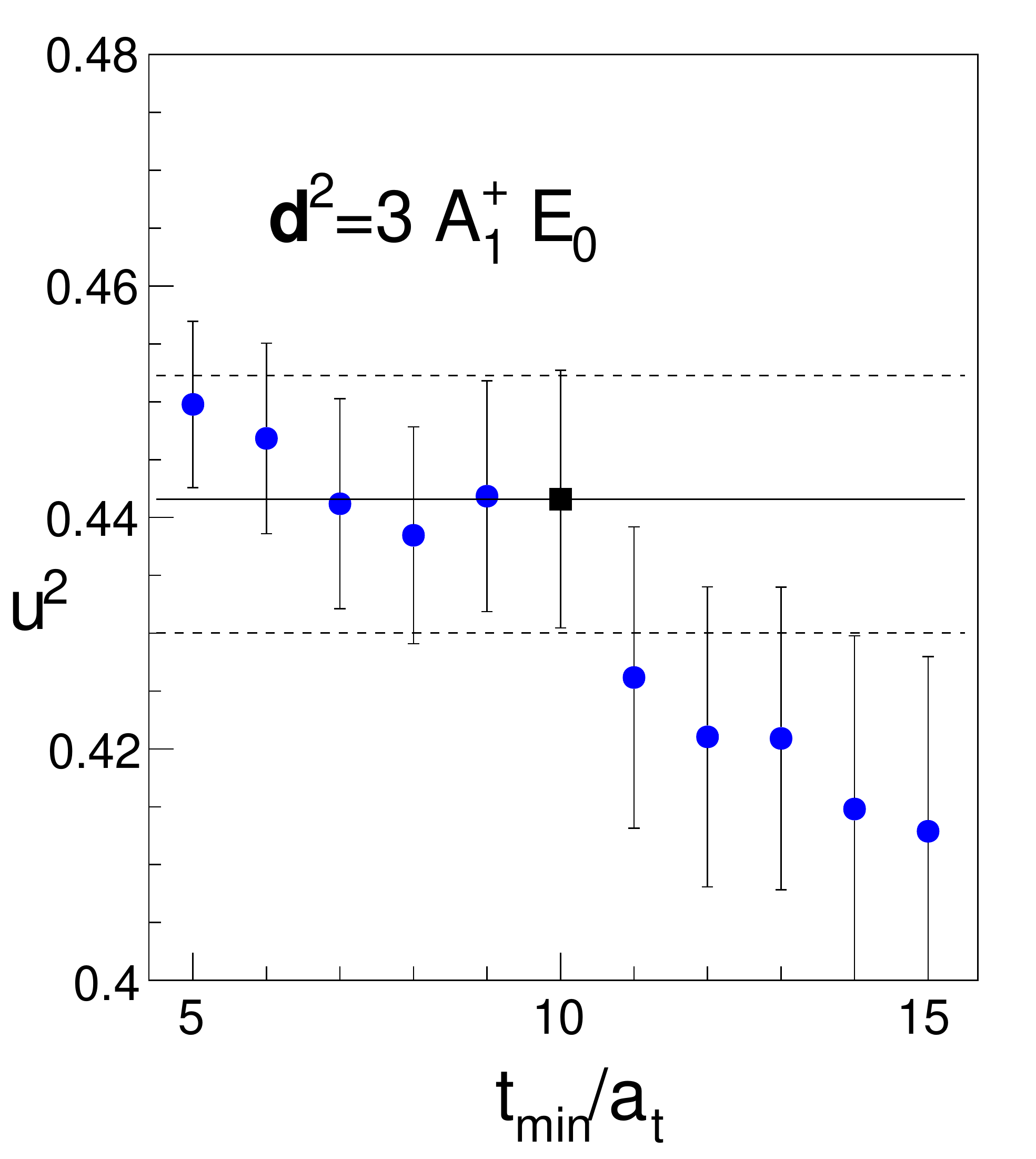}
\includegraphics[width=0.32\textwidth]{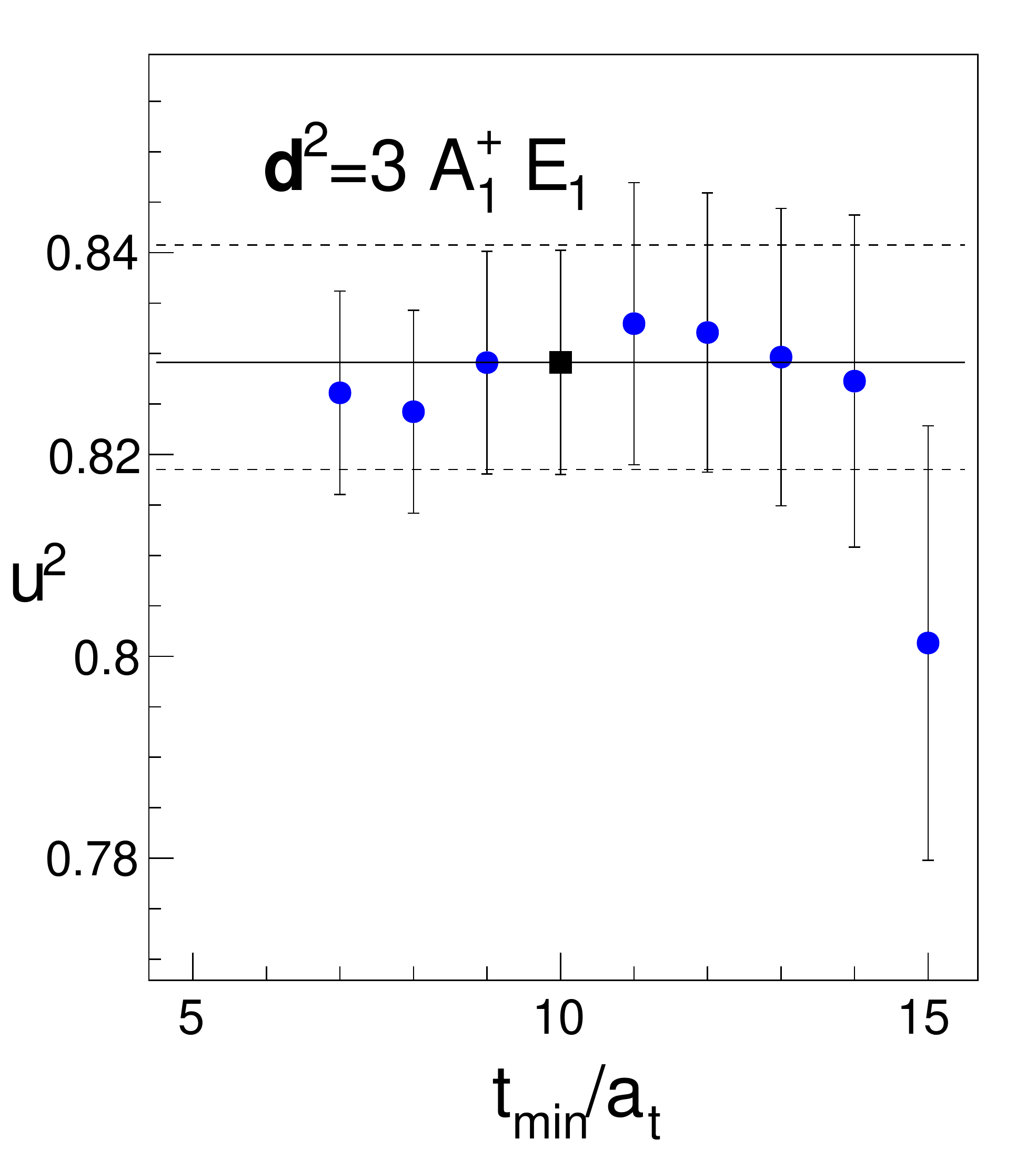}
	\caption{\label{f:i2d3}Same as Fig.~\ref{f:i2d0} but for   
$I=2$, $\boldsymbol{d}^2=3$.}
\end{figure}

As in App.~\ref{s:app1}, this appendix contains $t_{\mathrm{min}}$-plots for all finite volume energies
used in the determination of the $I=2$, $\ell=0$ scattering amplitude. They are shown in
Figs.~\ref{f:i2d0},~\ref{f:i2d1},~\ref{f:i2d2}, and~\ref{f:i2d3}. 
\begin{figure}[!htb]
\includegraphics[width=0.32\textwidth]{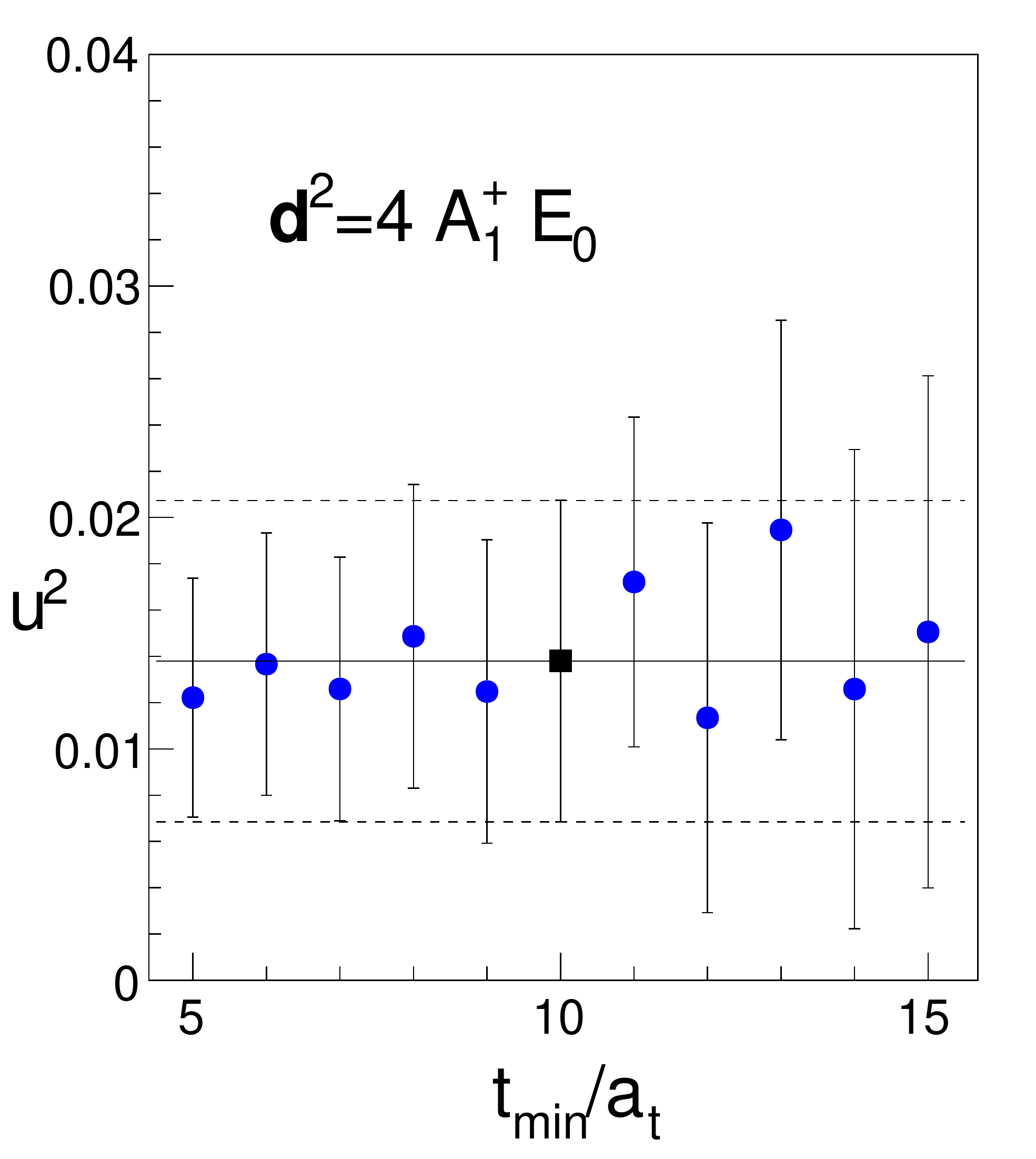}
\includegraphics[width=0.32\textwidth]{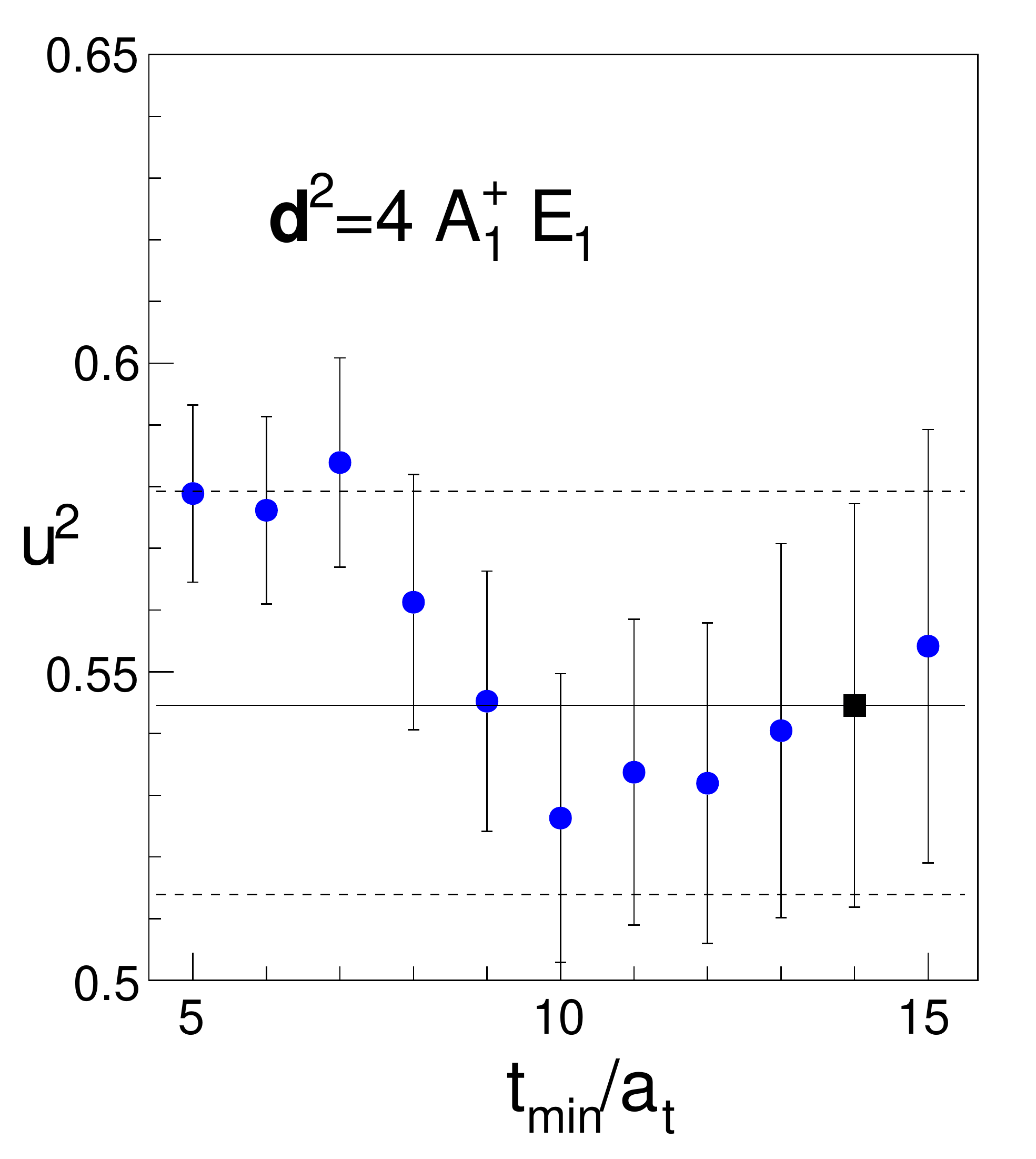}
\includegraphics[width=0.32\textwidth]{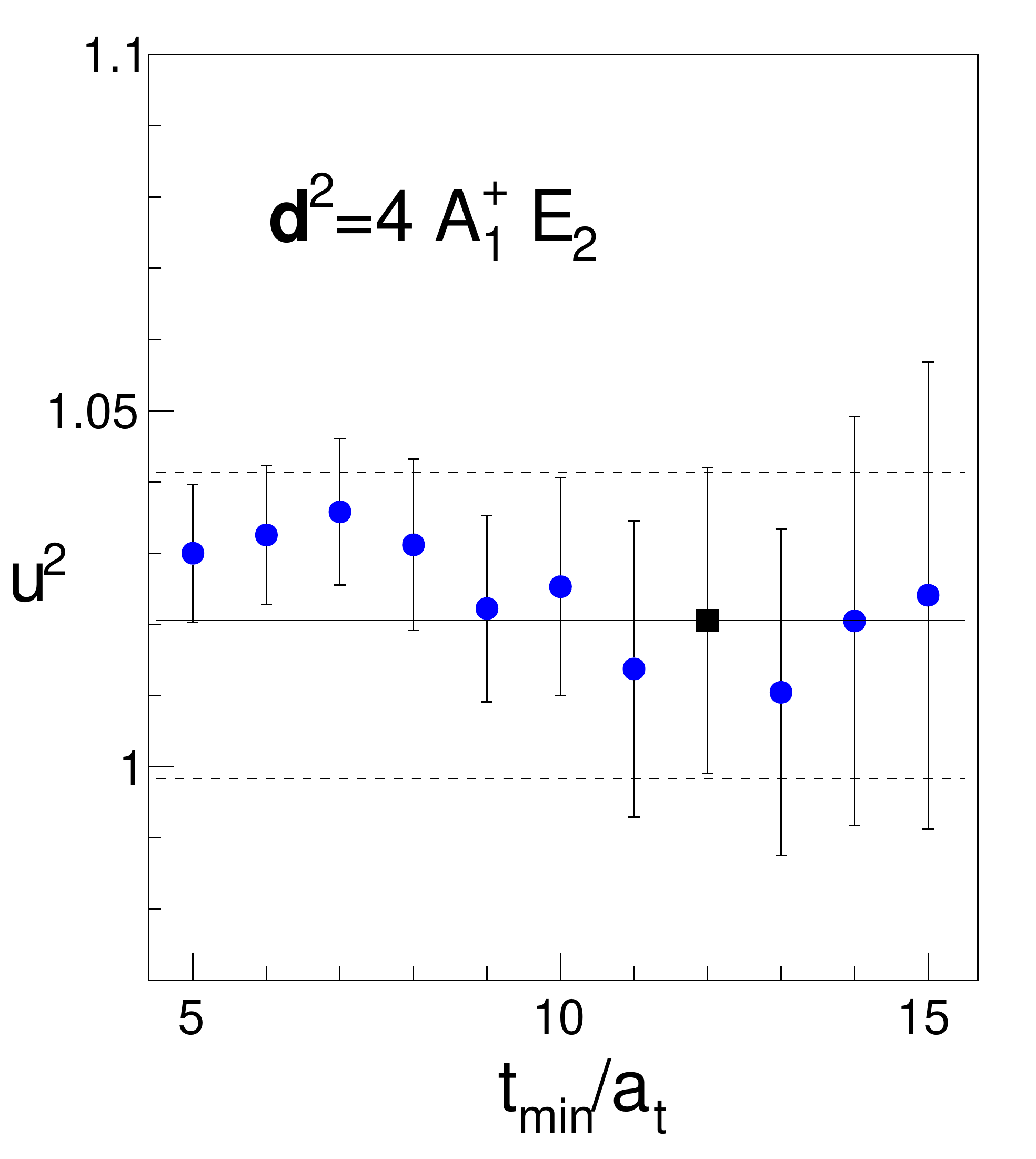}
	\caption{\label{f:i2d4}Plots showing the $t_{\mathrm{min}}$-dependence of 
the dimensionless center-of-mass momentum $u^2$ for $I=2$, 
$\boldsymbol{d}^2=4$.}
\end{figure}

%\input{app.tex}

% \bibliographystyle{style}   %if you use h-elsevier.bst
%\bibliography{jbulava2,latticen}           %or whatever your .bib file i
%\bibliography{latticen}           %or whatever your .bib file i

\end{document}